\documentclass[english,10pt]{article}
\usepackage[T1]{fontenc}
\usepackage[latin9]{inputenc}
\usepackage{color}
\usepackage{array}
\usepackage{multirow}
\usepackage{amsmath}
\usepackage{amssymb}
\usepackage{graphicx}
\usepackage{esint}
\usepackage[authoryear]{natbib}

\makeatletter

\providecommand{\tabularnewline}{\\}

\usepackage{multirow}
\usepackage{graphicx}

\usepackage{natbib}
\usepackage{graphicx}

\usepackage{float}
 
\usepackage{color} 
\usepackage[T1]{fontenc}
\usepackage{ae,aecompl}

\input glyphtounicode
\newsavebox{\astrutbox}
\sbox{\astrutbox}{\rule[-5pt]{0pt}{20pt}}

\usepackage{setspace}
\makeatother

\usepackage{babel}
\begin{document}


\title{Confined flow of suspensions modeled by a frictional rheology \thanks{Submitted to J. Fluid Mech. on Dec. 24, 2013, revised version July 10, 2014, accepted for publication Sept. 19, 2014}}

\author{Brice Lecampion$^1$
\thanks{Email for correspondence: blecampion@slb.com}, 
 Dmitry I. Garagash$^2$\thanks{Email for correspondence: garagash@dal.ca}
\\
\small{$^1$ Schlumberger, 1 cours du Triangle, 92936 Paris La Defense, France }
\\  
\small{$^2$ Department of Civil and Resource Engineering, Dalhousie University, Halifax, Canada }
}
\date{September 19, 2014}

\maketitle
\begin{abstract}
We investigate in detail the problem of confined pressure-driven laminar
flow of neutrally buoyant non-Brownian suspensions using a frictional
rheology based on the recent proposal of \citet{BoGu11}. The friction
coefficient (shear stress over particle normal stress) and solid volume
fraction are taken as functions of the dimensionless viscous number
$I$ defined as the ratio between the fluid shear stress and the particle
normal stress. We clarify the contributions of the contact and hydrodynamic
interactions on the evolution of the friction coefficient between
the dilute and dense regimes reducing the phenomenological constitutive
description to three physical parameters. We also propose an extension
of this constitutive framework from the flowing regime (bounded by
the maximum flowing solid volume fraction) to the fully jammed state
(the random close packing limit).

We obtain an analytical solution of the fully-developed flow in channel
and pipe for the frictional suspension rheology. The result can be
transposed to dry granular flow upon appropriate redefinition of the
dimensionless number $I$. The predictions are in excellent agreement
with available experimental results for neutrally buoyant suspensions,
when using the values of the constitutive parameters obtained independently
from stress-controlled rheological measurements. In particular, the
frictional rheology correctly predicts the transition from Poiseuille
to plug flow and the associated particles migration with the increase
of the entrance solid volume fraction.

We also numerically solve for the axial development of the flow from
the inlet of the channel/pipe toward the fully-developed state. The
available experimental data are in good agreement with our numerical
predictions, when using an accepted phenomenological description of
the relative phase slip obtained independently from batch-settlement
experiments. The solution of the axial development of the flow notably
provides a quantitative estimation of the entrance length effect in
pipe for suspensions when the continuum assumption is valid. Practically,
the latter requires that the predicted width of the central (jammed)
plug is wider than one particle diameter. A simple analytical expression
for development length, inversely proportional to the gap-averaged
diffusivity of a frictional suspension, is shown to encapsulate the
numerical solution in the entire range of flow conditions from dilute
to dense.
\end{abstract}

\section{Introduction}

Phenomenology of non-Brownian suspension flow distinguishes between
several regimes based on the value of the solid volume fraction $\phi$:
dilute, concentrated, and dense as the ``flowing'' limit $\phi\rightarrow\phi_{m}$
is approached \citep[e.g.][]{StPo05}. In the dilute regime $(\phi\lesssim0.2)$,
the mixture behaves as a Newtonian fluid whose apparent viscosity
is a slowly increasing function of the solid volume fraction \citep{Eins06,BaGr72},
and the normal stress components acting on the solid phase, hereafter
referred to as the particle or effective normal stresses, are negligible
\citep[e.g.][]{StPo05}. The upper bound of the dilute regime appears
to be related to the percolation threshold for a transient cluster
\citep{Dege79}. Above this threshold, i.e. in the concentrated regime,
suspension develops distinct compressive particle normal stress \citep{DeGa09,DbLo13},
and the apparent viscosity of the mixture increases significantly
with the solid volume fraction \citep{KrDo59}. No relative slip between
the fluid and solid phases for low Reynolds number flows has been
measured in this regime to the accuracy of the experimental methods
\citep{LyLe98a,IsBe10}. For yet larger solid volume fraction, approaching
the jamming/flowing transition $\phi\rightarrow\phi_{m}$, the apparent
viscosity diverges, and a finite frictional yield stress is approached
in this limit even when the base fluid is strictly Newtonian. The
reported values for the flowing limit of mono-disperse suspensions,
$\phi_{m}\approx0.585$ in the effective stress controlled experiments
of \citet{BoGu11} and $\phi_{m}\approx0.605$ from the local MRI
measurements of \citet{OvBe06}, are distinctly lower than the random
close packing value $\phi_{\mathrm{rcp}}\approx0.63-0.64$. Above
the flowing limit ($\phi>\phi_{m}$), the macroscopic solid velocity
is zero and the medium behaves as a porous granular solid, through
which the base fluid phase can percolate. In that limit, the relative
phase slip governs pore fluid pressure diffusion ``a la Darcy''
\citep{Bear72}. 

Recently, dense suspension rheology has been investigated experimentally
by \citep{BoGu11} under conditions of a simple shear flow at a fixed
applied particle normal stress $-\sigma_{n}'$ (with the convention
of positive stress in tension). Their results indicate that the suspension
exhibits an apparent shear-thinning behaviour under constant particle
normal stress conditions (i.e. the apparent viscosity decreases with
the shear-rate). This behaviour is to be contrasted with an apparent
Newtonian behaviour when the solid volume fraction is imposed and
the particle normal stress is allowed to vary (increase) with the
shear-rate. \citet{BoGu11} further show that suspensions, akin to
dry granular media, can be phenomenologically described by a viscoplastic
frictional rheology, fully characterized by the dependence of the
friction coefficient $\mu=\tau/(-\sigma_{n}')$ and the volume fraction
$\phi$ on a dimensionless viscous number \citep{CaNi05} 
\[
I=\frac{\eta_{f}\dot{\gamma}}{-\sigma_{n}'}
\]
which contrasts the relative magnitudes of the viscous shear stress
$\eta_{f}\dot{\gamma}$ ($\eta_{f}$ is the viscosity of the base
fluid and $\dot{\gamma}$ is the macroscopic shear rate) and the particle
normal stress $-\sigma_{n}'$, respectively. The friction coefficient
$\mu(I)$ evolves from a static value $\mu_{1}$ in the jamming limit
$I=0$ of a dense suspension to a diverging value $\sim I+\frac{5}{2}\phi_{m}I^{1/2}$
for large $I$ in a dilute flow. This evolution spans the range of
behaviors from the pressure-dependent frictional granular solid in
the dense limit, $\tau=-\mu_{1}\sigma_{n}'$, to the Newtonian fluid
in the dilute limit, $\tau=\eta_{f}\dot{\gamma}$.  

In this work, we review the frictional suspension rheology and consider
an extension from the flowing regime to the jammed, non-flowing state.
We point out that such an extension is necessary to model pressure-driven
suspension flows, which are characterized by the existence of a jammed
central ``plug'', and to our knowledge has not been explicitly recognized
in previous modeling attempts. We propose that non-flowing material
compacts with the decreasing stress ratio, from the maximum flowing
solid volume fraction $\phi_{m}$ at the flow threshold $\mu=\mu_{1}$
to the random close packing value $\phi_{rcp}$ at $\mu\sim0$. The
compaction of the non-flowing pack is formally similar to the dilatant/compactive
behavior of the flowing material. However, the driving mechanisms
behind the flowing and non-flowing compaction have to be distinct.
In the former, the macroscopic shear flow enables changes of the particle
pack, while in the latter, non-flowing packs, we hypothesize that
microscopic, ``in-cage'' particle rearrangements are enabled by
the velocity/pressure fluctuations in the surrounding flowing material.
Notwithstanding the origin of the fluctuations, the latter mechanism
appears to be similar to the compaction of a static granular pack
in tapping and cyclic shear experiments \citep[e.g. ][]{Knight95,Pouliquen03}.

We use the extended frictional rheology to obtain solution for pressure-driven
flow in a channel and a pipe of a non-Brownian, neutrally buoyant
suspension of hard mono-dispersed spheres in a Newtonian liquid under
the condition of negligible inertia. Such type of flows has been extensively
investigated both experimentally \citep{KaGo66,LeAc87,SiCho91,LyLe98a,HaMa97}
and theoretically with suspensions balance models \citep{NoBr94,MiSn95,FaMa02,MiMo06,DbLe13,RaLe08,Rama13}
among others. In such confined flow, the velocity profile transitions
from a Poiseuille to plug-like shape when the entrance solid volume
fraction increases. The frictional rheology has the advantage of combining
the existence of a yield stress, which value depends on the magnitude
of the particle normal stress, and the evolution of the shear and
normal viscosities with solid volume fraction, similar to suspension
balance models. It is therefore of interest to test its prediction
on pressure-driven flow for which experimental results for velocity
and solid volume fraction profiles across the gap are available in
the literature. We develop an analytic solution for the fully-developed
flow, and then study numerically how the flow evolves with distance
from the inlet of the channel or pipe towards the fully-developed
state. The predictions based on these solutions compare very well
to the published experimental results. Importantly, these comparisons
are drawn for the solutions which are devoid of any matching parameters,
i.e. the constitutive parameters of the frictional rheology are identified
independently from rheological experiments of \citet{BoGu11} and
a phenomenological description of the intrinsic solid-pack permeability
function (or, alternatively, sedimentation hindrance function) is
based on existing extensive compilations of fluidization and batch
settlement experiments \citep{GaAl77,DaAv85}.

To aid with navigating this paper's notation, we acknowledge that
starting with Section \ref{Sec:dev_flow} (formulation for flow in
a channel) and onwards, we make use of the \emph{normalized field
variables}, using scales defined by (\ref{s1})-(\ref{s3}). A recourse
back to the dimensional form of these variables, where not obvious,
is explicitly acknowledged.

\section{Continuum model for dense suspension/ wet granular media}

\subsection{Conservation laws}

As already defined, $\phi$ is the solid volume fraction, $\pmb{v}^{f}$
and $\pmb{v}^{s}$ are the local average fluid and solid Eulerian
velocities, respectively, and $\pmb{u}=\phi\pmb{v}^{s}+(1-\phi)\pmb{v}^{f}$
is the mixture velocity. We assume that both fluid and solid constituents
are incompressible (taken separately). The continuity equations for
the solid phase and the mixture are then, respectively, 
\begin{eqnarray}
\frac{\partial\phi}{\partial t}+\nabla\cdot(\phi\pmb{v}^{s}) & = & 0\label{eq:SolidContinuity}\\
\nabla\cdot\pmb{u}=\nabla\cdot(\pmb{q}+\pmb{v}^{s}) & = & 0\label{eq:BulkContinuity}
\end{eqnarray}
where a relative phase slip velocity (with respect to the solid velocity)
was introduced \citep{Bear72} 
\[
\pmb{q}=(1-\phi)\left(\pmb{v}^{f}-\pmb{v}^{s}\right)=\pmb{u}-\pmb{v}^{s}
\]

It will be convenient to also use an alternative form of the solid
phase continuity, which is referred to as the consolidation equation
in the porous media literature \citep[e.g.,][]{Bear72},
\begin{equation}
\frac{1}{\phi}\frac{d^{s}\phi}{dt}=\nabla\cdot\pmb{q}\label{consolidation}
\end{equation}
where $d^{s}\phi/dt=\partial\phi/\partial t+\pmb{v}^{s}\cdot\nabla\phi$
is the solid material time derivative.

\subsubsection{Balance of momentum, drag force}

Neglecting inertial terms, the balance of the mixture momentum, in
the absence of body forces, reduces to:
\begin{equation}
\nabla\cdot\pmb{\sigma}=\pmb{0}\label{bulk_stress}
\end{equation}
where the total stress tensor $\pmb{\sigma}=\pmb{\sigma}^{f}+\pmb{\sigma}'$
is the sum of the fluid $\pmb{\sigma}^{f}$ and particle (effective)
$\pmb{\sigma}'$ stress tensors \citep{Terza40}. In the remainder
of this paper, we will make use of the fluid ($p^{f}$), particle
($p'$), and mixture ($p=p^{f}+p'$) pressure, defined as the respective
mean stress value taken positive in compression, and of the mixture
stress-deviator tensor $\pmb{\tau}=\pmb{\sigma}+p\pmb{I}$. 

In addition to (\ref{bulk_stress}), the balance of momentum of either
the fluid or solid phases is needed to describe the two-phase continuum.
For the fluid phase (see \citep{Jack00} for more details),
\[
(1-\phi)\nabla\cdot\pmb{\sigma}^{f}-\pmb{F}=\pmb{0}
\]
where the force $\pmb{F}$ is the total interaction force between
the solid and fluid phases, besides the buoyancy $\phi\nabla\cdot\pmb{\sigma}^{f}$.
 For negligible Reynolds number, the latter is limited to drag force,
which is proportional to the phase slip velocity \citep{Bear72} 
\begin{equation}
\pmb{F}=(1-\phi)\frac{\eta_{f}}{k(\phi)}\pmb{q}\quad\text{with}\quad k(\phi)=a^{2}\kappa(\phi),\label{F}
\end{equation}
where $k$ is the intrinsic permeability of the solid particles assembly,
$\kappa$ its normalized form, $a$ the particle size (radius), and
$\eta^{f}$ the fluid viscosity. The combination of the fluid phase
balance of momentum and the expression for the drag force gives:
\begin{equation}
\pmb{q}=\frac{a^{2}\kappa(\phi)}{\eta_{f}}\nabla\cdot\pmb{\sigma}^{f}\label{eq:Q_law}
\end{equation}
which is further reduced to Darcy's law under the assumption of negligible
fluid shear stress ($\pmb{\sigma}^{f}\approx-p^{f}\pmb{I}$). For
granular porous media, a classic choice for the permeability dependence
on the solid volume fraction is given by the Kozeny-Carman law $\kappa(\phi)=(1-\phi)^{3}/45\phi^{2}$
\citep{Koze27,Carm37}.

In the suspension rheology literature, the balance of momentum for
the solid phase is used instead to derive an expression for the particle
slip velocity with respect to the bulk, $\pmb{v}^{s}-\pmb{u}=-\pmb{q}$,
\citep[e.g.,][]{Jack00}, which in the absence of body forces has
the form: 
\begin{equation}
-\pmb{q}=\frac{2a^{2}}{9\eta_{f}}\frac{f(\phi)}{\phi}\nabla\cdot\pmb{\sigma}'\label{eq:J_Law}
\end{equation}
where $f(\phi)$ is the sedimentation hindrance function, evaluated
in batch sedimentation ($\pmb{\sigma}'=\pmb{u}=0$) as the ratio of
the particle settling velocity in the suspension to the terminal settling
velocity of a single particle in clear fluid.

The two formalisms (equations (\ref{eq:Q_law}) and (\ref{eq:J_Law}))
are equivalent in view of (\ref{bulk_stress}) when the scaled permeability
$\kappa(\phi)$ is uniquely related to the hindrance function $f(\phi)$,
\begin{equation}
\kappa(\phi)=\frac{2}{9}\frac{f(\phi)}{\phi}\label{eq:Hindrance_Perm_Relation}
\end{equation}

\begin{figure}
\begin{centering}
\includegraphics[scale=0.5]{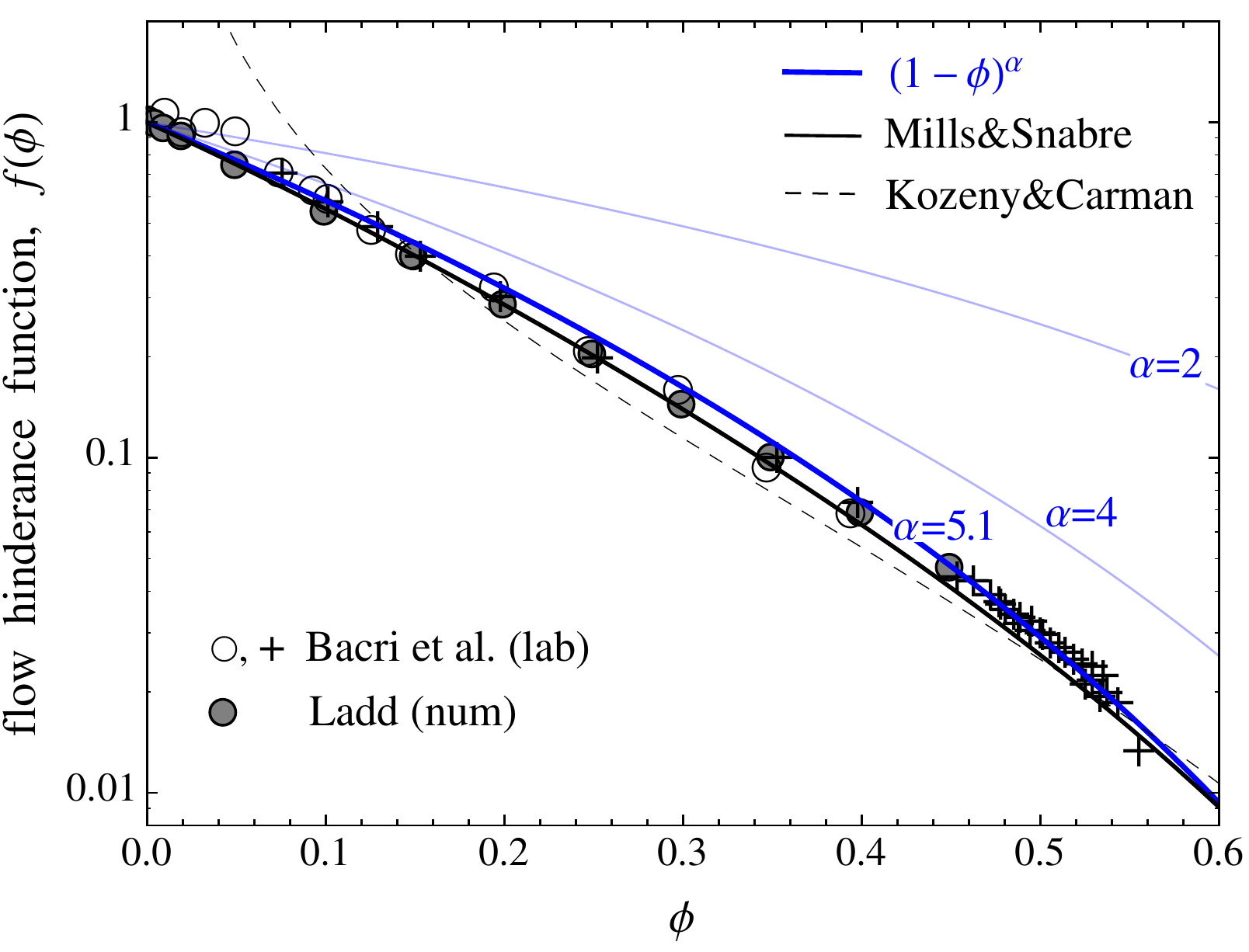}
\par\end{centering}

\caption{Different phenomenological expressions for the hindrance function
$f(\phi)$ in log-linear scale (with the corresponding normalized
permeability function given by $\kappa(\phi)=2f(\phi)/9\phi$) are
contrasted to the data from batch sedimentation experiments \citep{Bacri86},
and from the numerical simulations of the dispersion of hard spheres
\citep{Ladd90}. The \citet{RiZa54} empirical form, $f=(1-\phi)^{\alpha}$
with $\alpha=5.1$ \citep{GaAl77,DaAv85}, and \citet{MillsSnabre94}
theoretical expression, $f=(1-\phi)/(1+4.6\phi/(1-\phi)^{3})$, are
in excellent agreement with both the experimental and numerical data
over the entire (flowing) range of the solid volume fraction. Kozeny-Carman's
expression, $f=(1-\phi)^{3}/10\phi$, provides an adequate approximation
of the sedimentation data for all but dilute ($\phi\lesssim0.1$)
suspensions. \label{fig:PermModels}}
\end{figure}

A particularly simple empirical form of the hindrance function, $f(\phi)=\left(1-\phi\right)^{\alpha}$
with $\alpha=4.65$, was proposed by \citet{RiZa54} based on a number
of fluidization experiments at low Reynolds numbers. The value of
the exponent was later slightly revised to $\alpha=5.1$ by \citet{GaAl77}
(see also \citet{DaAv85}) based on the extensive study of fluidization
and batch settlement experimental data available at the time. Figure
\ref{fig:PermModels} exemplifies how remarkably well the Richardson-Zaki
phenomenology with $\alpha=5.1$ reproduces the experimental results
of \citet{Bacri86}, obtained using very accurate acoustic measurements
of sedimentation fronts%
\footnote{%
}, and the results of the direct numerical calculations of transport
properties from dispersion of hard spheres \citep{Ladd90}. The expressions
of Kozeny-Carman and \citet{MillsSnabre94}, respectively, as well
as the Richardson-Zaki relations with $\alpha=2$ and $\alpha=4$,
are also shown in figure \ref{fig:PermModels} for comparison purposes. 

\subsection{Frictional rheology\label{sec:fric}}

In the rheology of suspension, constitutive relations for the evolution
of the shear $\eta_{s}$ and normal $\eta_{n}$ viscosities of the
mixture as function of the solid volume fraction $\phi$ are often
used in so called suspension balance models \citep{MoBo99,ZaHi00,MiMo06}.
Here, we use the frictional rheology proposed by \citet{BoGu11} to
describe the constitutive behavior of dense suspension. Such a constitutive
model is akin to a frictional viscoplastic law. A similar frictional
framework has been successfully proposed for dry granular flow in
the liquid regime \citep[and references therein]{Midi04,FoPo08,JoFo06}. 

\begin{figure}
\noindent \begin{centering}
\includegraphics[scale=0.45]{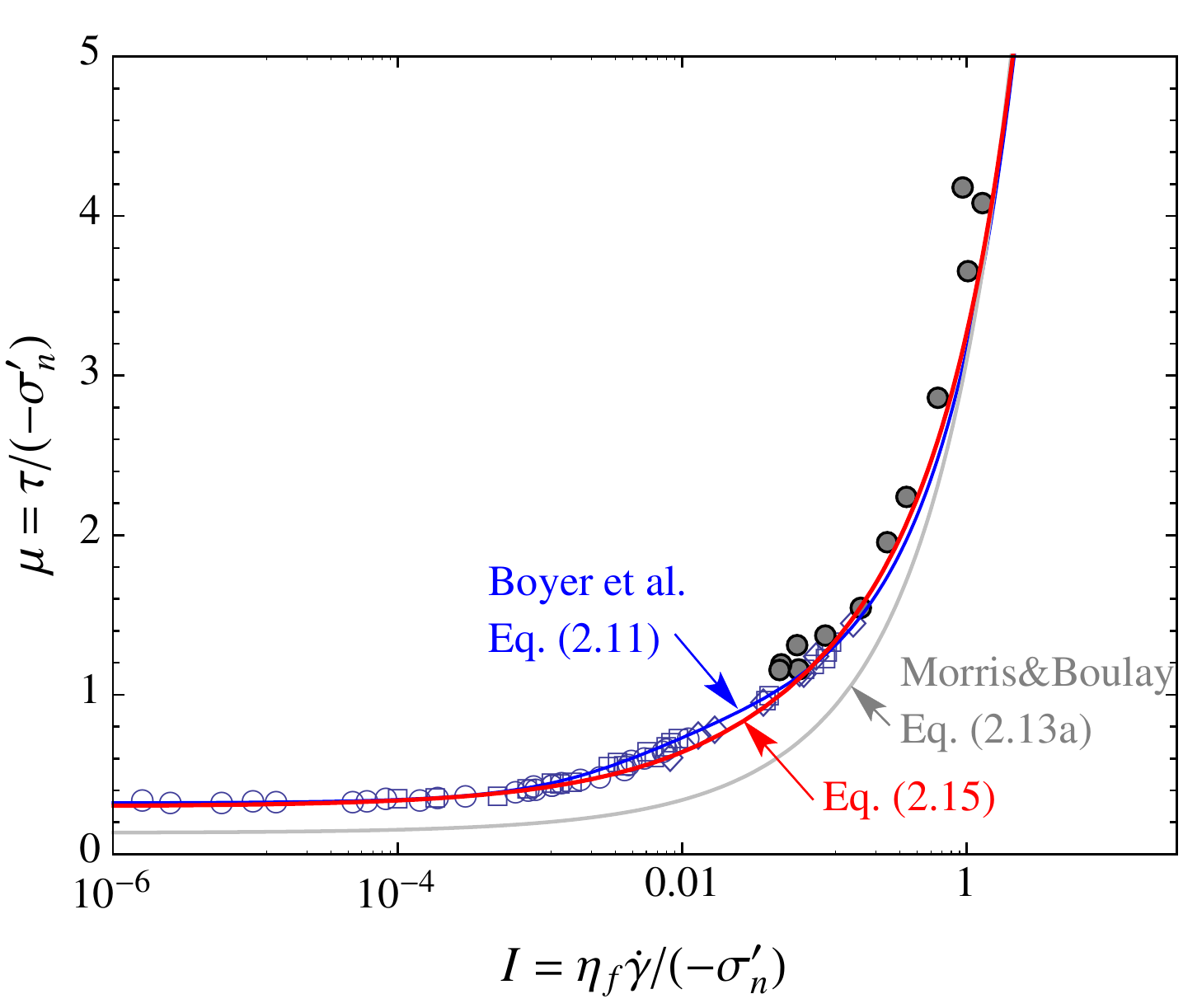}\enskip{}\includegraphics[scale=0.45]{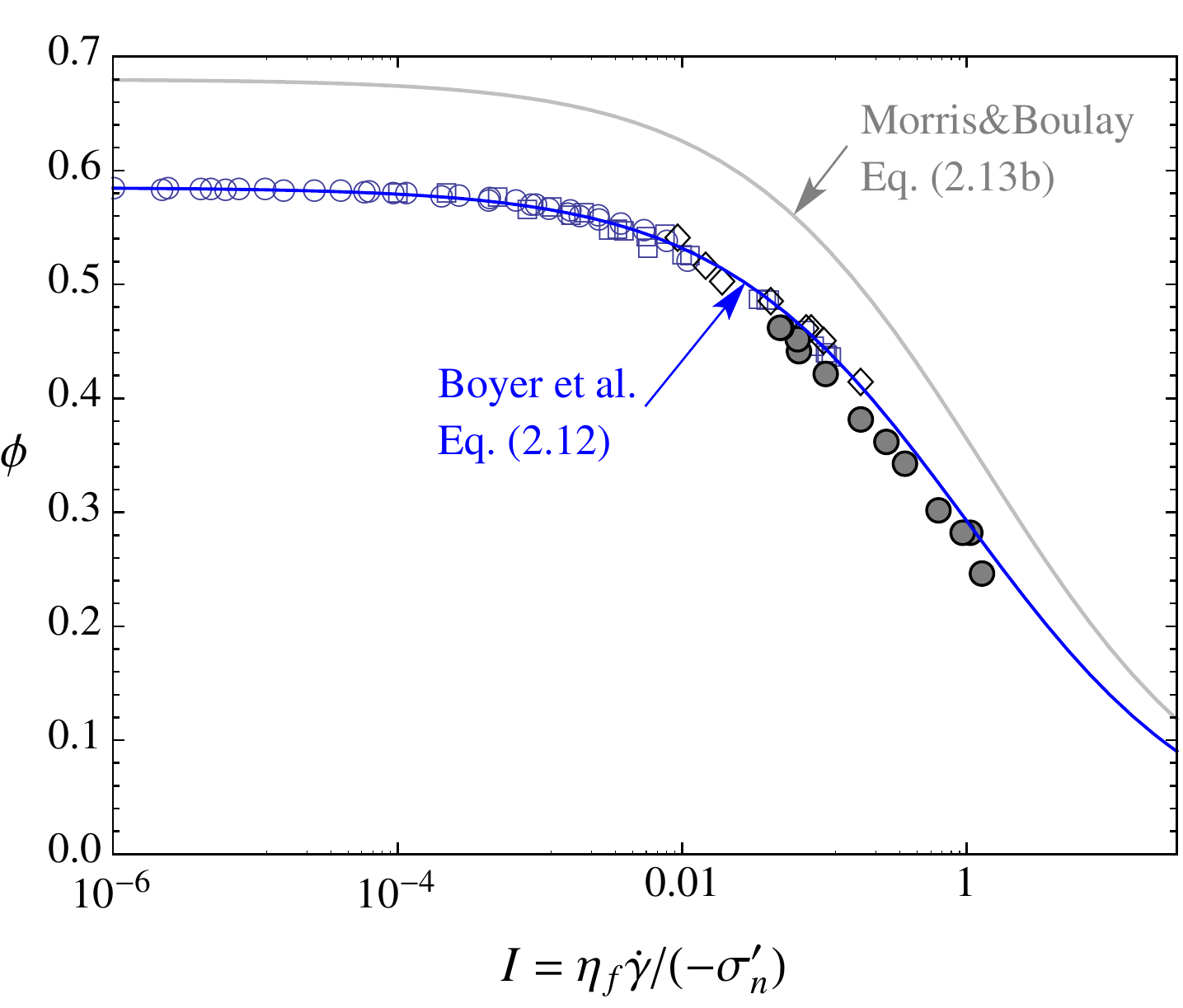}
\par\end{centering}

\caption{Selected frictional rheologies of a flowing suspension, which relate
the stress ratio / friction coefficient $\mu$, solid volume fraction
$\phi$, and the viscous number $I$, contrasted to experimental data
of \citet{BoGu11}, obtained for nearly mono-dispersed suspensions
in annular shear cell under effective stress control ($\mbox{\Large\ensuremath{\circ}}$,
$\mbox{\small\ensuremath{\square}}$, $\mbox{\Large\ensuremath{\diamond}}$),
and \citet{DbLo13}, obtained in a parallel plate rheometer with local
pressure measurements ($\mbox{\Large\ensuremath{\bullet}}$). \label{fig:rheology}}
\end{figure}

For such a complex two-phase fluid, the effective stress, shear stress,
shear rate, and solid volume fraction are intrinsically related, and
only two of these four field variables can be prescribed independently.
For a simple shear flow, \citet{BoGu11} write a frictional relation
for the  mixture shear stress and the effective (particle) confining
stress  and an evolution of the solid volume fraction as 
\begin{equation}
\tau=\mu(I)\,(-\sigma_{n}')\qquad\phi=\phi(I)\label{Friction}
\end{equation}
where macroscopic friction coefficient $\mu$ and the solid volume
faction $\phi$ are functions of the viscous number $I$, defined
as a ratio of the viscous shear stress in the fluid to the particle
confining stress, 
\begin{equation}
I=\frac{\eta_{f}\dot{\gamma}}{-\sigma'_{n}}\label{Iv}
\end{equation}
This number was originally proposed by \citet{CaNi05} based on a
micro-mechanical consideration of timescales controlling the solid
particles motion in a suspension. For large values of this dimensionless
number, the stress transmitted through\textbf{ }short-range particle
interactions (contacts, lubrication), which, we further refer to as
``contacts'' for brevity, is negligible compared to the fluid viscous
stress: hydrodynamics interactions dominate the behavior of the suspension.
This limit corresponds to the dilute regime. On the contrary, for
small $I$,  short-range particle interactions dominate the suspension
behavior and the macroscopic friction coefficient tends to a constant
thus defining an effective pressure-dependent yield stress. Based
on rheological experiments performed under effective normal stress
control on two different suspensions of mono-disperse spheres in Newtonian
fluid, \citet{BoGu11} propose the following phenomenological relation
for the friction coefficient as a function of the viscous number:
\begin{equation}
\mu(I)=\underbrace{\mu_{1}+\frac{\mu_{2}-\mu_{1}}{1+I_{0}/I}}_{\mu^{cont}}+\underbrace{I+\frac{5}{2}\phi_{m}I^{1/2}}_{\mu^{hydro}}\label{Mu}
\end{equation}
This law combines a contribution from particle contacts $\mu^{cont}(I)$
similar to the one reported for dry granular media and a hydrodynamic
contribution $\mu^{hydro}(I)$ designed to recover the behavior of
the dilute regime. The second constitutive equation relates the solid
volume fraction law to the viscous number $I$, and chosen by \citet{BoGu11}
in the following form 
\begin{equation}
\phi(I)=\frac{\phi_{m}}{1+I^{1/2}}\label{Phi}
\end{equation}
As can be seen from figure \ref{fig:rheology}, relations (\ref{Mu})-(\ref{Phi})
with $\mu_{1}=0.32$, $\mu_{2}=0.7$, $\phi_{m}=0.585$, and $I_{0}=0.005$
reproduce very well experimental results of \citet{BoGu11} and \citet{DbLo13}.

As shown by \citet{BoGu11}, a frictional rheology (\ref{Friction})
with (\ref{Iv})  is formally equivalent to suspension balance rheology
\citep[e.g.,][]{MoBo99,StPo05} 
\[
\tau=\eta_{f}\eta_{s}(\phi)\dot{\gamma}\qquad{{  \sigma'_{n}=-\eta_{f}\eta_{n}(\phi)\dot{\gamma}}}
\]
when the relative shear and normal viscosities are expressed as $\eta_{s}=\mu(I)/I$
and $\eta_{n}=1/I$, respectively, and $I=I(\phi)$, (\ref{Friction}).
The shear and normal viscosity functions of the solid volume fraction
proposed in the literature \citep[e.g.][]{KrDo59,MoBo99,BoGu11} diverge
when $\phi$ tends toward $\phi_{m}$ (i.e. when the shear rate tends
to zero), highlighting the presence of a frictional yield stress,
as intrinsically present in the frictional rheology. The ratio of
the two suspension viscosities defines the friction coefficient, i.e.
$\mu=\eta_{s}/\eta_{n}$. A frictional yield stress will thus be present
only if $\lim_{\phi\rightarrow\phi_{m}}\eta_{s}(\phi)/\eta_{n}(\phi)$
tends to a finite constant, i.e. if the two viscosities diverge at
the same rate $\sim1/I(\phi)$,e.g. $\sim(\phi_{m}-\phi)^{-2}$ per
(\ref{Phi}). 

Along the above lines, the rheology of \citet{MoBo99} \citep[see also][]{FaMa02,MiMo06}
provide an example of suspension-balance rheology characterized by
a finite frictional yield stress. Indeed, their rheology can be shown
to be equivalent to 
\begin{equation}
\mu(I)=\underbrace{\mu_{1}}_{\mu^{cont}}+\underbrace{I+\frac{5}{2}\phi_{m}\frac{I^{1/2}}{K_{n}^{1/2}}}_{\mu^{hydro}}\qquad\phi(I)=\frac{\phi_{m}}{1+K_{n}^{1/2}I^{1/2}}\label{MB}
\end{equation}
where $\mu_{1}=0.133$, $K_{n}=0.75$, and $\phi_{m}=0.68$ are used
by these authors to fit their model predictions to wide-gap Couette
flow data of \citet{Phillips92}. (Note that the authors make use
of the parameter $K_{s}=\mu_{1}K_{n}$ instead of $\mu_{1}$). The
rheology (\ref{MB}) is qualitatively similar to that of \citet{BoGu11}
(equations (\ref{Mu})-(\ref{Phi})) with one significant distinction
being in the contribution of particle contacts to friction ($\mu^{cont}$).
For the rheology of \citet{MoBo99}, $\mu^{cont}$ is a constant given
by the jamming value $\mu_{1}$, while for the experimentally-derived
rheology of \citet{BoGu11}, $\mu^{cont}$ is increasing with $I$
from the minimum value $\mu_{1}$ at $I=0$. Figure \ref{fig:rheology}
shows that, although the two rheologies are qualitatively similar,
the rheology of \citet{MoBo99} does not represent the experimental
data very well quantitatively, which can be tracked to their choice
of the jamming values of the solid volume fraction (overestimated
$\phi_{m}$) and the friction coefficient (underestimated $\mu_{1}$),
as well as the lack of dependence of their contact contribution $\mu^{cont}$
on the viscous number.

\subsubsection{Alternative form of the friction expression}

Although the functional form (\ref{Mu}) of the friction law $\mu(I)$
proposed by \citet{BoGu11} provides a very good match to the experimental
data, we do observe a slight inconsistency between their functional
form and the interpretation of $\mu^{cont}$ and $\mu^{hydro}$ as
the terms contributing to the total friction coefficient $\mu=\mu^{cont}+\mu^{hydro}$
from physically distinct ``contact'' and ``hydrodynamic'' interactions
between the particles in a suspension, respectively. (This distinction
may be blurred in the intermediate flowing regime, but is apparent
in the two end member regimes corresponding to $\phi\rightarrow\phi_{m}$
and $\phi\rightarrow0$, respectively). Specifically, the departure
of the friction coefficient from the \emph{jamming} limit ($\mu=\mu_{1}$
at $I=0$) with increasing shear rate in \emph{\citeauthor{BoGu11}}
framework, $\mu-\mu_{1}=\frac{5}{2}\phi_{m}I^{1/2}+O(I)$, is  given
by the \emph{Einstein}'s term. This appears to be at odds with the
physical origin of the \emph{Einstein}'s term which lies in the suspension's
\emph{dilute} limit.

\begin{figure}
\noindent \begin{centering}
(a)\hspace{-1em}\includegraphics[scale=0.44]{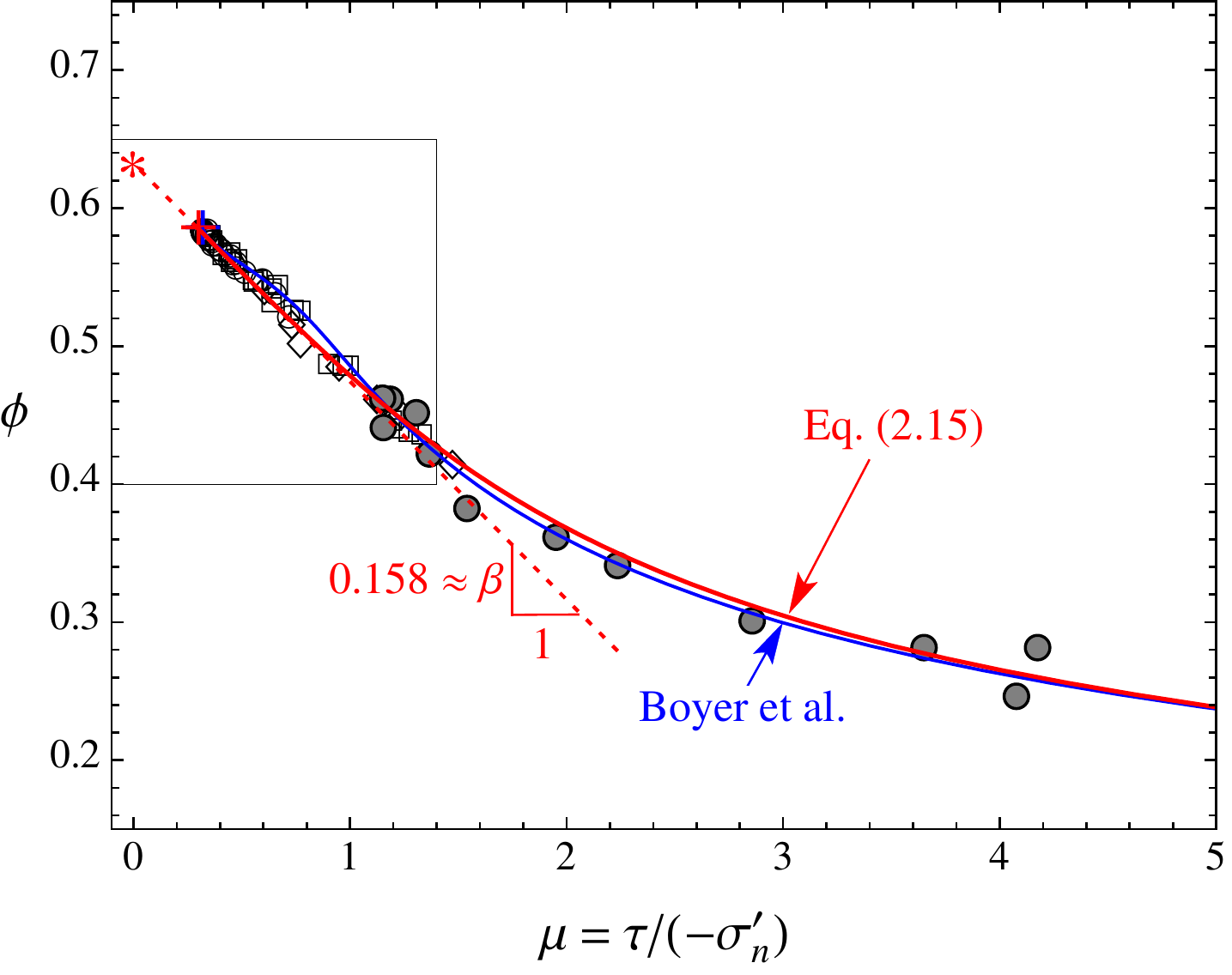}(b)\hspace{-1em}\includegraphics[scale=0.45]{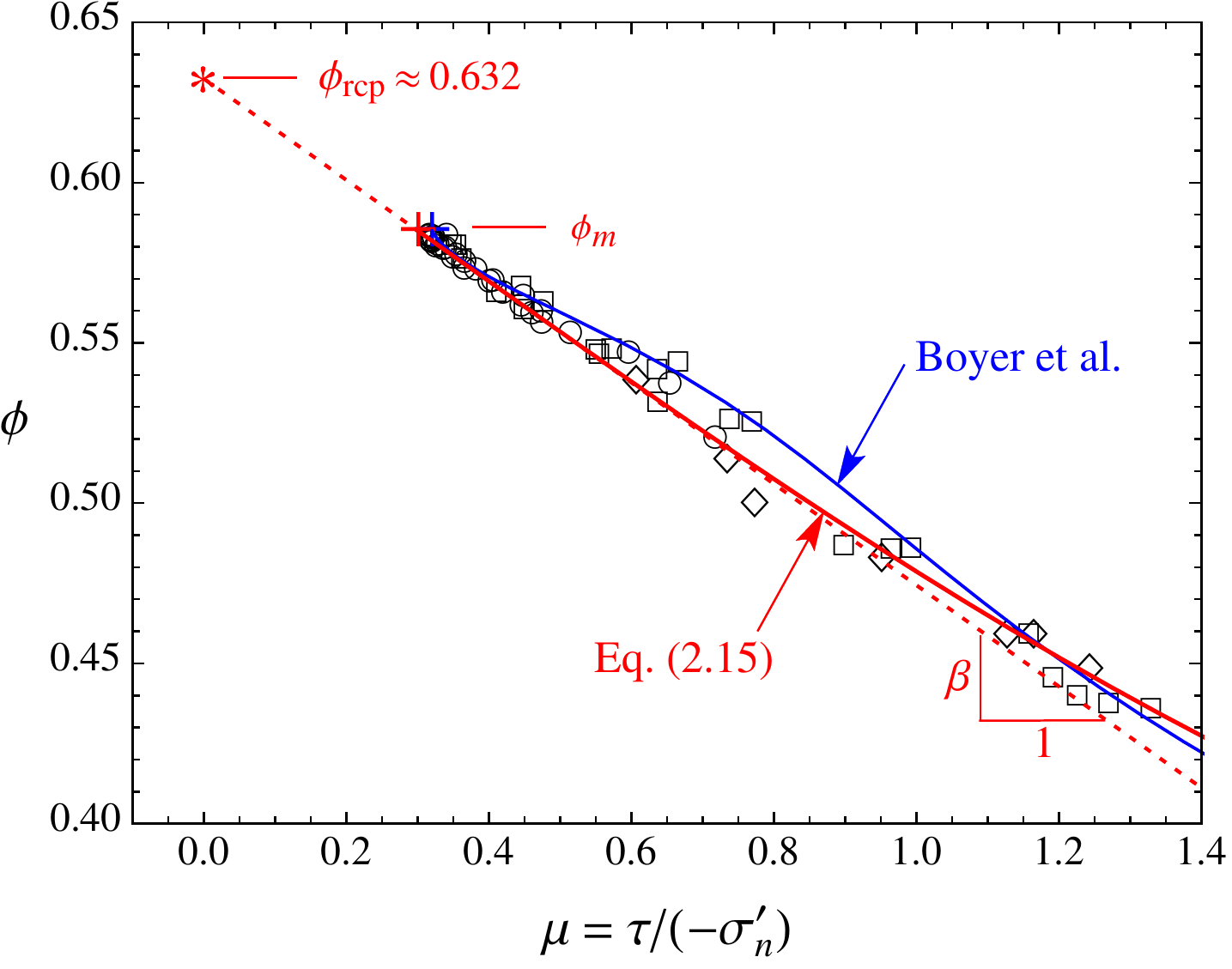}
\par\end{centering}

\caption{Solid volume fraction $\phi$ vs. stress ratio $\mu=\tau/{ (-\sigma_{n}')}$,
(a), for the selected frictional rheologies contrasted to the experimental
data of \citet{BoGu11} and \citet{DbLo13}. Dotted line shows a linear
fit with the slope of $\beta\approx0.158$ to the experimental data
in the dense regime (see also the close-up of this regime shown in
(b)), which is identified with the contribution of particle contacts
to friction, $\mu^{cont}(\phi)$, in rheology Eq. (\ref{mu_alt}),
shown by solid red line. The latter rheology allows simple (linear)
continuation into the jammed regime ($\phi_{m}<\phi<\phi_{rcp}$),
where solid fraction relaxes to the maximum value $\phi_{rcp}\approx0.632$
with either decreasing shear stress or increasing effective mean stress.\label{fig:rheo}}
\end{figure}

We suggest to model the contact dominated response of a dense suspension,
$\mu^{cont}$, by a simple linear dependence on the solid volume fraction
\begin{equation}
\mu^{cont}=\mu_{1}+\frac{\phi_{m}-\phi}{\beta}\label{mu_cont}
\end{equation}
where $\beta=-(d\phi/d\mu)^{cont}$ is a ``compressibility'' coefficient.
This linear relation with 
\[
\phi_{m}=0.585\qquad\mu_{1}=0.3\qquad\beta=0.158
\]
 provides an excellent approximation to the available data when recasted
onto the ($\phi,\mu$) plane on figure \ref{fig:rheo}. Importantly,
a linear relation between $\mu$ and $\phi$ in the dense regime has
also been corroborated for dense \emph{dry} granular media in numerical
2D simple shear experiments \citep{daCruz05,RognonRoux08} and in
the laboratory \citep{CraigBuckholz86}. A friction law, which is
not limited to the dense regime, is then put together similarly to
\citet{BoGu11} by adding a ``hydrodynamic'' interactions term to
the ``contact'' ones,
\begin{equation}
\mu(\phi)=\underbrace{\mu_{1}+\frac{\phi_{m}}{\beta}\left(1-\frac{\phi}{\phi_{m}}\right)}_{\mu^{cont}}+\underbrace{\left(I(\phi)+\left(\frac{5}{2}\phi_{m}+2\right)I(\phi)^{1/2}\right)\left(1-\frac{\phi}{\phi_{m}}\right)^{2}}_{\mu^{hydro}}\label{mu_alt}
\end{equation}
and adopting\textit{ \citeauthor{BoGu11}'s} relation (\ref{Phi})
between the viscous number and the solid volume fraction, 
\begin{equation}
I(\phi<\phi_{m})=(\phi_{m}/\phi-1)^{2}\quad\text{and}\quad I(\phi\ge\phi_{m})=0.\label{I_alt}
\end{equation}
We comment on the form of $\mu^{hydro}$ in (\ref{mu_alt}) which
is similar to the \textit{Boyer's} $\mu^{hydro}$ in the dilute limit
($\phi\rightarrow0$ or $I\rightarrow\infty$), which in itself is
equivalent to the \textit{Einstein's} correction. This dilute limit
is weighted by a quadratic prefactor $\sim(\phi_{m}-\phi)^{2}$ in
(\ref{mu_alt}) in order to enforce the dominance of the ``contacts''
interactions (\ref{mu_cont}) in the dense regime, i.e. $\mu=\mu^{cont}(\phi)+O(\phi_{m}-\phi)^{3}$.

\begin{figure}
\noindent \begin{centering}
\includegraphics[scale=0.5]{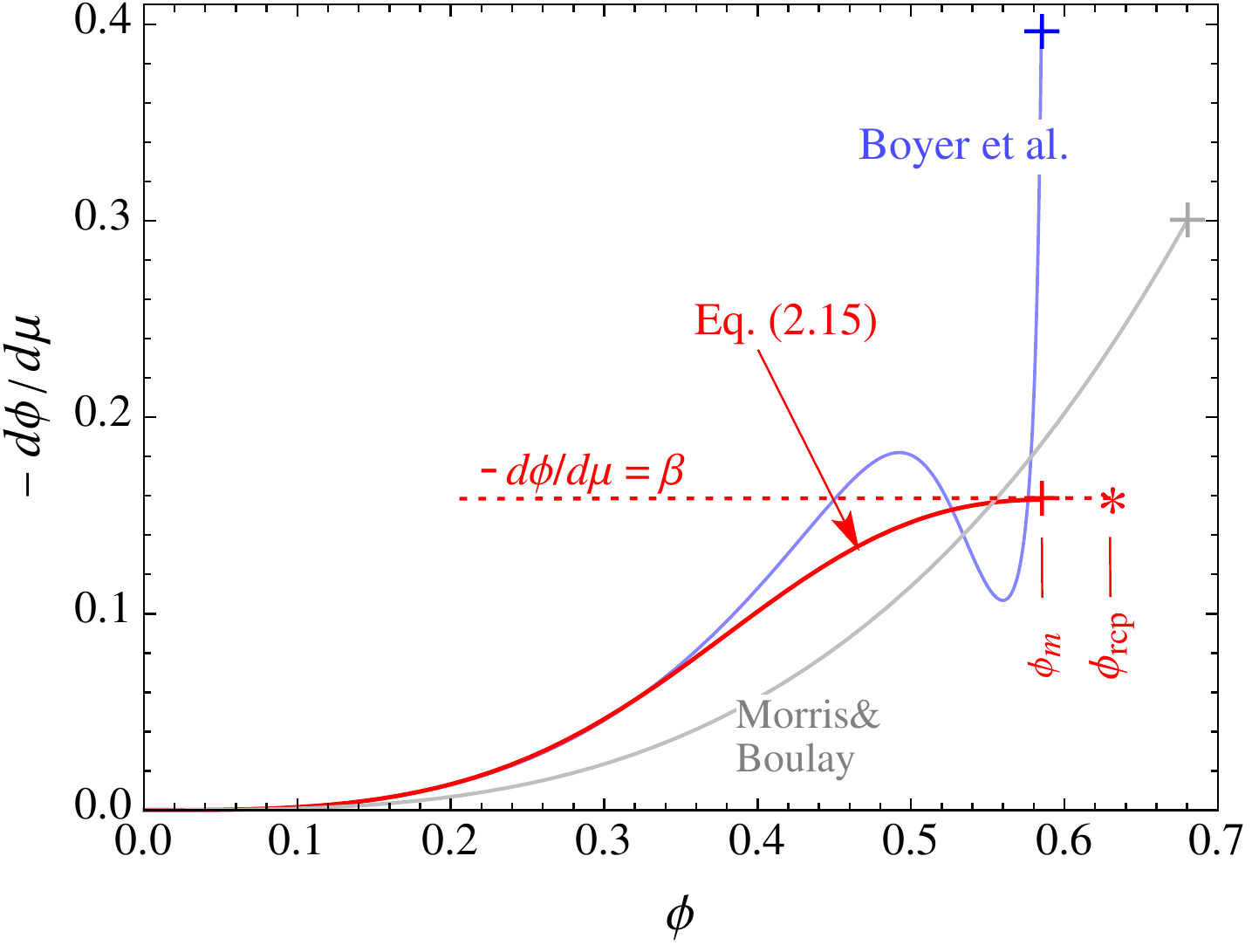}
\par\end{centering}

\caption{Inelastic ``compressibility'' parameter $(-d\phi/d\mu)$ as a function
of a flowing solid volume fraction $\phi<\phi_{m}$ for three frictional
rheologies, as in figure \ref{fig:rheology}. Corresponding jamming
points are shown by a cross. The modification of the Boyer et al.'s
rheology (Eq. (\ref{mu_alt}), red line) regularizes the latter in
the dense flowing regime, and allows simple continuation into the
jammed regime ($\phi_{m}<\phi<\phi_{rcp}$) using the asymptotic compressibility
$\beta=0.158$ (dotted line). \label{fig:dPhi}}
\end{figure}

Similarly to the original Boyer's rheological expression (\ref{Mu}),
the framework (\ref{mu_alt}) provides a very good match to the experimental
data (figures \ref{fig:rheology} and \ref{fig:rheo}), and, as to
be seen in Section \ref{Sec:dev_flow}, the two constitutive frameworks
yield close quantitative predictions for the fully-developed suspension
flow in a channel or a pipe for all but very dense flows (with average
$\phi\gtrsim0.5$). One can thus legitimately wonder whether this
``fine-tuning'' of the frictional rheology is at all necessary.
In fact, the advantage of the alternative frictional description becomes
fully apparent when considering the axial flow development. It is
to be shown in section \ref{sec:Axial-Flow-Developement} that, similarly
to previous studies using suspension balance models, the flow development
is driven by the cross diffusion of the particle normal stress. This
stress diffusion is governed by the ``hydraulic conductivity'' $k(\phi)/\eta_{f}$
and the ``compressibility'' parameter of the suspension $(-d\phi/d\mu)$.
The latter is shown on figure \ref{fig:dPhi} for the original \citet{BoGu11}
and the modified frictional models. Clearly, the expression (\ref{MB})
proposed by \citet{BoGu11} results in a seemingly degenerate behavior
of the compressibility parameter in the dense regime, which impacts
the flow axial development prediction and hinder the numerical treatment
of the problem.

\subsubsection{Extension to non-flowing state }

In writing constitutive equations in the form (\ref{mu_alt}-\ref{I_alt}),
we effectively extended the linear relation between the solid volume
fraction and the stress ratio $\mu$, found experimentally in the
dense flowing regime, to the jammed state ($\phi>\phi_{m}$). We can
find the corresponding maximum value of the solid fraction that can
be achieved if the stress ratio is allowed to vanish in this constitutive
description, $\phi\rightarrow\phi_{m}+\beta\mu_{1}\approx0.632\approx\phi_{rcp}$
as $\mu=-\tau/\sigma'_{n}\rightarrow0$ (where we have used $\phi_{m}=0.585$,
$\beta=0.158$, and $\mu_{1}=0.3$ as corroborated before). This value
is conspicuously similar to the random-close-packing value $\phi_{rcp}$
for mono-dispersed spheres, estimated in the range between 0.63 and
0.64 \citep[e.g.,][]{ScKi69,Berryman83}. We therefore \textit{hypothesize}
that the random close packing in the jammed part of otherwise flowing
granular system can be achieved by either increasing particle normal
stress(to infinity) while maintaining fixed, non-zero value of the
shear stress, or by decreasing the shear stress to zero while maintaining
a non-zero value of the particle normal stress. This response of the
model is not inconsistent with what we usually think of the relaxation
of the jammed solid fraction from an initially flowing state. In foreseen
applications of suspension flow in various geometries, the jammed
region of the flow is constantly excited by perturbations from the
nearby flowing region, similar to tapping, vibrations or particle
pressure / velocity fluctuations that are usually used to achieve
closed packing from an initially loose jammed state in experiments
\citep[e.g. ][]{Knight95,Pouliquen03}. Thus, we can infer with this
model that the jammed state subjected to  particle pressure fluctuations
will evolve to the solid volume fraction value defined by the imposed
macroscopic stress ratio, $\phi=\phi(\mu)$. We are not aware of any
laboratory or numerical experiments that would try to quantify the
dependence of the solid fraction in a jammed system of\emph{ }rigid\emph{
}particles on the imposed macroscopic $\mu$ (that would prove or
disprove our proposal), but this issue can be probed by looking at
solid fraction variation within a plug of otherwise flowing systems
(such as pressure or gravity driven flow in a channel or pipe). 
As we will show in Section \ref{sec:Comp}, when the resolution of
experimental methods allow \citep{HaMa97}, one can observe a solid
fraction gradient in the plug region of a pressure-driven flow consistent
with our jammed rheology and inferred values of the stress ratio gradient
across the plug. 

We also note that the random close packing has been approached in
the limit of non-zero stress ratio $\mu(\phi_{\mathrm{rcp}})\sim0.1$
in the numerical simulation of flowing (simple shear) dry granular
system of \emph{frictionless} particles \citep{PeyneauRoux08}. This
may indicate that our assumption of vanishing $\mu(\phi_{\mathrm{rcp}})$
may not be fully accurate. Accepting for a moment that $\mu(\phi_{\mathrm{rcp}})>0$,
would result in a \emph{finite} region at the random close packing
(where the stress ration is $0\le\mu\le\mu(\phi_{\mathrm{rcp}})$)
within the broader jammed plug ($0\le\mu\le\mu(\phi_{\mathrm{m}})$)
in a pressure-driven flow in a confined geometry. This is to be compared
to a single point where the random close packing is reached within
the plug when $\mu(\phi_{\mathrm{rcp}})=0$ is assumed. Resolution
of the solid volume fraction measurements within a plug in the existing
pressure-driven flow experiments is insufficient to eliminate either
possibility, and, thus, we settle for $\mu(\phi_{\mathrm{rcp}})=0$
in this work. 

Accounting for the inelastic compaction beyond the flowing regime,
suggests that formulation of the frictional rheology using the solid
volume fraction $\phi$, as the main state variable, i.e. $\mu=\mu(\phi)$
and $I=I(\phi)$, where $\phi$ spans both flowing ($I>0,\,\phi<\phi_{m}$)
and non-flowing ($I=0,\,\phi_{m}<\phi<\phi_{rcp}$) regimes, may be
a preferred form over the one using $I$ as the main state variable.
Equations (\ref{mu_alt}-\ref{I_alt}), which provide a particular
form of this rheology, contain only three independent parameters,
namely, values of the solid fraction, stress ratio, and compressibility
$(-d\phi/d\mu)$ at the jamming transition, $\phi_{m}$, $\mu_{1}$,
and $\beta$, respectively.  Furthermore, in view of the proposed
relation between these parameters and the random-close-packing limit,
the system can be alternatively parametrized by values of the solid
volume fraction and compressibility at random-close-packing%
\footnote{We assume, as previously, that jammed-value of compressibility is
a constant, independent of $\phi$ in the jammed range $\phi_{m}\le\phi\le\phi_{rcp}$.%
}, $\phi_{rcp}$ and $\beta$, respectively, and by $\phi_{m}$ (or
$\mu_{1}=(\phi_{rcp}-\phi_{m})/\beta$) at the jamming transition.

\subsubsection{Transient Effects}

The constitutive framework discussed so far pertains to a ``steady-state''
flow, which assumes that the internal relaxation time of the system
in response, for example, to a change of the macroscopic shear rate
is small compared to the timescale of this change. One such relaxation
process is transient inelastic dilatancy/contraction, as may occur,
for example, in the beginning or cessation of the flow. As inspired
by critical state soil mechanics \citet{Muir90}, the evolution of
the system to a new steady-state can be described by an evolution
law for the transient solid volume fraction development of the form
\citep{PaPo09} 
\begin{equation}
\frac{1}{\phi}\frac{d^{s}\phi}{dt}\propto\dot{\gamma}\left[\phi_{cs}(I)-\phi\right]\label{state_evol}
\end{equation}
where $\phi_{cs}(I)$ refers here to the ``critical-state'' solid
volume fraction given by (\ref{Phi}). For a pressure-driven flow
in a channel or a pipe, the macroscopic timescale of axial flow development
$t_{macro}\sim(L/H)/\dot{\gamma}$, where $H$ is the channel half-width
(pipe radius) and $L\sim H^{3}/a^{2}\gg H$ is the axial development
length \citep{NoBr94}. Consequently, the relaxation time $\sim1/(\phi\dot{\gamma})$
associated with state evolution (\ref{state_evol}) is negligibly
small compared to $t_{macro}$, pointing to the critical-state nature
of this flow.

\subsubsection{Normal Stress Differences }

Neutrally buoyant non-Brownian suspensions are known to develop normal
stress differences with increasing solid concentration \citep[see, for example,][]{ZaHi00,BoPo11,GaGa13,DbLo13}.
 In the case of channel flow, normal stress differences do not impact
the flow predictions when a frictional rheology linking the shear
stress to the particle confining stress is assumed (see developments
of Section \ref{sec:Formulation-channel-flow}). They may however
play a role in pipe flow, especially at elevated bulk values of the
solid volume fraction (see \citet{MiMo06,Rama13}, and Appendix \ref{sec:Pipe-Geometry}).
In our treatment of the pipe flow, to the first approximation, we
neglect the effect of normal stress differences, while leaving more
complete treatment of the problem to a future work.

\subsubsection{Choice of kinematic variables}

In the following, we will use the solid velocity $\pmb{v}^{s}$ and
the relative phase slip velocity $\pmb{q}$ as the kinematic variables.
Moreover, we express the frictional rheology as function of the solid
shear rate. The choice of the solid shear rate to describe the kinematic
of suspension deformation stems from the fact that in the dense regime,
near jamming it is the only acceptable one in order to define a jammed
state unambiguously while allowing the fluid to percolate through
the jammed medium. Moreover, for dilute suspensions, the distinction
between the fluid and solid shear rates is negligible for the case
of non-inertial flow: i.e. phase slip is small in the dilute regime.
We write thus the total stress in terms of the solid phase strain
rate and write the law for the relative flux vector $\pmb{q}$ neglecting
the fluid shear stress compared to the ``pore'' fluid pressure $p^{f}$
(as it is typically done in porous media flow analysis). For the
remainder of this paper, we thus drop the index ``$s$'' in the
solid velocity and in the corresponding material time derivative for
clarity and write: $\pmb{v}^{s}=\pmb{v}$ and $d^{s}/dt=d/dt$.

\section{Formulation of channel flow\label{sec:Formulation-channel-flow}}

We now turn to the case of pressure-driven Stokesian flow of a suspension
characterized by the previously described frictional rheology in a
channel.  Suspension flow in a circular pipe is amenable to a similar
method of solution, which details are given in Appendix \ref{sec:Pipe-Geometry}.

\subsection{Scaling}

We denote as $U_{0}$ a characteristic axial velocity (set here to
the entrance velocity value) of the flow in a channel with half-width
$H$, a characteristic axial length $L$, and a characteristic aspect
ratio of the channel (presumably small)
\begin{equation}
\delta=H/L\label{s0}
\end{equation}
We assume plane flow such that the velocity is given $\pmb{v}=v_{x}\pmb{e_{x}}+v_{y}\pmb{e_{y}}$,
where the $y$ coordinate denotes the direction perpendicular to the
channel axis. Since the constitutive laws for the suspension described
previously are incrementally akin to a Newtonian fluid, we will use
the classical Newtonian lubrication scaling (e.g., \citet{FrRy04}).
We therefore introduce the following kinematic scales
\begin{equation}
t_{*}=\frac{L}{U_{o}}\qquad x_{*}=L\qquad y_{*}=H\qquad(v_{x})_{*}=U_{o}\qquad(v_{y})_{*}=\delta\, U_{o}\qquad\dot{\gamma}_{*}=\frac{U_{o}}{H}\label{s1}
\end{equation}
shear/deviatoric stress ($\tau_{*}$), particle stress ($p_{*}'$),
total stress ($p_{*}$), and fluid pressure ($p_{*}^{f}$) scales
\begin{equation}
\tau_{*}=p_{*}'=\frac{\eta_{f}U_{o}}{H}\qquad p_{*}=p_{*}^{f}=\frac{\tau_{*}}{\delta}\label{s2}
\end{equation}
and relative phase velocity scale
\begin{equation}
q_{*}\equiv\frac{a^{2}}{\eta_{f}}\frac{p_{*}^{f}}{L}=\left(\frac{a}{H}\right)^{2}U_{o}\label{s3}
\end{equation}
This scaling reflects the expectation that the velocity component
across the channel, $v_{y}$, is much smaller (by $O(\delta)$) than
the axial velocity, while the relative phase flux is more effective
across the channel than along it. The stress scales suggest that the
shear stress for pressure-driven flows is much smaller than the normal
components of total stress and the pore pressure.

Finally, we choose the lengthscale $L$ to scale the axial flow development
length, the entrance length of the flow over which the shear-driven
particle migration across the channel leads to the fully-developed
state. Significant particle migration across the channel during flow
development requires that the relative phase cross-flux is comparable
to the solid cross-flux, $q_{y}\sim v_{y}$, or, in view of their
scales, (\ref{s1}) and (\ref{s3}), that aspect ratio $\delta=H/L$
is comparable to $(a/H)^{2}$. This is therefore equivalent to the
\citet{NoBr94} scaling argument prescribing the development lengthscale
in the form 
\begin{equation}
L=H^{3}/a^{2}\label{s4}
\end{equation}
which is equivalent to setting $\delta=H/L=(a/H)^{2}$.

Hereafter, we will make use of the \emph{normalized field variables},
using scales (\ref{s1}-\ref{s3}), while a recourse back to the dimensional
form of these variables, where not obvious, will be explicitly acknowledged.

\subsection{Normalized equations in scaling (\ref{s1}-\ref{s3})}

In the adopted scales, the normalized two component momentum balance
for the mixture becomes
\begin{eqnarray}
0 & = & \frac{\partial\tau_{xy}}{\partial y}-\frac{\partial p}{\partial x}+\delta\frac{\partial\tau_{xx}}{\partial x}\label{eq:Balance1}\\
0 & = & -\frac{1}{\delta}\frac{\partial p}{\partial y}+\frac{\partial\tau_{yy}}{\partial y}+\delta\frac{\partial\tau_{xy}}{\partial x}\label{eq:Balance2}
\end{eqnarray}
where both equations have been multiplied by $\delta$.

The expressions for the normalized components of the relative phase
slip vector reduce to:
\begin{eqnarray}
q_{x} & = & -\kappa(\phi)\frac{\partial p^{f}}{\partial x}=-\kappa(\phi)\left(\frac{\partial p}{\partial x}-\delta\frac{\partial p'}{\partial x}\right)=-\kappa(\phi)\frac{\partial p}{\partial x}+O(\delta)\label{qx}\\
q_{y} & = & -\frac{\kappa(\phi)}{\delta}\frac{\partial p^{f}}{\partial y}=-\kappa(\phi)\left(\frac{1}{\delta}\frac{\partial p}{\partial y}-\frac{\partial p'}{\partial y}\right)=-\kappa(\phi)\frac{\partial\sigma_{yy}'}{\partial y}+O(\delta)\label{qy}
\end{eqnarray}
where, in the second equation, we used (\ref{eq:Balance2}) to substitute
for $\partial p/\partial y$, and then substituted $\sigma_{yy}'+p'$
for $\tau_{yy}$.

The solid and mixture continuity equations become
\begin{eqnarray}
\frac{1}{\phi}\frac{d\phi}{dt} & = & \left(\delta\frac{\partial q_{x}}{\partial x}+\frac{\partial q_{y}}{\partial y}\right)\label{eq:SolidContinuity_Adim}\\
\frac{\partial v_{x}}{\partial x}+\frac{\partial v_{y}}{\partial y} & = & -\left(\delta\frac{\partial q_{x}}{\partial x}+\frac{\partial q_{y}}{\partial y}\right)\label{eq:MixtureContinuity_Adim}
\end{eqnarray}
where the scaled solid material time derivative retains the exact
form of its dimensional original, $d\phi/dt=\partial\phi/\partial t+v_{x}\partial\phi/\partial x+v_{y}\partial\phi/\partial y$.
Finally, the boundary conditions at the channel walls are 
\begin{equation}
q_{y}=v_{y}=0,\quad v_{x}=0\quad\text{at}\quad y=\pm1\label{bc}
\end{equation}
where the latter ($v_{x}$) condition can be relaxed to account for
finite wall slip velocity. The boundary conditions at the channel
entrance for a pressure driven flow are that for uniform axial velocity
and solid volume fraction across the gap and zero relative phase flux,
i.e.
\begin{equation}
v_{x}=v_{o}\,(=1),\quad\phi=\phi_{o},\quad q_{x}=0\quad\text{at}\quad x=0\label{bc0}
\end{equation}
 Global continuity allows to relate the entrance boundary conditions
to the gap-averages (accounting for the channel symmetry) $\left\langle \cdot\right\rangle =\int_{0}^{1}(\cdot)\text{d}y$
of profiles at a given location $x>0$ along the channel:
\begin{equation}
\left\langle v_{x}\right\rangle +\delta\left\langle q_{x}\right\rangle =v_{o},\qquad\left\langle \phi v_{x}\right\rangle =\phi_{o}v_{o}\label{eq:GlobalMixture_Adim}
\end{equation}

In the following, we consider a \emph{continuum (macroscopic) approximation}
($\delta=(a/H)^{2}\ll1$), which allows, to the first order, to set
$\delta=0$ in the above governing equations and boundary conditions%
\footnote{In view of the axial scale $L=H^{3}/a^{2}$ adopted to normalize the
equations, this approximation also implies that the channel is at
least as long as $L$. %
}.  Reduced momentum balance equations can be integrated, and accounting
for the symmetry, to yield linear shear stress and uniform mean stress
across the gap
\begin{equation}
\tau_{xy}=\left|\frac{\mbox{\ensuremath{\partial}}{ p}}{\partial x}y\right|\qquad p=p(x)\qquad(|y|<1)\label{mom}
\end{equation}

\section{Fully-developed flow\label{Sec:dev_flow}}

We first consider the fully-developed flow ($\partial/\partial x=0$)
which is expected to be reached for large normalized distances $x\gg1$
from the channel entrance, or, in dimensional terms for $x\gg L=H^{3}/a^{2}$,
(\ref{s4}).

\subsection{General solution \label{sub:General-solution}}

The solid and mixture volume balance equations for the fully-developed
flow reduce to $\partial\phi v_{y}/\partial y=\partial(v_{y}+q_{y})/\partial y=0$.
In light of the no-cross-flow boundary condition at the channel wall,
this leads to $v_{y}=q_{y}=0$ everywhere in the gap, and, therefore,
uniform fluid pressure $p^{f}=p^{f}(x)$ and effective normal stress
$\sigma'_{yy}=\sigma'_{yy}(x)$ (see (\ref{qy})) in a channel cross-section.

The frictional constitutive law with shear stress $\tau=\tau_{xy}$
and effective normal stress $\sigma'_{n}=\sigma'_{yy}$ components,
combined with the predicted linear shear stress distribution across
the gap 
\begin{equation}
\tau=\left|\frac{\mbox{\ensuremath{\partial}}p}{\partial x}y\right|=-\mu(\phi)\sigma'_{n}(x)\label{tau}
\end{equation}
leads to an expected conclusion that both the mean stress gradient
driving the flow and the effective normal stress are constant, independent
of position in the fully-developed channel flow, 
\[
\partial p/\partial x=\mathrm{const}\qquad\sigma'_{n}=\mathrm{const}
\]
Consequently, we can rephrase (\ref{tau}) as
\begin{equation}
\mu(\phi)=\mu_{\mathrm{w}}|y|\label{mu1}
\end{equation}
where
\begin{equation}
\mu_{\mathrm{w}}=\mu(\phi_{\mathrm{w}})={\displaystyle \frac{\left|\partial p/\partial x\right|}{-\sigma'_{n}}}\label{muw}
\end{equation}
is the wall friction (at $y=\pm1$), and $\phi_{\mathrm{w}}$ is the
corresponding wall value of the solid volume fraction.

The solid volume fraction profile $\phi(y)$ across the channel is
given implicitly by (\ref{mu1}). For the functional form of frictional
rheologies discussed in so far (e.g., equation (\ref{mu_alt})), one
can analytically invert (\ref{mu1}) for $\phi$. The resulting expression
is omitted here for brevity, but for the simpler expression in the
central plug, corresponding to   the linear jammed rheology, 
\begin{equation}
\phi(y)=\phi_{m}+\beta\,(\mu_{1}-\mu_{\mathrm{w}}|y|),\qquad|y|\leq y_{\mathrm{plug}}=\frac{\mu_{1}}{\mu_{\mathrm{w}}}\label{eq:Phi_plug}
\end{equation}
where $y=\pm y_{\mathrm{plug}}$ are the plug boundaries. 

The profile of the dimensionless viscous number $I$ in the flowing
part of the channel follows directly from that for $\phi$ (equation
(\ref{mu1})) by means of the constitutive relation (\ref{I_alt}).
In view of the expression for $I=-\left|\partial v_{x}/\partial y\right|/\sigma'_{n}$
and (\ref{muw}), the shear rate profile can be expressed as a multiple
of the driving total pressure gradient 
\[
\frac{\partial v_{x}}{\mbox{\ensuremath{\partial}}y}=\mathrm{sgn}(y)\frac{I(\phi(y))}{\mu_{\mathrm{w}}}\times\frac{\partial p}{\partial x}
\]
Integrating for the velocity profile with a no-slip condition at
the walls, and using substitution $\text{d}y=\text{d}\mu/\mu_{\mathrm{w}}$,
(\ref{mu1}), we can obtain
\begin{equation}
v_{x}(y)=-h(\phi(y))\times\frac{\partial p}{\partial x}\label{eq:VelocityProfile}
\end{equation}
where
\begin{align}
h(\phi) & =\frac{1}{\mu_{\mathrm{w}}^{2}}\int_{\phi}^{\phi_{\mathrm{w}}}I(\phi)\,\frac{\mbox{d}\mu}{\mbox{d}\phi}\mbox{ d}\phi,\label{h}
\end{align}
and, as introduced before, $\phi_{\mathrm{w}}$ and $\mu_{\mathrm{w}}$
are the wall values of $\phi$ and $\mu$, respectively. We note that
since $I(\phi\geq\phi_{m})=0$, both $h(\phi)$ and velocity $v_{x}$
are, as expected, uniform in the central plug ($|y|<y_{\text{plug}}$)
and given by their values at the plug boundary, $h(\phi_{m})$ and
$-h(\phi_{m})(\partial p/\partial x)$, respectively.

\subsection{Cross-sectional averages\label{sub:Cross-sectional-averages}}

We can use a similar ansatz to evaluate the gap-averaged axial velocity:
\begin{equation}
\left\langle v_{x}\right\rangle =-\left\langle h\right\rangle \times\frac{\partial p}{\partial x}\label{eq:AverageVelocity}
\end{equation}
where integrating separately over the plug and the flowing part, and
using substitution $\text{d}y=\text{d}\mu/\mu_{w}$ in the latter,
we can write for $\left\langle h\right\rangle $ from (\ref{h})
\begin{equation}
\left\langle h\right\rangle =\frac{1}{\mu_{\mathrm{w}}^{3}}\int_{\phi_{m}}^{\phi_{\text{w}}}I(\phi)\,\mu(\phi)\,\frac{\mbox{d}\mu}{\mbox{d}\phi}\mbox{ d}\phi\label{<h>}
\end{equation}
$\left\langle h\right\rangle $ can be seen as a dimensionless gap-averaged
fluidity accounting for the geometrical effect of channel flow. For
a Newtonian fluid, which is the large shear rate limit of the frictional
rheology: $\left\langle h\right\rangle =1/3$. 

Similarly, the gap-average of the solid volume fraction is obtained
as:
\begin{equation}
\left\langle \phi\right\rangle =\frac{1}{\mu_{\mathrm{w}}}\int_{\phi_{rcp}}^{\phi_{\text{w}}}\phi\frac{\mbox{d}\mu}{\mbox{d}\phi}\mbox{ d}\phi\label{<phi>}
\end{equation}
where $\phi_{rcp}=\phi_{m}+\beta\mu_{1}$. Evaluating the part of
the integral over the plug using (\ref{eq:Phi_plug}) allows to further
write 
\begin{equation}
\left\langle \phi\right\rangle =\frac{\mu_{1}}{\mu_{\mathrm{w}}}\left\langle \phi\right\rangle _{\text{plug}}+\frac{1}{\mu_{\mathrm{w}}}\int_{\phi_{m}}^{\phi_{\text{w}}}\phi\frac{\mbox{d}\mu}{\mbox{d}\phi}\mbox{ d}\phi,\qquad\left\langle \phi\right\rangle _{\text{plug}}=\frac{\phi_{m}+\phi_{rcp}}{2}\label{<phi>-1}
\end{equation}

\subsection{Solution for imposed entrance velocity and solid volume fraction}

The two global continuity equations provide relations (\ref{eq:GlobalMixture_Adim})
to be solved for $\partial p/\partial x$ and the effective normal
stress $\sigma'_{n}$ given velocity $v_{o}$ and the entrance volume
fraction $\phi_{o}$. Specifically, for the fully developed flow,
(\ref{eq:GlobalMixture_Adim}) reduces to $\left\langle v_{x}\right\rangle =v_{o}$
and $\left\langle \phi v_{x}\right\rangle =\phi_{o}v_{o}$, allowing
to write 
\begin{equation}
\phi_{o}=\frac{\left\langle \phi v_{x}\right\rangle }{\left\langle v_{x}\right\rangle }\label{cond1}
\end{equation}
The right hand side is a function of $\phi_{\mathrm{w}}$ (or $\mu_{\mathrm{w}}=\mu(\phi_{\mathrm{w}})$)
only, which can be evaluated, similarly to other gap-averages evaluated
so far, as follows 
\begin{equation}
\frac{\left\langle \phi v_{x}\right\rangle }{\left\langle v_{x}\right\rangle }=\frac{1}{\mu_{\mathrm{w}}\left\langle h\right\rangle }\int_{\phi_{rcp}}^{\phi_{\mathrm{w}}}\phi h(\phi)\frac{\mbox{d}\mu}{\mbox{d}\phi}\mbox{ d}\phi\label{ratio}
\end{equation}
where $h(\phi)$ and $\left\langle h\right\rangle $ are given by
(\ref{h}) and (\ref{<h>}), respectively. Similarly to (\ref{<phi>-1}),
one can expand (\ref{ratio}) by evaluating the part of the integral
over the plug. 

It is important to note that the integrals (\ref{h}), (\ref{<h>}),
(\ref{<phi>}) and (\ref{ratio}) in the fully-developed solution
can be obtained analytically for particular sets of rheological functions,
$\mu(\phi)$ and $I(\phi)$, discussed in so far.  The resulting
lengthy expressions are omitted here for brevity.

Eqs. (\ref{cond1}-\ref{ratio}) establish $\phi_{\mathrm{w}}$ as
an implicit function of $\phi_{o}$, which is shown on figure \ref{fig:<phi>}
for fully-developed channel and pipe flows, as amount of dilution
at the wall, $\phi_{o}-\phi_{\text{w}}$, vs. $\phi_{o}$. Since the
average particle concentration $\left\langle \phi\right\rangle $
(equation (\ref{<phi>})) and the half-plug $y_{\text{plug}}=\mu_{1}/\mu_{\mathrm{w}}$
are functions of $\phi_{\mathrm{w}}$ only (which in turn is a function
of $\phi_{o}$), they are completely defined by the entrance concentration
$\phi_{o}$ (figures \ref{fig:<phi>} and \ref{fig:plug}) and independent
of the flow rate $v_{o}$ or the driving stress gradient. This is
in line with experimental observations which reported the independence
of the scaled velocity profiles with respect to the flow rate for
all $\phi_{o}$.

Knowing the wall value $\phi_{\text{w}}$ as a function of the entrance
value $\phi_{o}$ of particle concentration and the mean velocity
$\left\langle v_{x}\right\rangle =v_{o}$, the fully-developed flow
problem is completely resolved. Indeed, since the normalized tangent
fluidity $\left\langle h\right\rangle $ in (\ref{eq:AverageVelocity})
is uniquely in terms of $\phi_{\mathrm{w}}$, the normalized total
stress gradient driving the flow is evaluated from the mean velocity
($v_{o}=1$) as $\partial p/\partial x=-1/\left\langle h\right\rangle $.
The corresponding fully-developed value of the normalized particle
normal stress follows from (\ref{muw}), $-\sigma'_{n}=1/{\displaystyle (\mu_{\mathrm{w}}\left\langle h\right\rangle )}$.

\subsection{Results \label{sec:Results}}

We now examine salient features of the fully-developed flow solution
by making use of the particular frictional rheology (\ref{mu_alt})-(\ref{I_alt})
with the laboratory-constrained values of the constitutive parameters,
$\phi_{m}=0.585$, $\mu_{1}=0.3$, and $\beta=0.158$. 

\begin{figure}
\noindent \begin{centering}
\includegraphics[scale=0.55]{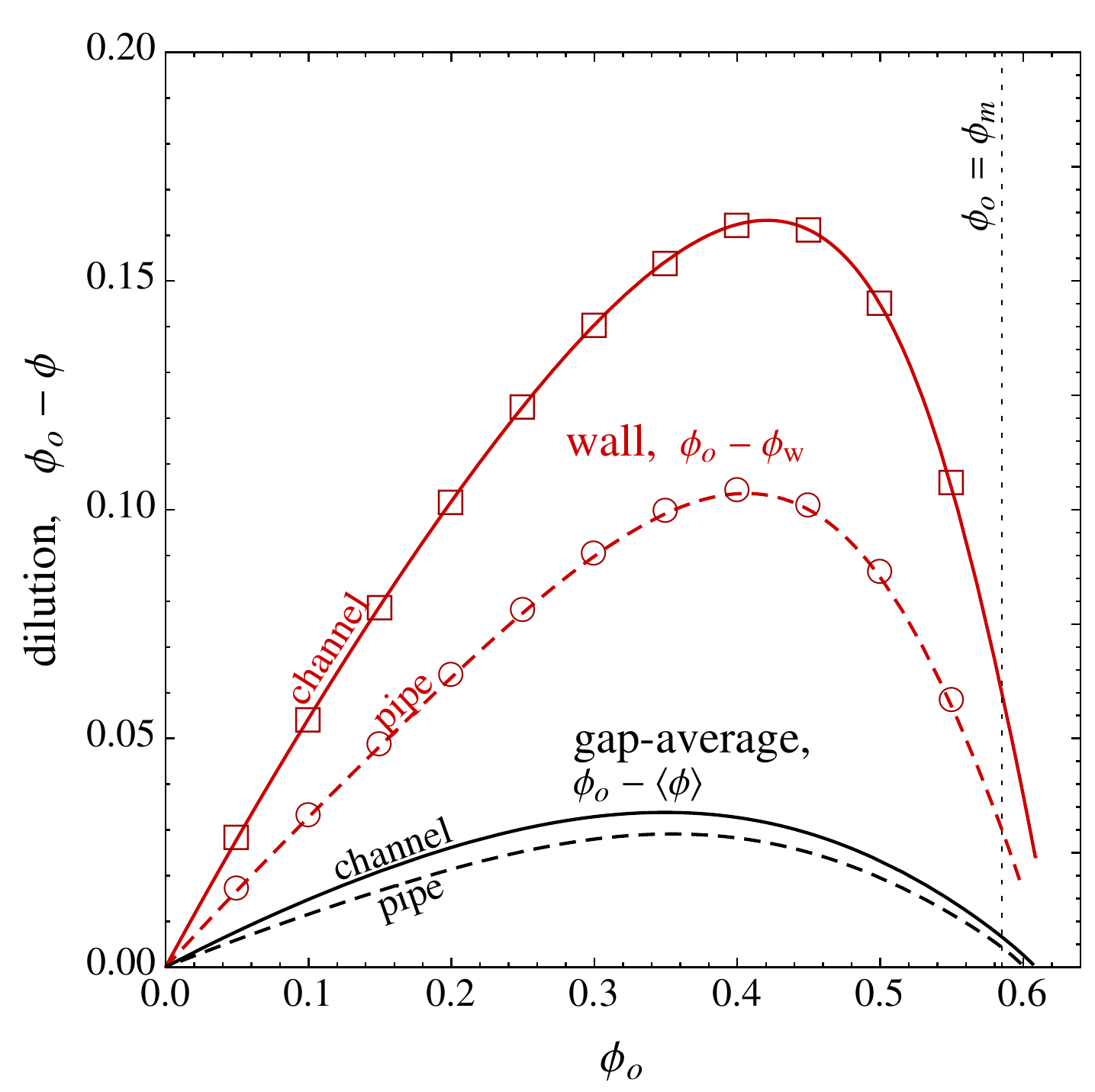}
\par\end{centering}

\caption{Dilution: change of the wall $\phi_{\text{w}}$ and the gap-average
$\left\langle \phi\right\rangle $ values of the solid volume fraction
from its entrance value $\phi_{o}$, in the fully-developed channel
(solid lines) and pipe (dashed lines) flow. The symbols show the fully-developed
limit of the dilution at the wall in the numerical solutions for axial
flow development (section \ref{sec:Axial-Flow-Developement}). \label{fig:<phi>}}
\end{figure}

Figure \ref{fig:<phi>} shows dilution, defined as the difference
of the fully-developed value of particle concentration from the entrance
value, $\phi_{o}-\phi$, as a function of $\phi_{o}$, both locally
at the channel wall and as the gap-average value. We find a small
amount of gap-average dilution in the well-developed flow, i.e. $\left\langle \phi\right\rangle $
always somewhat smaller than the entrance value $\phi_{o}$. The dilution
is larger in the channel than in a pipe, with the maximum gap-average
values of dilution, 0.034 and 0.029, occurring for $\phi_{o}\approx0.35$
in the channel and pipe flow, respectively. The amount of dilution
relative to the entrance value $\phi_{o}$, i.e. $(\phi_{o}-\left\langle \phi\right\rangle )/\phi_{o}$,
 continuously decreases with increasing concentration from the maximum
value of about 17\% for vanishing $\phi_{o}$. These observations
are consistent with previous theoretical and experimental studies
\citep{SeSu68,NoBr94,MiMo06}, where the dilution is attributed to
the faster flow in the central part of the gap where local particle
concentration is high and slower flow of less concentrated suspension
near the walls. 

We also note that the fully-developed solution has the well-defined
maximum flowing solid volume fraction, $\text{max}\phi_{o}=\text{max}\left\langle \phi\right\rangle $,
given by the plug-average value $\left\langle \phi\right\rangle _{\text{plug}}=(\phi_{m}+\phi_{rcp})/2\approx0.609$
for the channel and $(2\phi_{m}+\phi_{rcp})/3\approx0.601$ for the
pipe flow. These values correspond to the termination points of the
dilution curves on figure \ref{fig:<phi>}.

\begin{figure}
\noindent \begin{centering}
\includegraphics[scale=0.55]{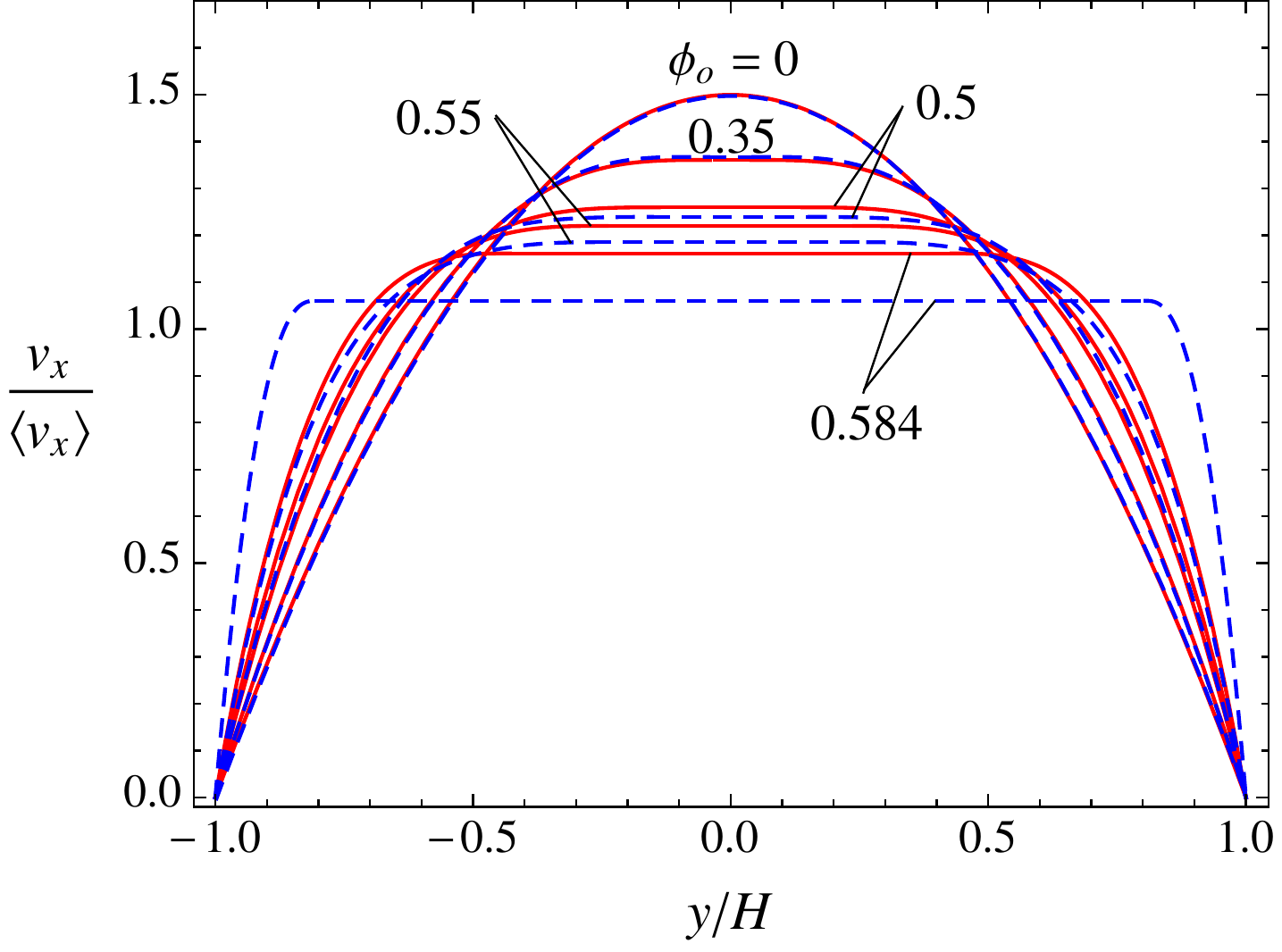}\includegraphics[scale=0.52]{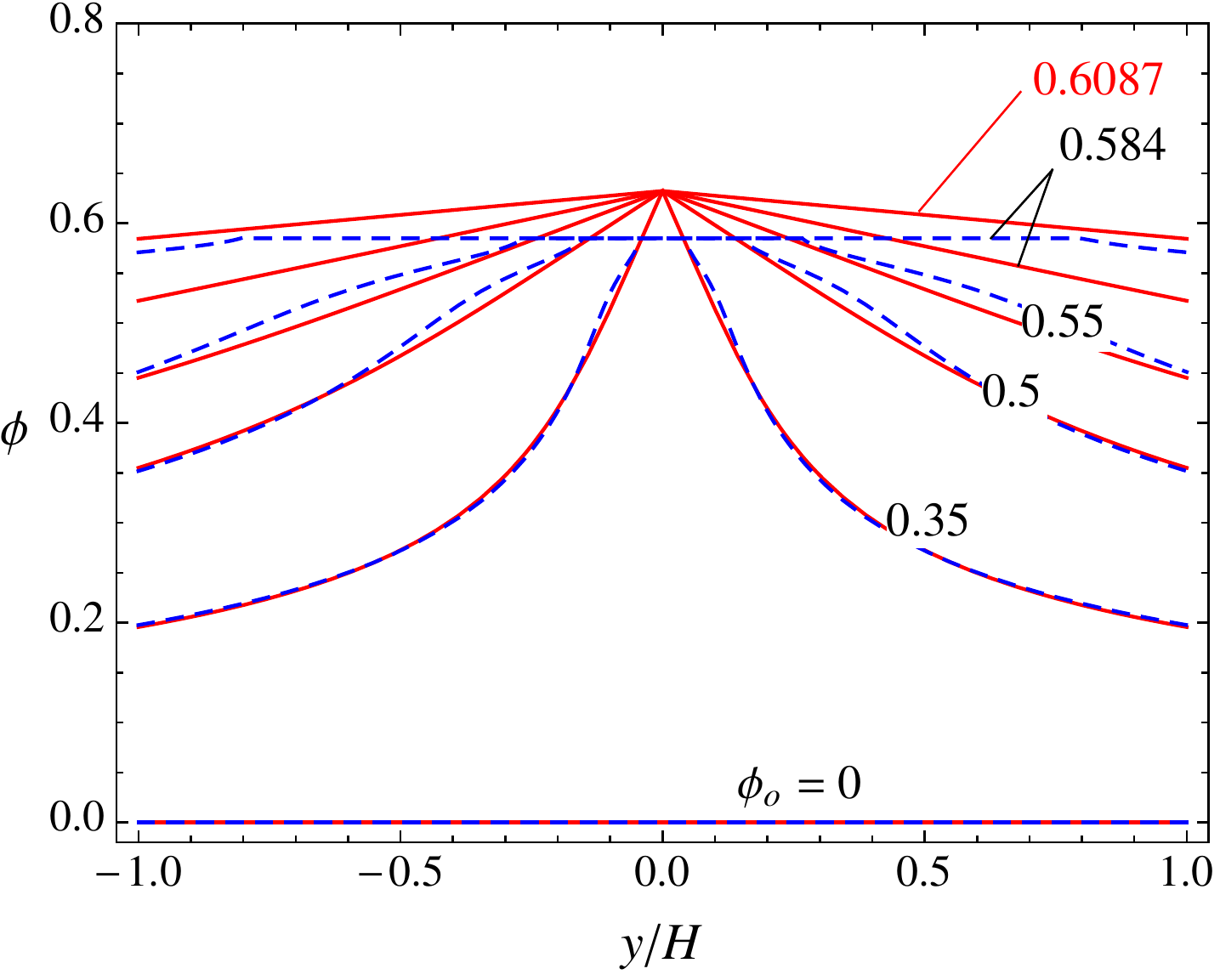}
\par\end{centering}

\caption{Transition from Poiseuille to plug flow in a channel with increasing
entrance solid volume fraction $\phi_{o}$: fully developed velocity
profiles scaled by the mean velocity (left) and solid volume fraction
profile (right). The results obtained using the frictional rheology
with compressible plug (\ref{mu_alt}) are plotted in solid (red)
lines whereas the original Boyer et al.'s rheology with an incompressible
plug are plotted in dash (blue) lines. \label{fig:Transition-Poiseuille-Plug}}
\end{figure}

\begin{figure}
\begin{centering}
\includegraphics[scale=0.5]{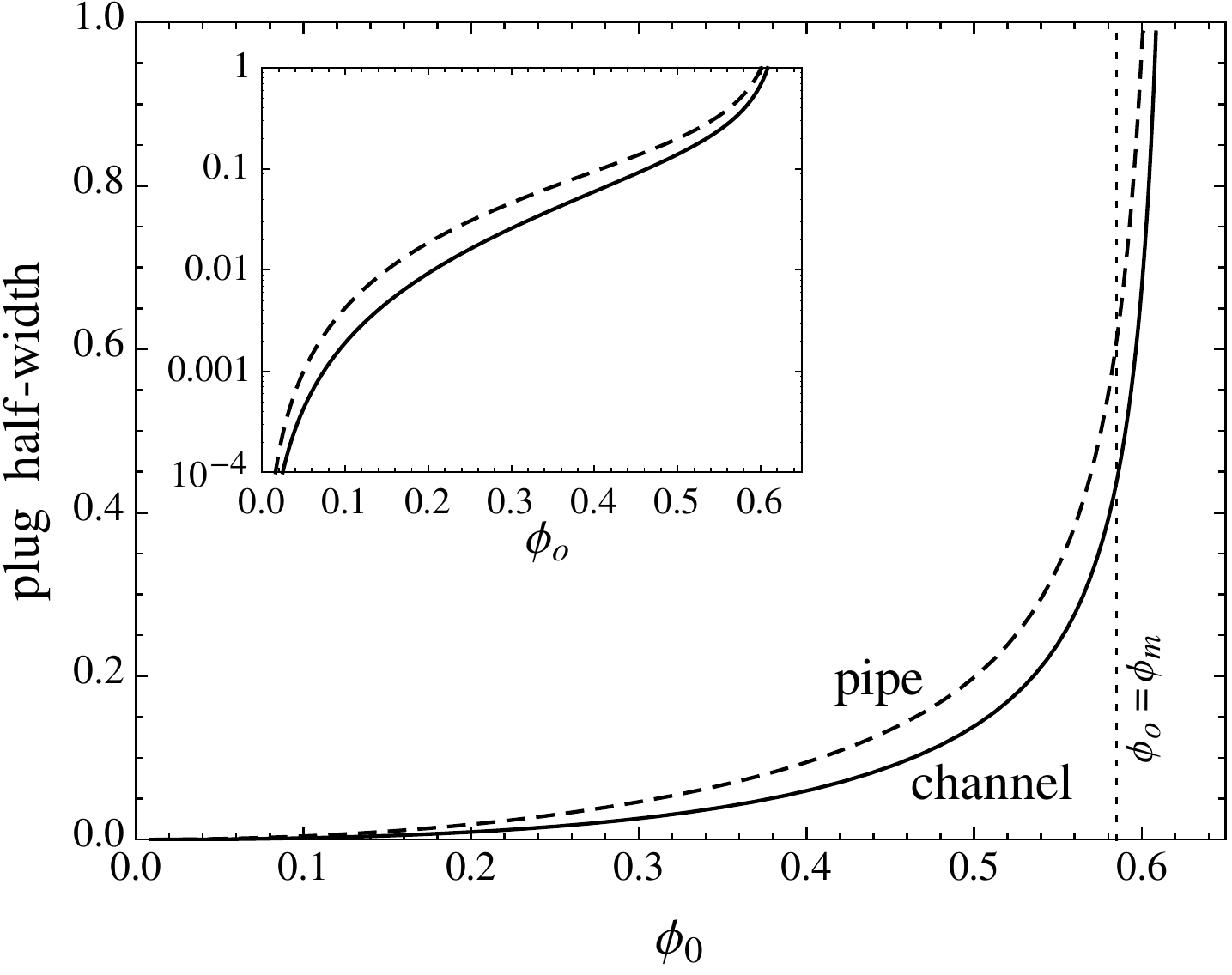} 
\par\end{centering}

\caption{Fully-developed plug half-width (slot, continuous line) and radius
(pipe, dashed line) as a function of the entrance solid volume fraction
$\phi_{o}$. The inset shows the details of vanishing plug in low
solid fraction flows on a semi-logarithmic plot. The plug dimension
is scaled by the slot half-width $H$ and pipe radius $R$, respectively.
\label{fig:plug}}
\end{figure}

Figure \ref{fig:Transition-Poiseuille-Plug} illustrates profiles
of the scaled velocity ($v_{x}/\left\langle v_{x}\right\rangle =h/\left\langle h\right\rangle $)
and particle concentration, respectively, across the channel for various
values of the entrance particle concentration $\phi_{o}$. A transition
from Poiseuille to plug flow as the entrance solid volume fraction
increases is evident in figure \ref{fig:Transition-Poiseuille-Plug},
and can be further quantified from the plot of the increasing plug
half-width (slot) or radius (pipe) with $\phi_{o}$ in figure \ref{fig:plug}.
The lower limit $\phi_{o}\approx0.25$ where a flatten velocity profile
has been detected experimentally for pipe flow \citep{CoMa71,HaMa97}
corresponds to a theoretical plug size of about 3\% of the pipe radius. 

It is also worthwhile to note that the predicted velocity/concentration
profiles (based on rheology (\ref{mu_alt})-(\ref{I_alt}) proposed
in section \ref{sec:fric}) are very similar to the predictions based
on original frictional rheology of \citet{BoGu11}, shown on figure
\ref{fig:Transition-Poiseuille-Plug} by dashed lines for comparison,
as long as $\phi_{o}<0.55$. The main difference between the two models
lie in the linear compaction in the central plug allowed for in the
former, but not in the latter. The plug compaction starts to impact
the velocity profile for value of $\phi_{o}$ close to $\phi_{m}$:
the compaction significantly reduces the plug size and allows a higher
velocity (see the case $\phi_{o}=0.584$ on figure \ref{fig:Transition-Poiseuille-Plug}).
(In fact, plug compaction allows for fully-developed flows with average
concentration exceeding the jamming value, i.e. $\left\langle \phi\right\rangle >\phi_{m}$).

\begin{figure}
\noindent \begin{centering}
\includegraphics[scale=0.48]{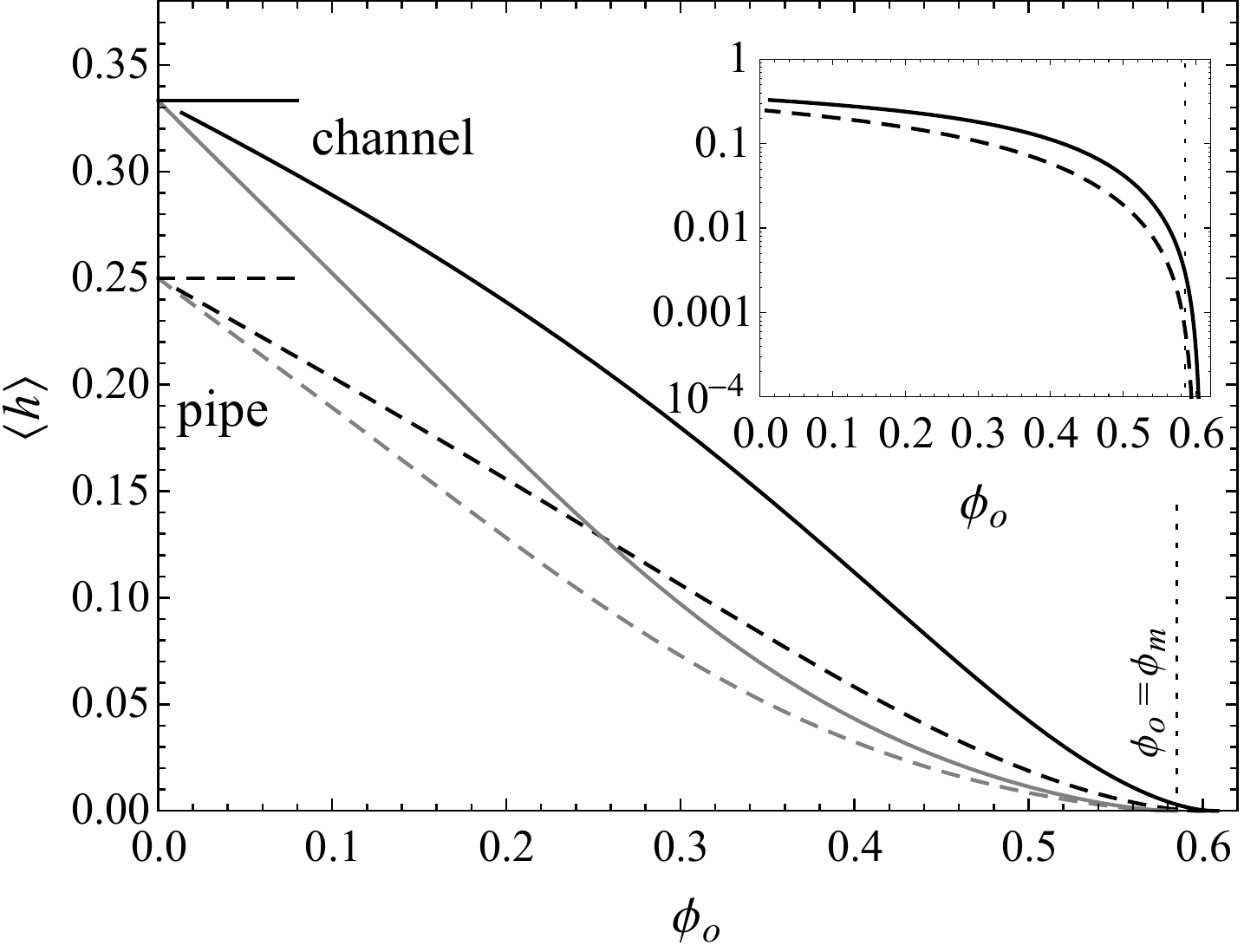}\enskip{}\includegraphics[scale=0.5]{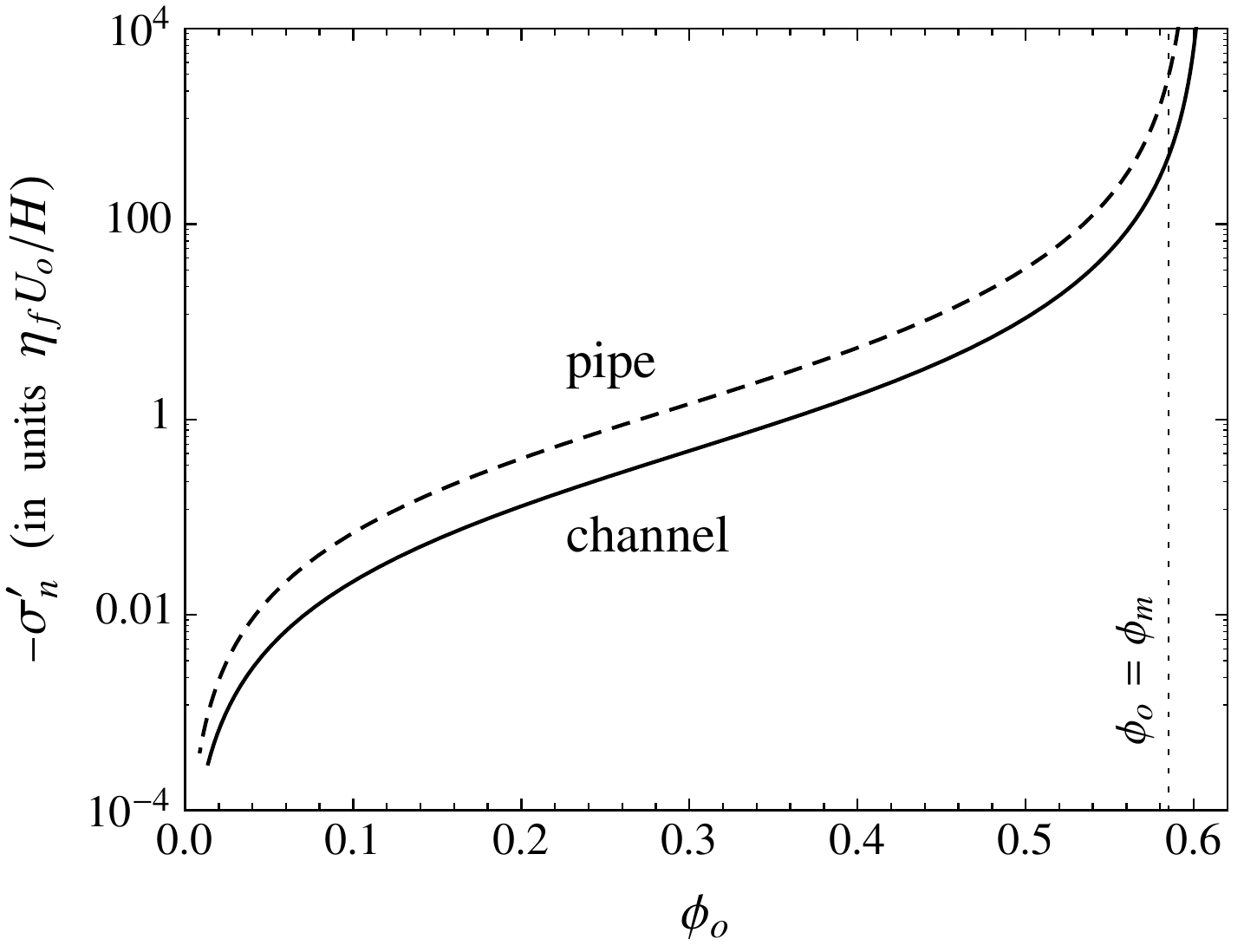}
\par\end{centering}

\caption{Tangent gap-averaged ``fluidity'' $\left\langle h\right\rangle $
= the inverse of the mean stress gradient $1/|\nabla p|^{\text{SLOT}}$
or $2/|\nabla p|^{\text{PIPE}}$, when expressed in units of $\eta_{f}U_{o}/H^{2}$
or $\eta_{f}U_{o}/R^{2}$, (left), and effective normal stress $-\sigma_{n}'$
in units of $\eta_{f}U_{o}/H$ or $\eta_{f}U_{o}/R$ (right) for fully-developed
flow in a slot (solid line) and a pipe (long dash line) as a function
of the entrance solid volume fraction $\phi_{o}$. The limiting Newtonian
values $1/3$ (slot), $1/4$ (pipe) of $\left\langle h\right\rangle $
for $\phi_{o}=0$ are highlighted, while the inset displays the behavior
of $\left\langle h\right\rangle $ in semi-log in the region close
to $\phi_{m}$. The approximations $\left\langle h\right\rangle \approx\frac{1}{3\eta_{s}(\phi_{o})}$
(slot), $\frac{1}{4\eta_{s}(\phi_{o})}$ (pipe) obtained by directly
taking the relative shear viscosity in a Newtonian parabolic profile
are also displayed (light gray continuous and dash lines). \label{fig:<h>-Pp}}
\end{figure}

The evolution of the gap-average tangent fluidity $\left\langle h\right\rangle $
as a function of $\phi_{o}$ is an important quantity to estimate
friction pressure in engineering applications. Figure \ref{fig:<h>-Pp}
displays the evolution of $\left\langle h\right\rangle $ from the
Newtonian limit at $\phi_{o}=0$ to the limit of no flow $\left\langle h\right\rangle =0$
at $\phi_{o}=\left\langle \phi\right\rangle _{\text{plug}}>\phi_{m}$.
It is interesting to point out that estimating $\left\langle h\right\rangle $
by directly using the shear viscosity as a function of $\phi_{o}$
in a Newtonian (parabolic) velocity profile results in a poor approximation.
Finally, the corresponding evolution of the dimensionless effective
normal stress $-\sigma_{n}'$ as a function of $\phi_{o}$ is also
displayed in figure \ref{fig:<h>-Pp}. We can again note that below
$\phi_{o}\approx0.25$, the effective stress appears negligible ($-\sigma_{n}'{\normalcolor <0.25}$)
in line with experimental observations of \citet{DeGa09,GaGa13},
but it is seen to drastically increase for $\phi_{o}>0.5$.

\subsection{Comparison with experiments\label{sec:Comp}}

We now compare the solution for fully developed flow of suspensions
governed by a frictional rheology to experimental results available
in the literature for low Reynolds number flows of neutrally buoyant,
mono-dispersed suspensions in channels and pipes. As previously in
figures \ref{fig:<phi>}-\ref{fig:<h>-Pp}, and for the remainder
of this paper, we make use of frictional rheology (\ref{mu_alt})-(\ref{I_alt})
with parameters $\phi_{m}=0.585$, $\mu_{1}=0.3$, and $\beta=0.158$,
which values were determined from matching with rheological measurements
of \citet{BoGu11} and \citet{DbLo13} (see figure \ref{fig:rheology}).
We therefore emphasize that no attempts are taken to ``tune'' the
rheological parameters to the pressure-driven flow experiments data
used in the comparisons.

\subsubsection{Channel flow}

\begin{figure}
\begin{centering}
\includegraphics[scale=0.5]{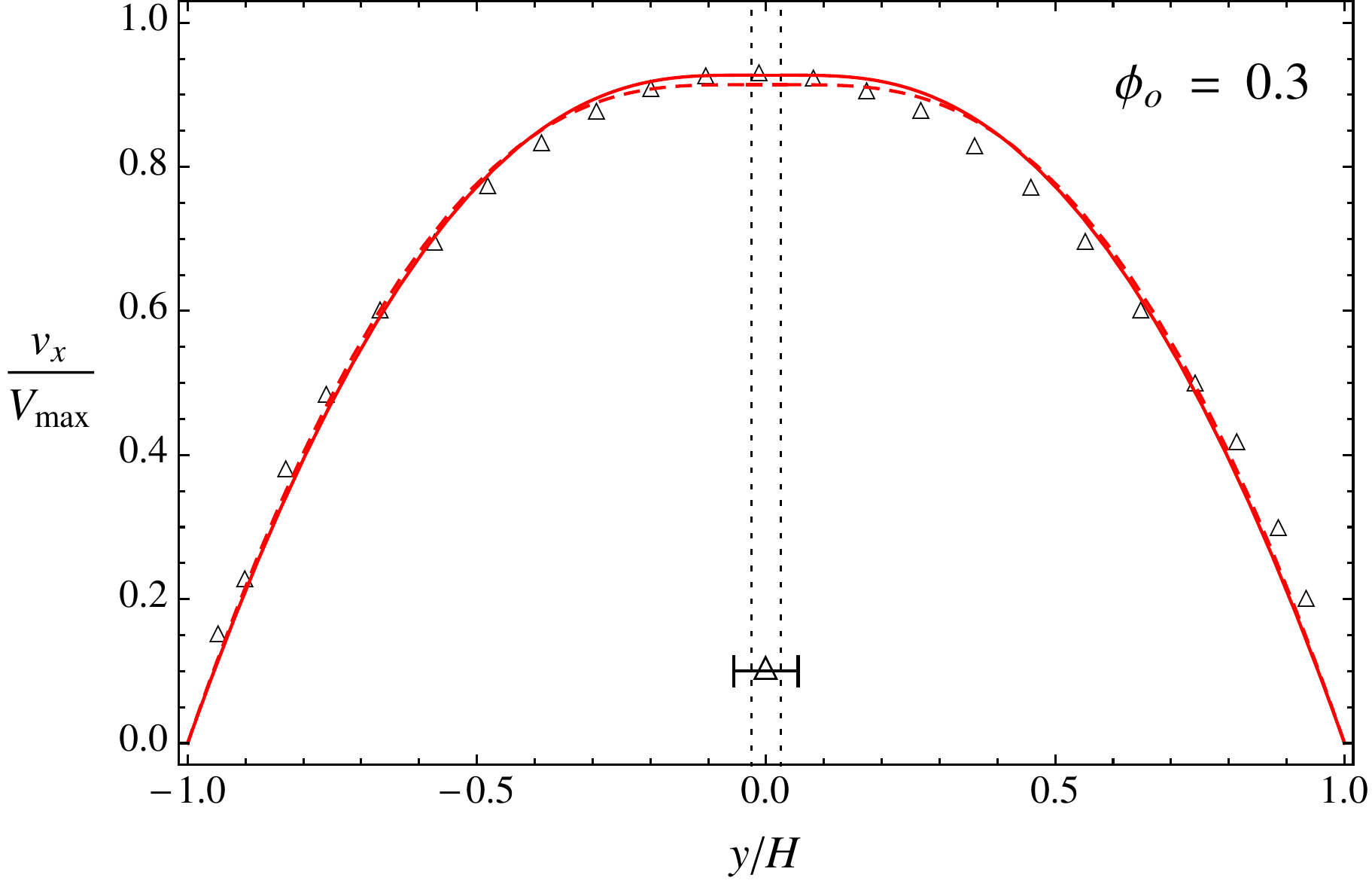}
\par\end{centering}

\begin{centering}
\hspace{0.5cm}\includegraphics[scale=0.5]{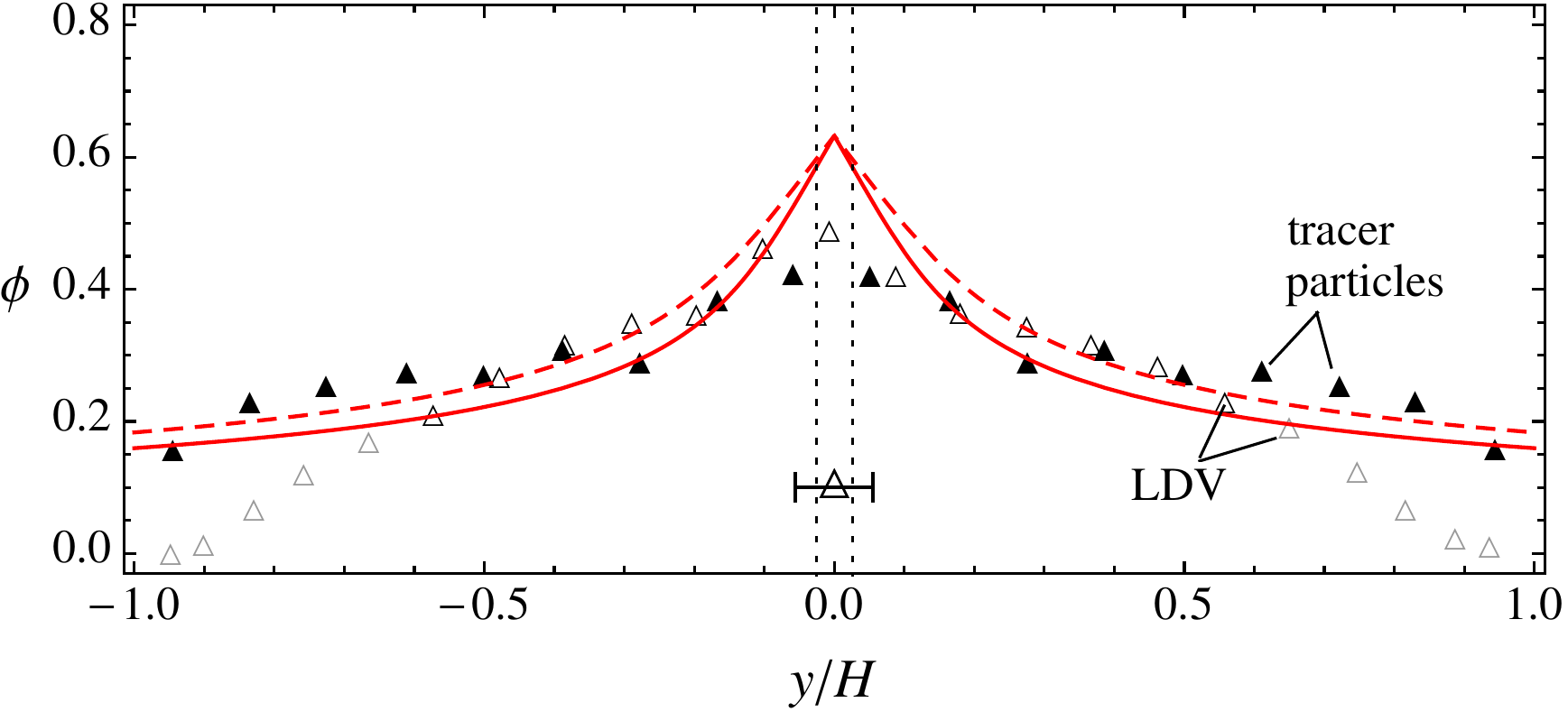}
\par\end{centering}

\caption{Comparison of the theoretical predictions (solid lines) with the experiment
of \citet{LyLe98a,LyLe98b} (Table \ref{tab:LL_Experiments}) for
channel flow of a suspension with $\phi_{o}=0.3$: scaled velocity
profile (top), solid volume fraction profile measured using the LDV
$(\vartriangle)$ and tracer-particles $(\blacktriangle)$ methods
(bottom). (The solid volume fraction measurements with LDV in the
outer 20\% of the channel are biased by wall-effects, and are shown
in faint gray). Bar shows scaled particle diameter. Predicted diluted
gap-averaged particle concentration is $\left\langle \phi\right\rangle \approx0.267$
(Table \ref{tab:LL_Experiments_SomeNumbers}). Theoretical profiles
which neglect dilution, i.e. assume $\left\langle \phi\right\rangle =0.3$,
are also shown by dashed lines.  \label{fig:Channel_Phi03}}
\end{figure}

\begin{figure}
\begin{centering}
\includegraphics[scale=0.5]{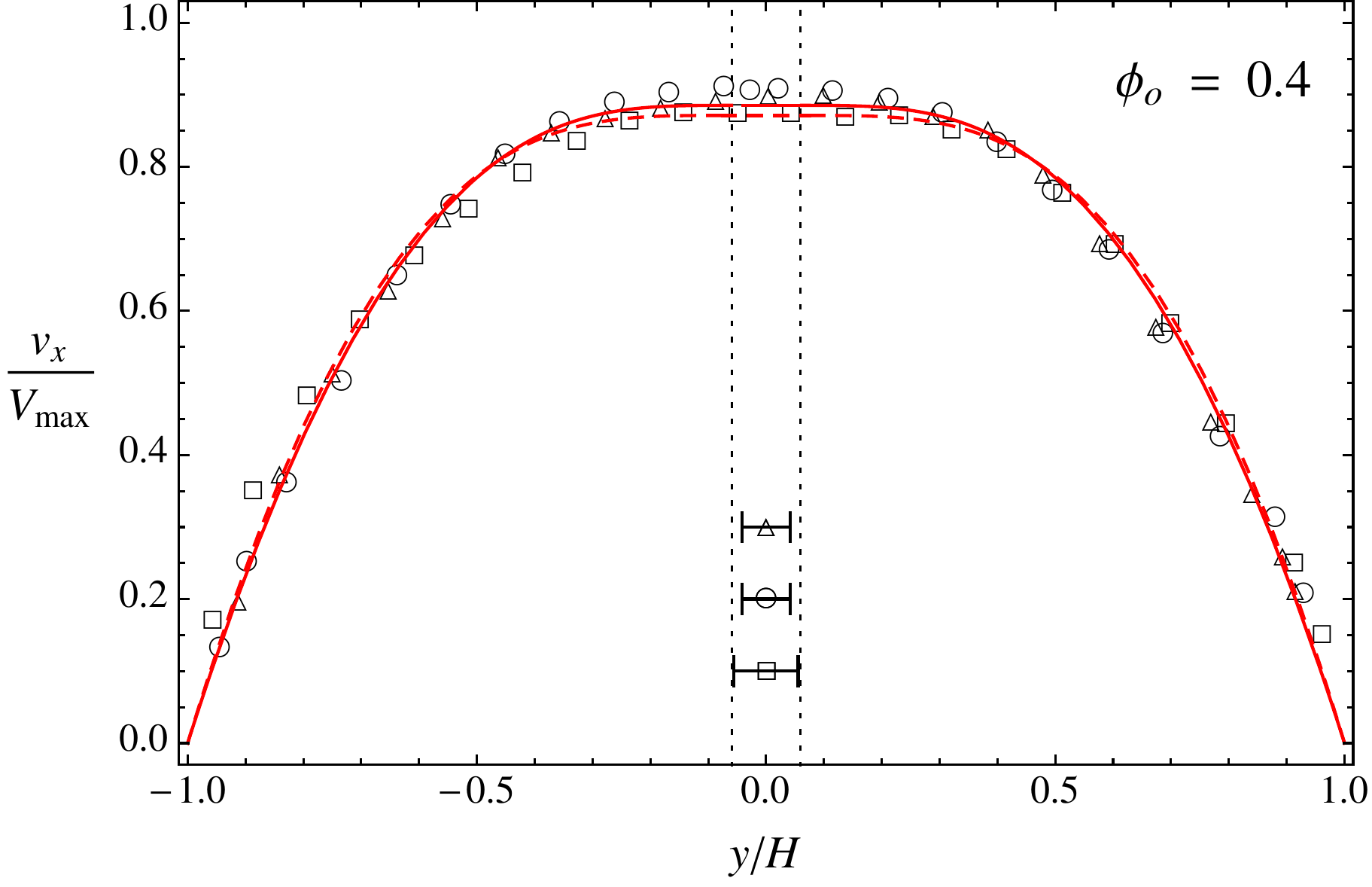}
\par\end{centering}

\begin{centering}
\hspace{0.5cm}\includegraphics[scale=0.5]{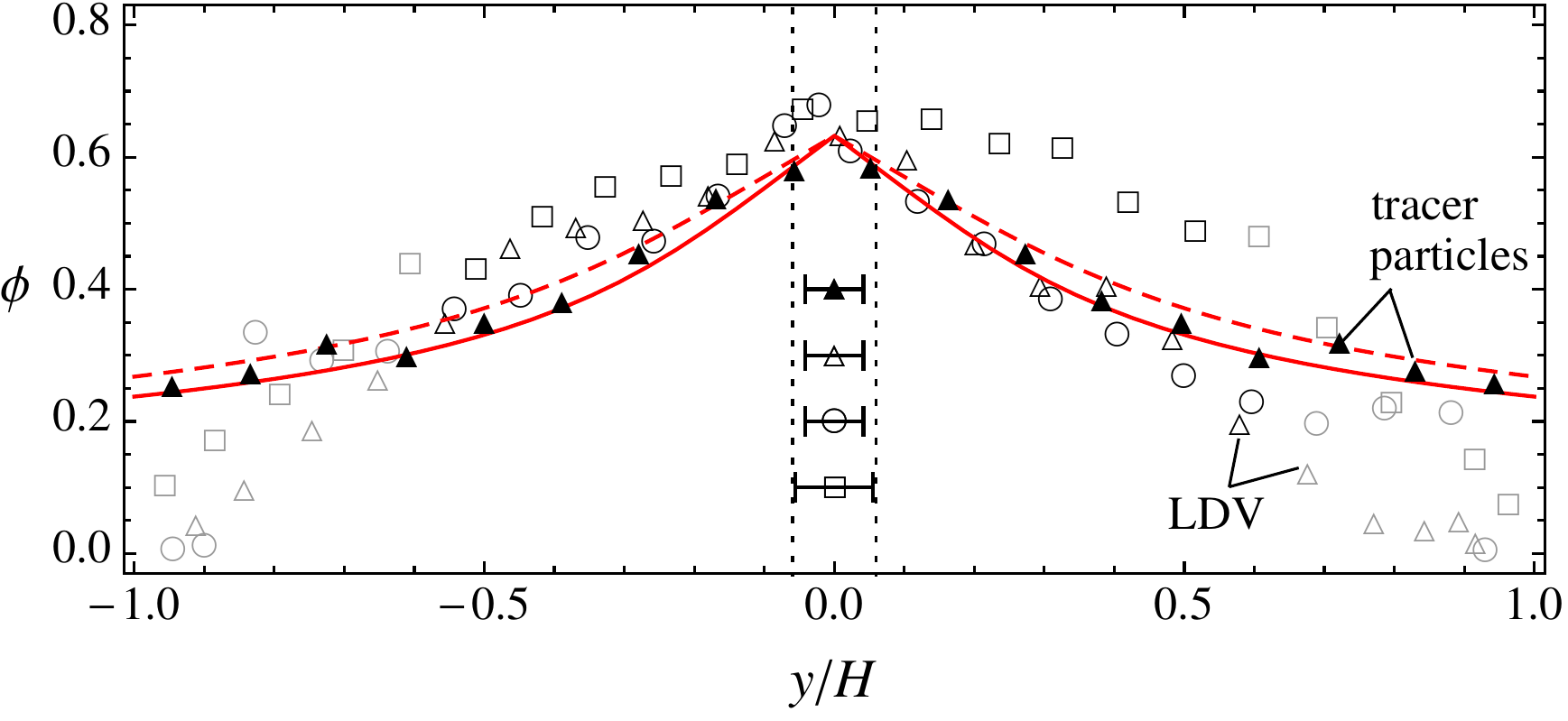}
\par\end{centering}

\caption{As in figure \ref{fig:Channel_Phi03} but for $\phi_{o}=0.4$ (predicted
diluted $\left\langle \phi\right\rangle =0.367$). \label{fig:Channel_Phi04}}
\end{figure}

\begin{figure}
\begin{centering}
\includegraphics[scale=0.5]{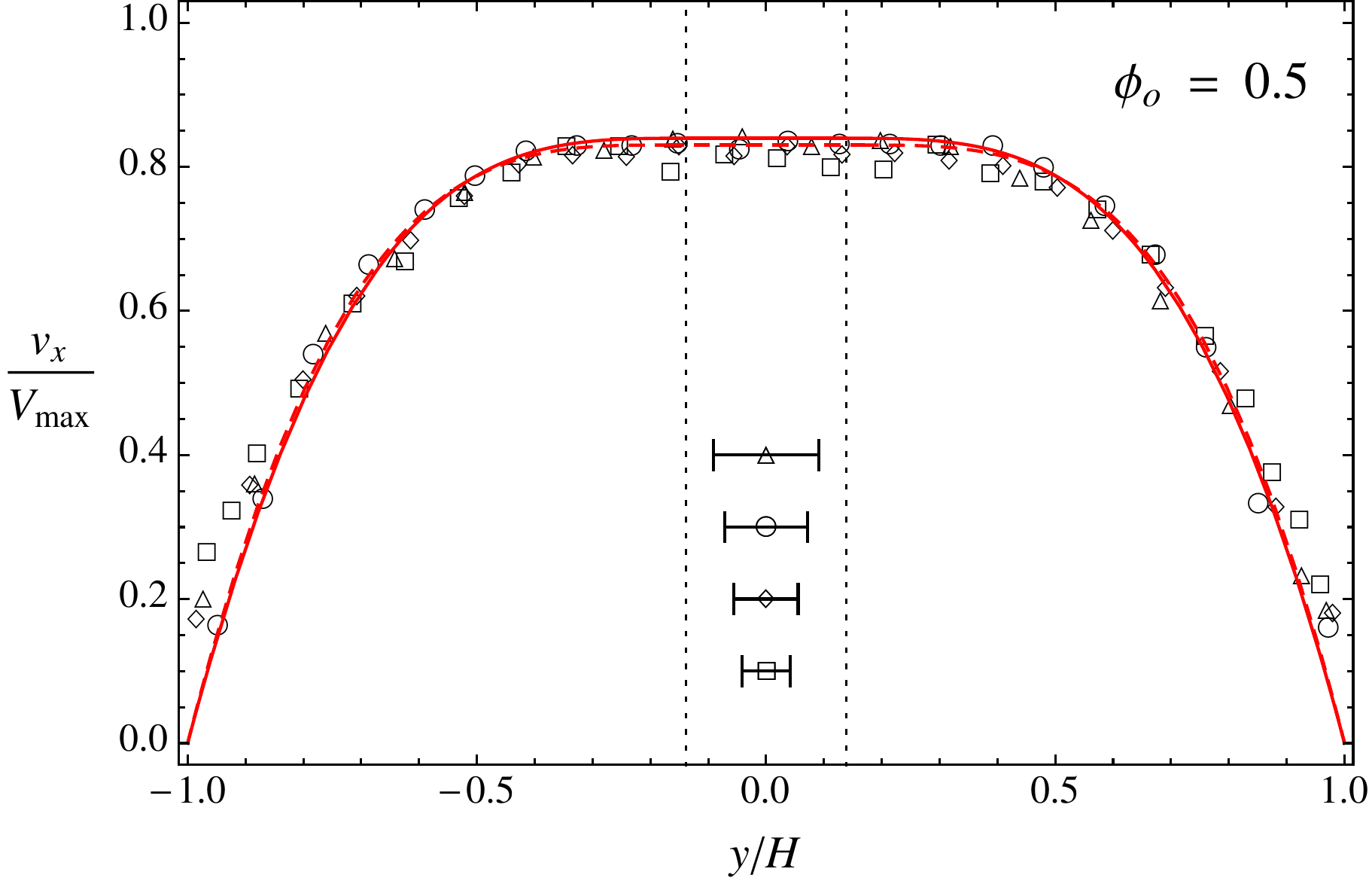}
\par\end{centering}

\begin{centering}
\hspace{0.5cm}\includegraphics[scale=0.5]{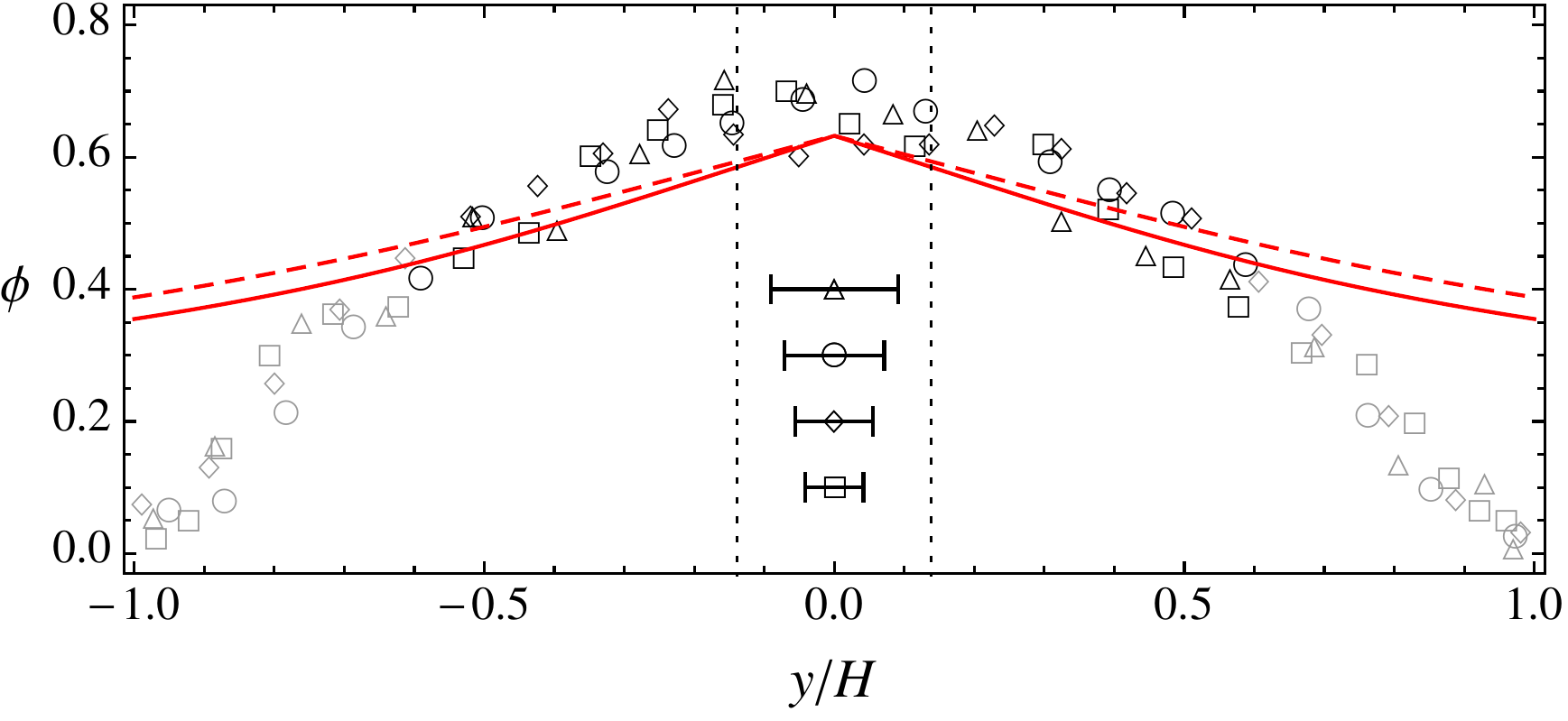}
\par\end{centering}

\caption{As in figure \ref{fig:Channel_Phi03} but for $\phi_{o}=0.5$ (predicted
diluted $\left\langle \phi\right\rangle =0.477$). \label{fig:Channel_Phi05} }
\end{figure}

\citet{LyLe98a,LyLe98b} report a series of pressure driven flow experiments
at low Reynolds number in a channel on suspensions of mono-disperse
PMMA spheres ($2a=70\pm6$ $\mu$m or $95\pm12$ $\mu$m) in a Triton
X-100/ 1,6-dibromohexane / UCON 75-H oil mixture ($\eta_{f}=4.8\mbox{ Poise})$.
Laser Doppler Velocimetry (LDV) was used to image both the velocity
and solid volume fraction profiles across the channel gap at a distance
sufficiently far from the inlet, where the flow can be considered
nearly fully-developed (see \citet{LyLe98a} for discussion, and our
estimates of the degree of the flow development in Table \ref{tab:LL_Experiments}).
Tracer-particles optical microscopy was used in some of the experiments
\citep{LyLe98b} and provided better local solid volume fraction measurements
compared to those made with the LDV. In particular, the LDV measurements
of the solid volume fraction appear to be inaccurate in the outer
20\% of the gap due to the channel wall effects \citep{LyLe98b}.
When presenting these measurements in figures \ref{fig:Channel_Phi03}-\ref{fig:Channel_Phi05},
we will show the wall-biased LDV data by a shade of gray to differentiate
from higher-confidence measurements in the core of the flow. 

\begin{table}
\noindent \centering{}%
\begin{tabular}{|c|c|c|c|c|c|c|c|c|c|}
\hline 
Exp. \# & $\phi_{o}$ & $U_{o}$ & $Re_{p}$ & $2H$ & $2a$ & $H/a$ & $\left[x/H\right]_{\text{exp}}$  & $[L_{\phi}/H]$ & development\tabularnewline
\hline 
 & - & {[}mm/s{]} & {[}$10^{-6}${]} & {[}mm{]} & {[}microns{]} & - & - & - & \%\tabularnewline
\hline 
\hline 
562 $(\blacktriangle,\vartriangle)$ & 0.3 & 4.7 & 3.6 & 1.7 & 95 & 18 & 224 & 415 & 81\tabularnewline
\hline 
482 $(\blacktriangle,\vartriangle)$ & \multirow{3}{*}{0.4} & 4.7 & 1.5 & 1.7 & 70 & 24 & 280 & 450 & 86\tabularnewline
\cline{1-1} \cline{3-10} 
483 $(\scalebox{1.3}{\ensuremath{\circ}})$ &  & 9.1 & 2.9 & 1.7 & 70 & 24 & 280 & 450 & 86\tabularnewline
\cline{1-1} \cline{3-10} 
553 $(\scalebox{0.75}{\ensuremath{\square}})$ &  & 4.7 & 3.6 & 1.7 & 95 & 18 & 224 & 244 & 94\tabularnewline
\hline 
550 $(\vartriangle)$ & \multirow{4}{*}{0.5} & 5.4 & 12. & 1 & 95 & 11 & 380 & 26 & \multirow{4}{*}{100}\tabularnewline
\cline{1-1} \cline{3-9} 
551 $(\scalebox{1.3}{\ensuremath{\diamond}})$ &  & 4.7 & 3.6 & 1.7 & 95 & 18 & 224 & 75 & \tabularnewline
\cline{1-1} \cline{3-9} 
575 $(\scalebox{1.3}{\ensuremath{\circ}})$ &  & 5.4 & 4.9 & 1 & 70 & 14 & 380 & 48 & \tabularnewline
\cline{1-1} \cline{3-9} 
008 $(\scalebox{0.75}{\ensuremath{\square}})$ &  & 4.7 & 1.4 & 1.7 & 70 & 24 & 220 & 137 & \tabularnewline
\hline 
\end{tabular}\caption{Summary of the channel flow experiments of \citet{LyLe98a,LyLe98b}:
$\phi_{o}$ is the entrance solid volume fraction, $U_{o}$ the entrance
velocity (computed from the reported flow rate and slot dimensions
of $2H$ by 2 inches). $Re_{p}=\frac{4}{3}\frac{\rho}{\eta_{f}}\frac{a^{3}}{H^{2}}V_{max}$
is the particle Reynolds number. The carrier fluid and particles are
density matched ($\rho=1190$$\mbox{kg}/\mbox{m}^{3}$), and $\eta_{f}=4.8\mbox{ Poise}$.
The last three columns show the axial distance of the measurements
from the flow entrance ($x_{\text{exp}}$), the axial distance $L_{\phi}$
at which the flow is predicted to be 95\% developed, and the predicted
actual percent of flow development at $x_{\text{exp}}$ (based on
the numerical solution for axial flow development of Section \ref{sec:Axial-Flow-Developement}).
\label{tab:LL_Experiments}}
\end{table}

\begin{table}
\noindent \begin{centering}
\begin{tabular}{|c|c|c|c||c|c|c|c|}
\hline 
Experiments & $\phi_{o}$ & \multicolumn{2}{c||}{$\left\langle \phi\right\rangle $} & $\phi_{\text{{w}}}$ & $\mu_{\mathrm{w}}$ & $-\sigma_{n}'$ {[}Pa{]} & $-\partial_{x}p$ {[}kPa/m{]}\tabularnewline
\hline 
\hline 
 \# &  & Measured & Theory & \multicolumn{4}{c|}{Theory}\tabularnewline
\hline 
\hline 
\multirow{2}{*}{562 $(\blacktriangle,\vartriangle)$} & \multirow{2}{*}{0.3} & 0.28$^{\text{a}}$ & \multirow{2}{*}{0.267} & \multirow{2}{*}{0.160} & \multirow{2}{*}{11.6} & \multirow{2}{*}{1.27} & \multirow{2}{*}{17.4}\tabularnewline
\cline{3-3} 
 &  & 0.24 &  &  &  &  & \tabularnewline
\hline 
\multirow{2}{*}{482 $(\blacktriangle,\vartriangle)$} & \multirow{4}{*}{0.4} & 0.38$^{\text{a}}$ & \multirow{4}{*}{0.367} & \multirow{4}{*}{0.238} & \multirow{4}{*}{5.04} & \multirow{2}{*}{4.72} & \multirow{2}{*}{28.0}\tabularnewline
\cline{3-3} 
 &  & 0.35 &  &  &  &  & \tabularnewline
\cline{1-1} \cline{3-3} \cline{7-8} 
483 $(\scalebox{1.3}{\ensuremath{\circ}})$ &  & 0.32 &  &  &  & 9.13 & 54.1\tabularnewline
\cline{1-1} \cline{3-3} \cline{7-8} 
553 $(\scalebox{0.75}{\ensuremath{\square}})$ &  & 0.36 &  &  &  & 4.72 & 28.0\tabularnewline
\hline 
550 $(\vartriangle)$ & \multirow{4}{*}{0.5} & 0.41 & \multirow{4}{*}{0.477} & \multirow{4}{*}{0.355} & \multirow{4}{*}{2.17} & 56.3 & 244\tabularnewline
\cline{1-1} \cline{3-3} \cline{7-8} 
551 $(\scalebox{1.3}{\ensuremath{\diamond}})$ &  & 0.42 &  &  &  & 29.0 & 73.9\tabularnewline
\cline{1-1} \cline{3-3} \cline{7-8} 
575 $(\scalebox{1.3}{\ensuremath{\circ}})$ &  & 0.43 &  &  &  & 56.3 & 244\tabularnewline
\cline{1-1} \cline{3-3} \cline{7-8} 
008 $(\scalebox{0.75}{\ensuremath{\square}})$ &  & 0.42 &  &  &  & 29.0 & 73.9\tabularnewline
\hline 
\end{tabular}
\par\end{centering}

\caption{Gap-averaged solid volume fraction $\left\langle \phi\right\rangle $
in the channel flow experiments (Table \ref{tab:LL_Experiments})
measured using the LDV \citep{LyLe98a} and tracer ($^{\text{a}}$,
$\blacktriangle$) particle \citep{LyLe98b} methods, and its theoretical
prediction. (Underestimation of $\left\langle \phi\right\rangle $
when using the LDV method compared to the tracer-particle and theoretically
predicted values is due to the poor resolution of the LDV method near
the channel walls). Theoretical values of the solid volume fraction
$\phi_{\text{w}}$ and friction $\mu_{\text{w}}$ at the walls in
the fully-developed flow, measured and theoretical estimate of the
normalized velocity width-average. Different set of experiments for
similar injected volume fraction $\phi_{o}$. \label{tab:LL_Experiments_SomeNumbers}}
\end{table}

Table \ref{tab:LL_Experiments} summarizes different experimental
conditions tested by \citet{LyLe98a}, which included tests at different
values of the entrance solid volume fraction, particle size, flow
rate, and channel width. Table \ref{tab:LL_Experiments_SomeNumbers}
lists values of the mean solid volume fraction across the gap obtained
from the reported experimental profiles (by trapezoidal integration)
for each series of experiments, as well as the predicted theoretical
values accounting for dilution. The experimental values of $\left\langle \phi\right\rangle $
obtained by LDV all appear smaller than the predicted theoretical
values, whereas a good match is obtained for the cases where $\phi$
was measured using the tracer particle method. As discussed in \citet{LyLe98b},
poor resolution of particles by the LDV technique close to the channel
walls explains the observed differences of the $\left\langle \phi\right\rangle $-values.
In addition, Table \ref{tab:LL_Experiments_SomeNumbers} lists theoretical
predictions for other essential parameters describing the flow for
different experiments series, such as wall values of the stress ratio
and solid volume fraction (the latter parametrizes the normalized
solution for fully-developed flow), the particle normal stress, and
the total pressure gradient.

The velocity profiles reported by \citet{LyLe98b} were scaled by
the maximum (centerline) velocity for a Newtonian flow profile with
an identical flow rate (i.e. $V_{max}=3/2\left\langle v_{x}\right\rangle $).
The theoretical prediction follows in the form of:
\[
\frac{v_{x}}{V_{max}}=\frac{2}{3}\frac{h(\phi(y))}{\left\langle h\right\rangle }
\]
where $\phi(y)$, $h(\phi)$, and $\left\langle h\right\rangle $
are given in sections \ref{sub:General-solution} and \ref{sub:Cross-sectional-averages}
as functions of a single parameter, $\phi_{\text{w}}$ or $\mu_{\text{w}}=\mu(\phi_{\text{w}})$.
The latter is a unique function of the entrance concentration $\phi_{o}$
given by (\ref{cond1}-\ref{ratio}), and plotted in figure \ref{fig:<phi>}.


Figures \ref{fig:Channel_Phi03}, \ref{fig:Channel_Phi04}, and \ref{fig:Channel_Phi05}
display the comparison between the experimental and theoretical profiles
of the scaled velocity and the solid volume fraction for $\phi_{o}=0.3$,
$0.4$, and $0.5$, respectively. The profiles predicted assuming
zero dilution (i.e. $\left\langle \phi\right\rangle =\phi_{o}$) are
also shown for comparison (dashed lines). Particle sizes used in various
experiments are also shown. 

The experimental and theoretical velocity profiles agree very well
for all three entrance solid volume fractions tested: the predictions
are actually within the experimental measurements error. The frictional
suspension rheology is notably able to correctly predict the development
of the plug region at the channel centerline as the entrance solid
volume fraction increases (see dotted lines marking the predicted
plug boundaries).

Examination of the solid volume fraction profiles show a striking
overall agreement between the theory and experiment, when the optical
particle-tracking method was used to measure $\phi$ (see $\blacktriangle$-symbols
in figures \ref{fig:Channel_Phi03} and \ref{fig:Channel_Phi04} for
$\phi_{o}=0.3$ and 0.4, respectively). The discrepancy near the center
of the channel for the case with $\phi_{o}=0.3$ stems from the loss
of the continuum approximation there, as the particle size in this
experiment is about twice the predicted size of the central plug (figure
\ref{fig:Channel_Phi03}). Remarkably, examination of the case with
$\phi_{o}=0.4$, where the experimental $\phi$-profile (obtained
using optical particle-tracking method) is matched by the theoretical
profile everywhere in the gap, including the central plug region,
which predicted width spans only about one particle diameter. In the
other words, it appears that the continuum approximation for this
type of flow holds down to the scale of a single particle.

Comparison to the LDV measured $\phi$-profiles shows expected discrepancy
in the outer (adjacent to the wall) region of the flow, due to the
previously discussed limitations of the LDV method there. Away from
the walls, LDV $\phi$-profiles, although more scattered than the
tracer-particle profiles, are in a reasonable agreement with the theory.
The most notable deviation from the theoretical $\phi$-profile is
observed in the core of the flow with $\phi_{o}=0.5$ (figure \ref{fig:Channel_Phi05}),
where high experimental values of the particle concentration (for
some measurements, exceeding the random-close-packing value) may be
indicative of partial crystallization in the mono-dispersed suspension.

\subsubsection{Pipe Flow}

Experimental investigations of suspension flow in a pipe have been
performed by \citet{KaGo66,CoMa71,SiCho91} among others. Here, we
focus on the results obtained by \citet{HaMa97} using Nuclear Magnetic
Resonance (NMR) method. They conducted measurements in the flow of
mono-dispersed ($2a=3175$ $\mu$m) and slightly polydispersed ($650\pm110$
$\mu$m) suspensions of PMMA spheres in pipes with internal diameter
$2R=50.8$ mm and $25.4$ mm, respectively, at various values of the
entrance solid volume fraction. Tested particle-to-pipe radius ratios
were $a/R=1/16$ and $1/39$, respectively. The liquid solution of
UCON oil (H-9500), polyalkylene glycol and tetrabomoethane with a
reported viscosity of $2.1$ Pa$\cdot$s was used as the carrying
fluid.

\begin{table}
\noindent \begin{centering}
\begin{tabular}{|c|c|c||c|c|c|c|}
\hline 
$\phi_{o}$ & \multicolumn{2}{c||}{$\left\langle \phi\right\rangle $} & $\phi_{\text{{w}}}$ & $\mu_{\mathrm{w}}$ & $-\sigma_{n}^{\prime}$ {[}Pa{]} & $-\partial_{x}p$ {[}kPa/m{]}\tabularnewline
\hline 
\hline 
 & Measured & Theory & \multicolumn{4}{c|}{Theory}\tabularnewline
\hline 
\hline 
\multirow{2}{*}{0.2} & 0.17$^{\text{a}}$ & \multirow{2}{*}{0.179} & \multirow{2}{*}{0.137} & \multirow{2}{*}{16.1} & \multicolumn{1}{c|}{6.6} & \multicolumn{1}{c|}{16.7}\tabularnewline
\cline{2-2} \cline{6-7} 
 & 0.16$^{\text{b}}$ &  &  &  & 3.3 & 4.2\tabularnewline
\hline 
\multirow{2}{*}{0.3} & 0.27$^{\text{a}}$ & \multirow{2}{*}{0.272} & \multirow{2}{*}{0.210} & \multirow{2}{*}{6.51} & \multicolumn{1}{c|}{24} & \multicolumn{1}{c|}{24.5}\tabularnewline
\cline{2-2} \cline{6-7} 
 & 0.25$^{\text{b}}$ &  &  &  & 12 & 6.1\tabularnewline
\hline 
\multirow{2}{*}{0.45} & 0.41$^{\text{a}}$ & \multirow{2}{*}{0.425} & \multirow{2}{*}{0.350} & \multirow{2}{*}{2.24} & 90 & 71\tabularnewline
\cline{2-2} \cline{6-7} 
 & 0.41$^{\text{b}}$ &  &  &  & 45 & 17.7\tabularnewline
\hline 
\end{tabular}
\par\end{centering}

\caption{Gap-averaged solid volume fraction $\left\langle \phi\right\rangle $
in the pipe flow experiments measured using the NMR method \citep{HaMa97},
and the corresponding theoretical prediction (independent of particle
size). The two different sets of experimental conditions correspond
to ($^{\text{a}}$) $2R=25.4$ mm, $a/R=1/39$; and ($^{\text{b}}$)
$2R=50.8$ mm, $a/R=1/16$. We also give theoretical predictions of
the solid volume fraction and the stress ratio at the pipe wall, as
well as the particle normal stress and the total pressure gradient
for an assumed value of the average flow velocity ($U_{o}=100$ mm/s).
\label{tab:pipe_experiment}}
\end{table}

\begin{figure}
\noindent \begin{centering}
\includegraphics[scale=0.36]{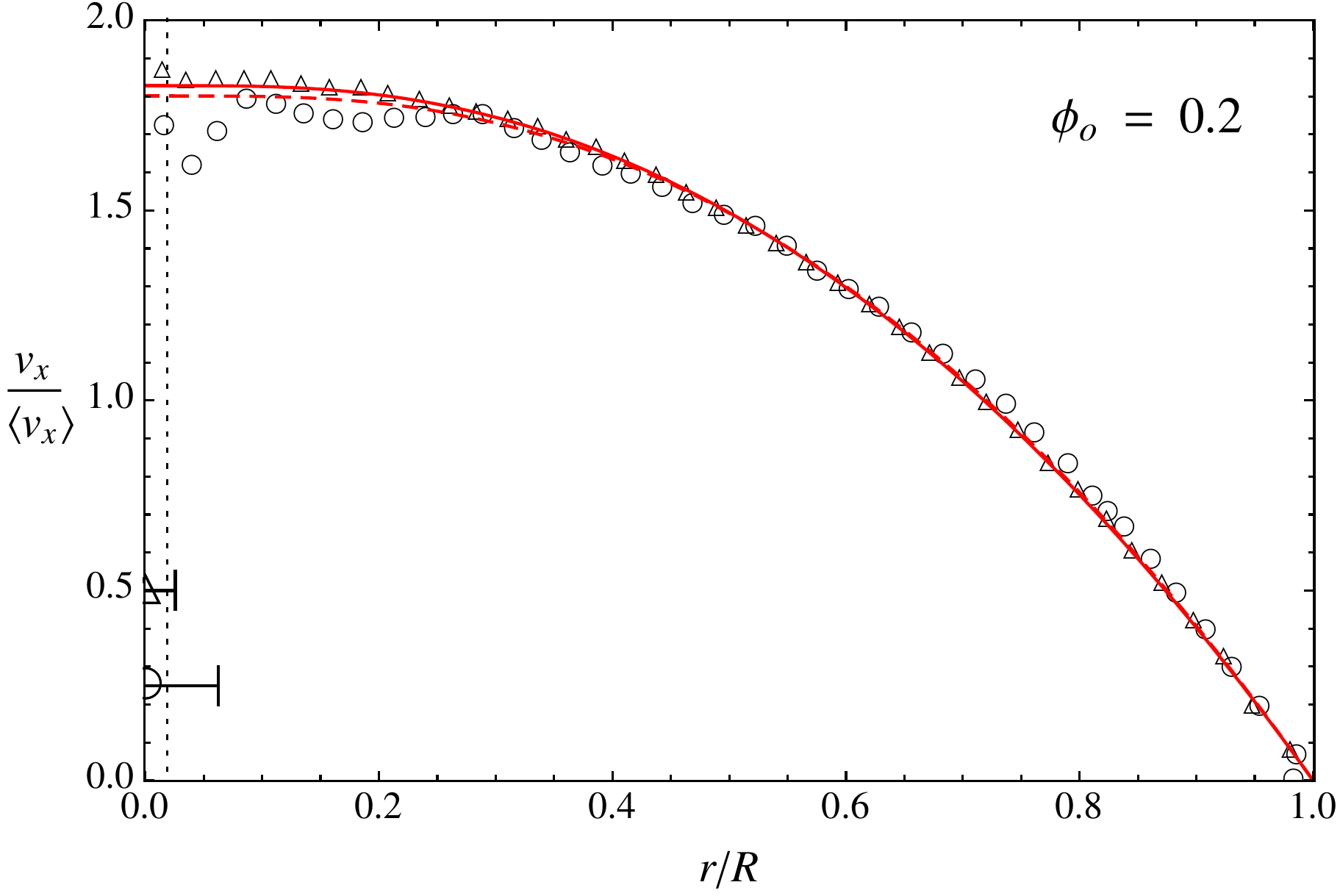}\includegraphics[scale=0.37]{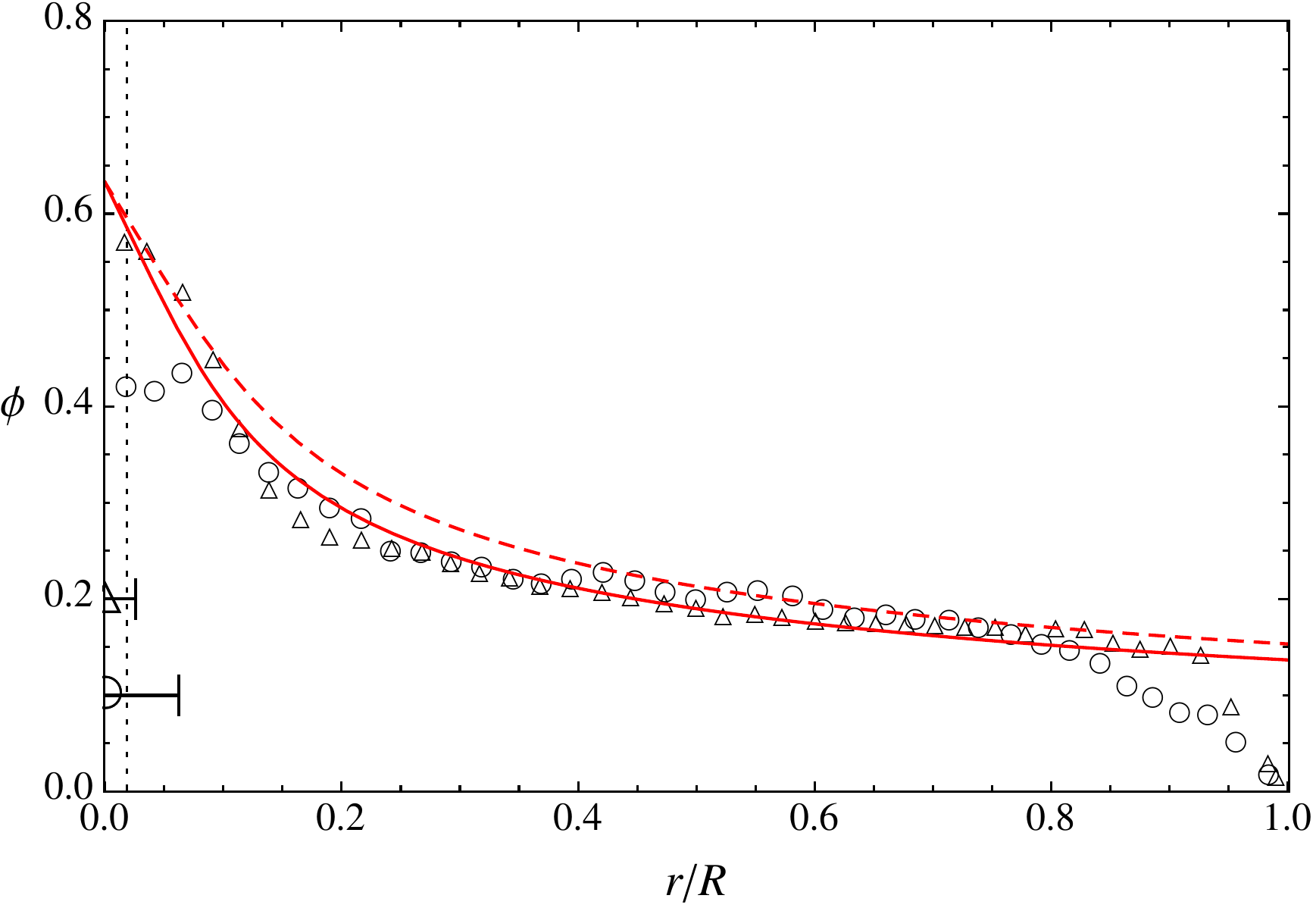}\\
\includegraphics[scale=0.36]{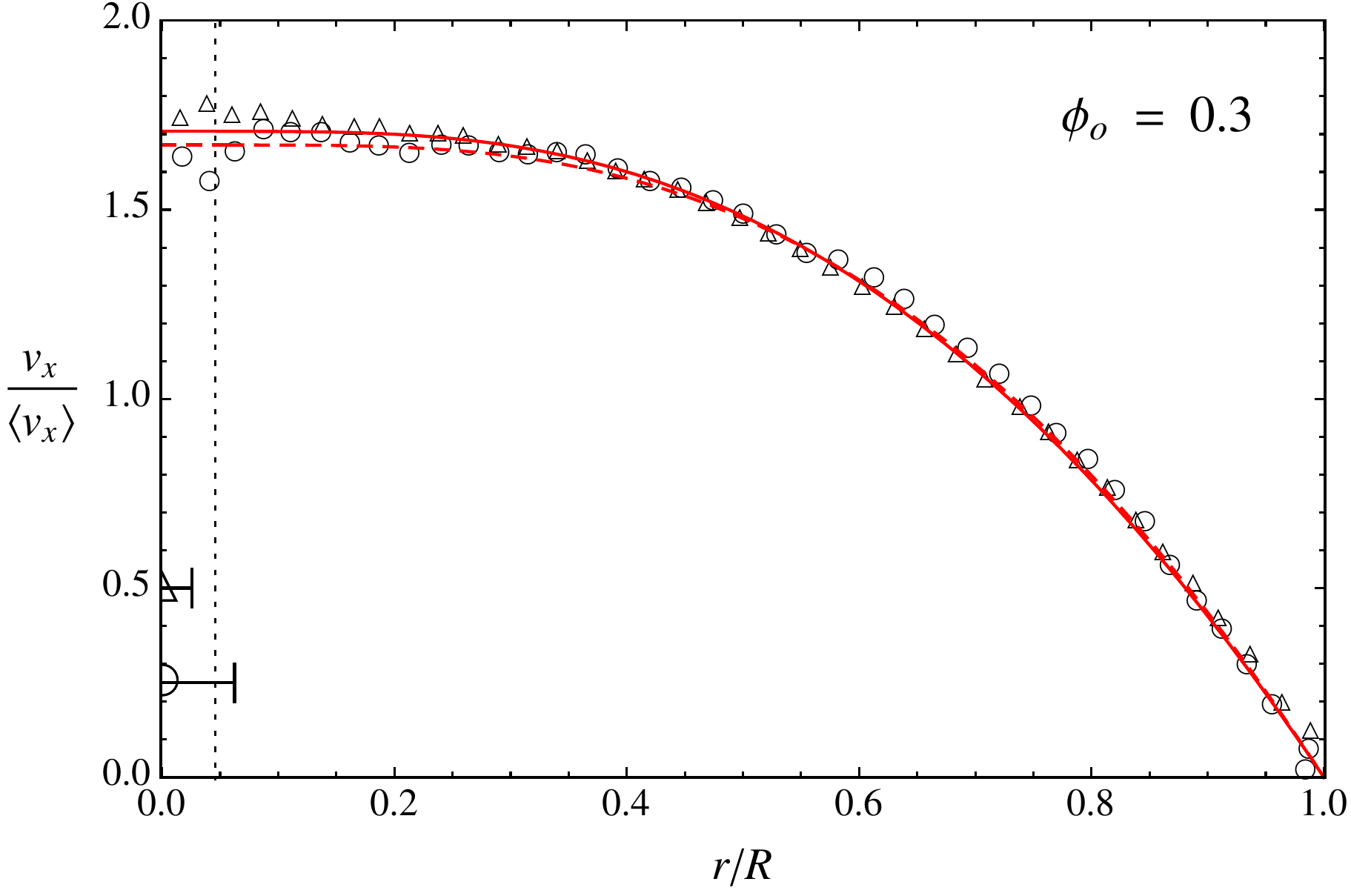}\includegraphics[scale=0.37]{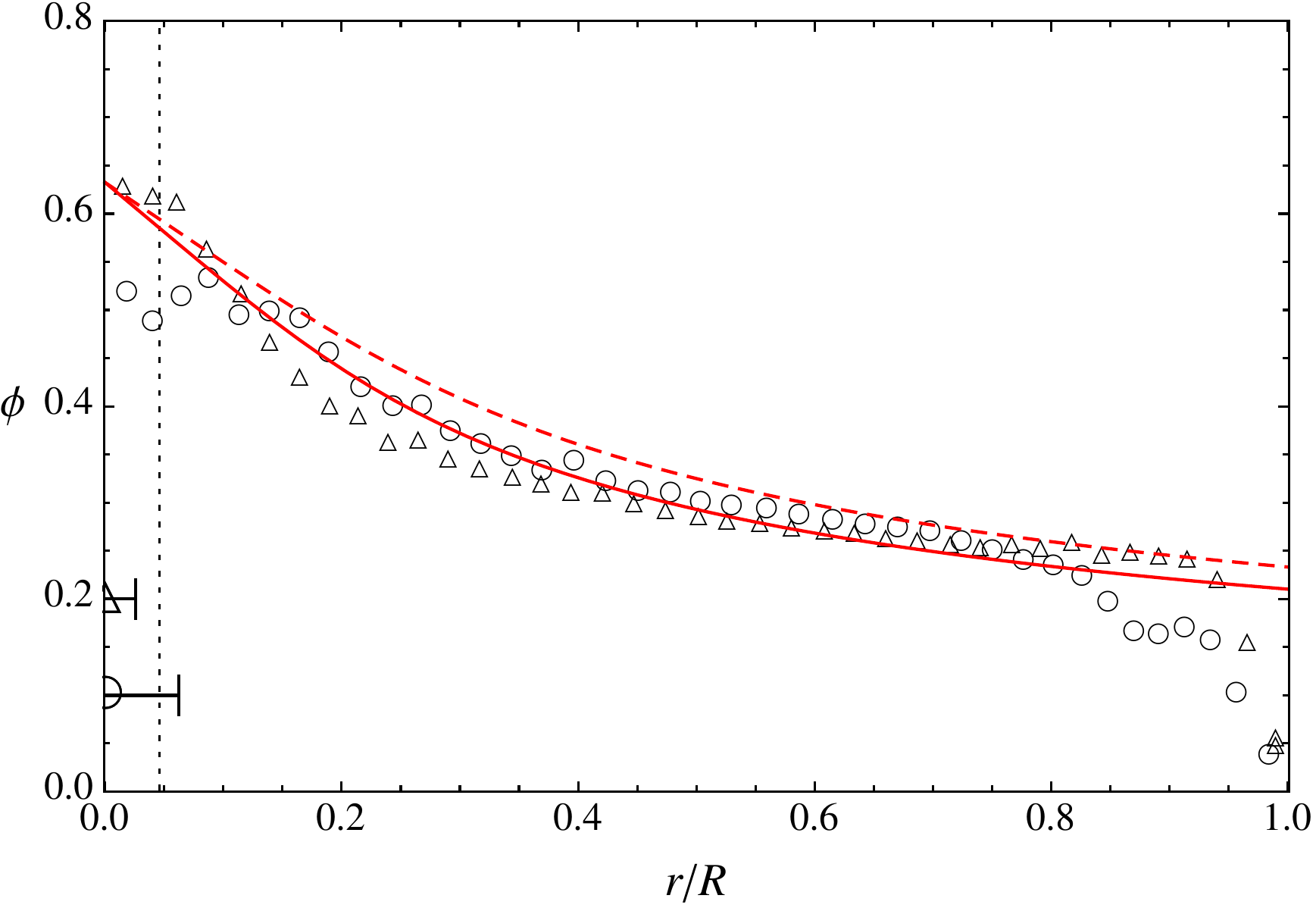}\\
\includegraphics[scale=0.36]{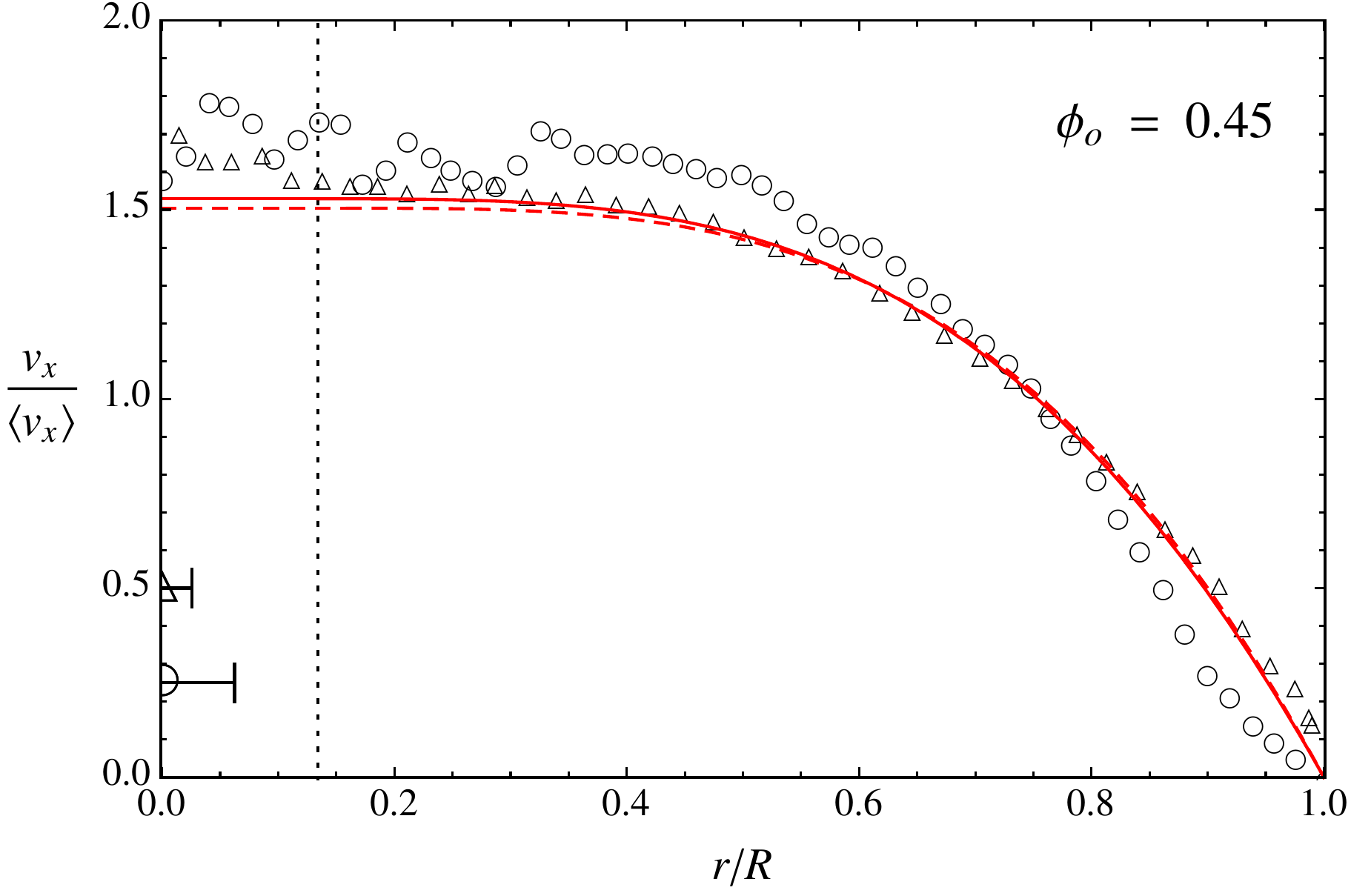}\includegraphics[scale=0.37]{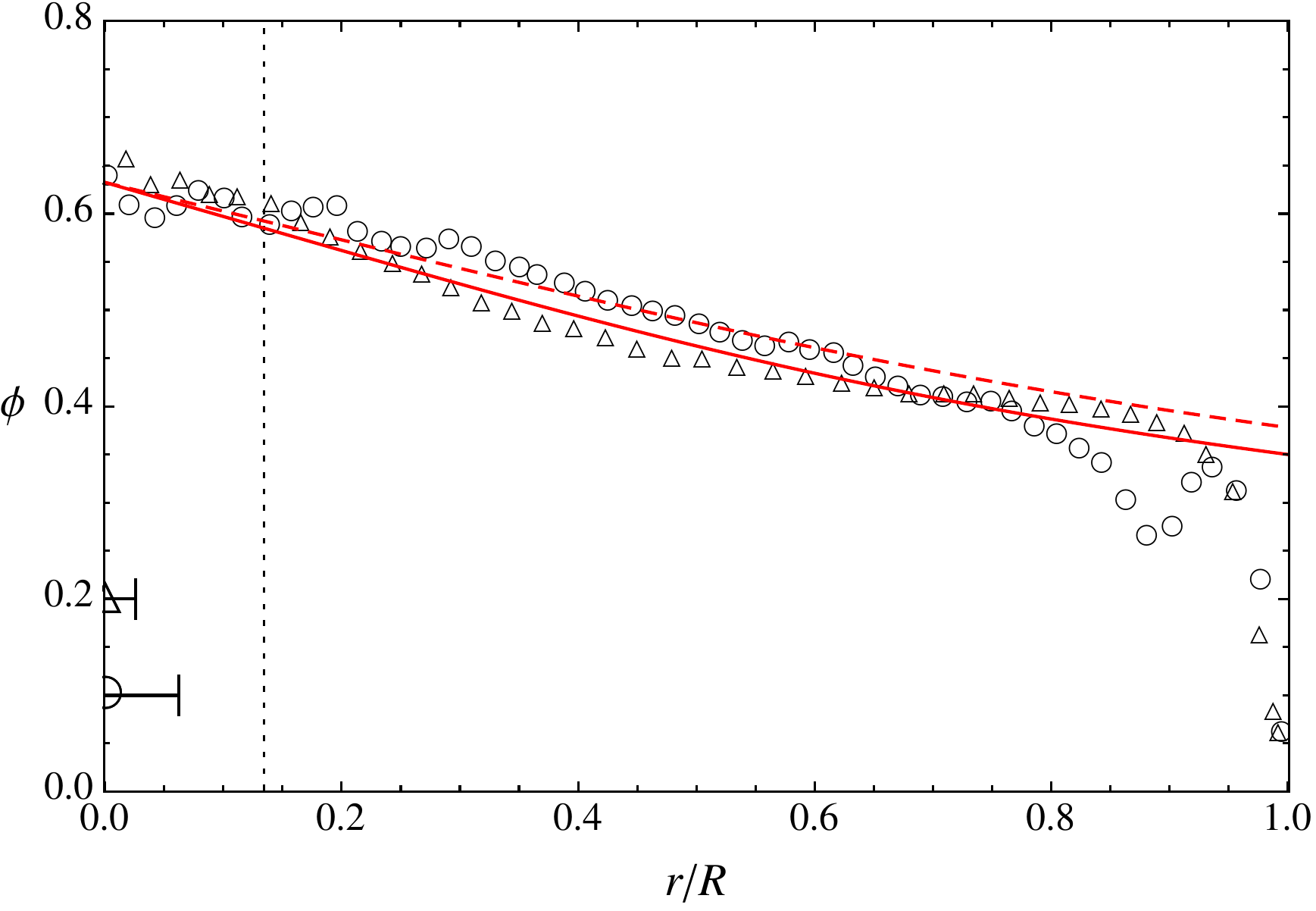}
\par\end{centering}

\caption{Comparison of the theoretical predictions (solid lines) of the scaled
velocity (left) and solid volume fraction (right) profiles with experiments
of \citet{HaMa97} for pipe flow of suspensions with $\phi_{o}=0.2$
(top), $\phi_{o}=0.3$ (middle), and $\phi_{o}=0.45$ (bottom) and
two different particle-to-pipe radius ratios, $a/R$ = $1/39\,(\vartriangle)$
and $1/16$ ($\mbox{\Large\ensuremath{\circ}}$). Bars show scaled
values of the particle radius in the experiments. Dotted lines show
the predicted plug boundary. Theoretical values of the diluted, gap-averaged
particle concentration are $\left\langle \phi\right\rangle =0.179$,
0.272, and 0.425 for $\phi_{o}=0.2$, 0.3, and 0.45, respectively
(Table \ref{tab:pipe_experiment}). Theoretical predictions which
neglect dilution are also shown (dashed lines) for comparison.\label{fig:Pipe-Ha-Phi}}
\end{figure}

Velocity and solid volume fraction profiles for three different values
of the entrance solid volume fraction ($\phi_{o}=0.2,\,0.3$, and
$0.45$) are compared in figure \ref{fig:Pipe-Ha-Phi} to the theoretical
predictions based on the frictional rheology. Scaled radii of small
and large particles are also indicated.

Similar to the comparisons drawn for channel flow in the preceding
section, the theoretical velocity profiles for pipe flow are in excellent
agreement with the experimental results, especially so for the suspensions
with the smaller particle size ($\vartriangle$). The only notable
discrepancy between experimental and theoretical velocity profiles
is observed for the suspension with the larger particle size ($\mbox{\Large\ensuremath{\circ}}$)
at the highest solid volume fraction ($\phi_{o}=0.45$) studied experimentally.
This may have resulted from partial crystallization of the mono-dispersed
suspension in the high concentration flow (as measured $\phi$-values
in this case slightly exceed the theoretical ones in the bulk of the
flow, see bottom plot of figure \ref{fig:Pipe-Ha-Phi}). 

The experimental solid volume fraction profiles compare well with
the theoretically predicted ones with the exception of a particle-size
boundary layer at the pipe's wall, where experimental values are lower.
Gap-averaging of the fully-developed experimental profiles indicates
dilution from the corresponding entrance values, and compares well
to the predicted values of $\left\langle \phi\right\rangle $ (Table
\ref{tab:pipe_experiment}).

Similarly to our observations for channel flow, measured $\phi$-values
near the axis of the pipe agree very well with the predicted values
when the (predicted) plug diameter exceeds the particle size, $r_{plug}>a$,
(see the cases with the smaller particle size for $\phi_{o}=0.2$
and $\phi_{o}=0.3$, and the case with either large or small particles
for $\phi_{o}=0.45$, figure \ref{fig:Pipe-Ha-Phi}). When the predicted
plug size is smaller than the particle size ($r_{plug}<a$), the experimental
$\phi$-profile is flattened (compared to the theoretical prediction)
over a particle-sized region which embeds the predicted plug (see
the cases with the larger particles size for $\phi_{o}=0.2$ and $\phi_{o}=0.3$,
figure \ref{fig:Pipe-Ha-Phi}).

It is also interesting to observe an approximate linearity of the
solid volume fraction profile within the central plug when the latter
is resolved on a particle scale ($r_{plug}>a$), e.g. for all cases
with the smaller particles (figure \ref{fig:Pipe-Ha-Phi}). Such a
linear compaction  is captured by the proposed rheology, which extends
the linear relation between the solid volume fraction and the stress
ratio from the dense flowing regime into the fully-jammed state.

\section{Axial flow development\label{sec:Axial-Flow-Developement}}

We now examine the axial development of the flow from the inlet of
the channel toward its fully-developed state.

\subsection{Numerical Solution}

Setting $\delta=a/H$ to zero, the solid continuity equation (\ref{eq:SolidContinuity_Adim})
together with the expression for the cross-component of the relative
flux (\ref{qy}) become
\begin{equation}
\frac{1}{\phi}\frac{d\phi}{dt}=\frac{\partial q_{y}}{\partial y},\qquad q_{y}{\normalcolor =-\kappa(\phi)\frac{\partial\sigma'_{yy}}{\partial y}}\label{diff}
\end{equation}
where $d/dt=v_{x}\partial/\partial x+v_{y}\partial/\partial y$ for
a steady flow ($\partial/\partial t=0$). The solid volume change
can be related to that of the stress ratio $\mu=-\tau/\sigma_{n}'$
via the inelastic ``compressibility'' $d\phi/d\mu$, defined by
a unique function of the stress ratio across jammed ($0<\mu<\mu_{1}$)
and flowing ($\mu>\mu_{1}$) states of the suspension (figure \ref{fig:dPhi}).
This leads to a specialization of (\ref{diff}), which can be viewed
as a non-linear consolidation equation in terms of the particle normal
stress $\sigma'_{n}$,  
\begin{equation}
\frac{S(\phi)}{\sigma'_{n}}\frac{d\sigma'_{n}}{dt}-\frac{S(\phi)}{\tau}\frac{d\tau}{dt}=\frac{\partial q_{y}}{\partial y},
\qquad q_{y}=-\kappa(\phi)\frac{\partial\sigma'_{n}}{\partial y},\label{diff1}
\end{equation}
where 
\begin{equation}
S(\phi)=-\frac{\mu}{\phi}\frac{\mbox{d}\phi}{\mbox{d}\mu}>0\label{S}
\end{equation}
is the inelastic storage coefficient. 

The reduced ($\delta=0$) form of the mixture continuity equation
(\ref{eq:MixtureContinuity_Adim}) is: 
\begin{equation}
\frac{\partial v_{x}}{\partial x}+\frac{\partial(v_{y}+q_{y})}{\partial y}=0\label{mix1}
\end{equation}

Continuity equations (\ref{diff1}-\ref{mix1}), momentum balance,
$\tau=|\nabla p|y$ (equation (\ref{mom})), and the expression 
\begin{equation}
\frac{\partial v_{x}}{\partial y}=\sigma_{n}'\,\mathcal{I}(-\tau/\sigma_{n}'),\label{rheo}
\end{equation}
where $\mathcal{I}(\mu)$ is the rheological dependence of $I(\phi)$
(equation (\ref{I_alt})) on $\mu(\phi)$ (equation (\ref{mu_alt})),
are solved numerically together with the boundary conditions (\ref{bc}-\ref{bc0})
for the axial development of the unknown particle normal stress $\sigma_{n}'(x,y)$,
particle velocity components $v_{x}(x,y)$ and $v_{y}(x,y)$, and
the total pressure gradient $|\nabla p(x)|$ with $x\ge0$ and $0\le y\le1$
(using the channel symmetry).

We start with formulating the entrance conditions ($x=0$) for the
unknowns. For a uniform particle concentration profile at the flow
entrance, $\phi(x=0,y)=\phi_{o}$, we have for the particle stress
and velocity there 
\begin{equation}
{\normalcolor -\sigma_{n}'=\frac{\left|\nabla p_{o}\right|y}{\mu_{o}},\quad}v_{x}=-\int_{y}^{1}\sigma_{n}'I_{o}dy=\left|\nabla p_{o}\right|\frac{I_{o}}{\mu_{o}}\frac{1-y^{2}}{2}\quad\text{at}\quad x=0,\label{ini}
\end{equation}
respectively, where $\mu_{o}=\mu(\phi_{o})$ and $I_{o}=I(\phi_{o})$
are the corresponding rheological values of $\mu$ and $I$. Given
the gap-average value $v_{o}=1$ of the entrance velocity, we find
the entrance value of the mean stress gradient to be
\begin{equation}
\left|\nabla p_{o}\right|=\frac{3\mu_{o}}{I_{o}}\label{ini1}
\end{equation}

We adopt the following iterative approach to the numerical solution. 
\begin{itemize}
\item At the start (the zeroth iteration), all unknowns are assigned to
their entrance values (\ref{ini}-\ref{ini1}), i.e. $v_{x}^{(0)}(x,y)\equiv v_{x}(x=0,y)$,
$\sigma_{n}'^{(0)}(x,y)\equiv\sigma_{n}'(x=0,y)$, etc, except for
the mean stress gradient, which initial guess is assigned to vary
smoothly along the channel from the entrance value $\left|\nabla p_{o}\right|$,
(\ref{ini1}), to the fully-developed value $\left|\nabla p_{\infty}\right|=1/\left\langle h\right\rangle $
(Section \ref{Sec:dev_flow}) at the end $x=x_{\text{end}}$ of the
computational interval. The latter is chosen to be a finite multiple
of our estimate of the flow development length, as discussed below
in Section \ref{Sec:dev_length}. The partial differential equation
(\ref{diff1}) is then solved for the 1st iteration of the particle
stress, $\sigma_{n}'^{(1)}(x,y)$, using the method of lines PDE solver
(Mathematica, ver. 9). Corresponding numerical error can be assessed
from contrasting the left and right hand sides of the consolidation
equation (\ref{diff1}), which can be evaluated from the obtained
solution at various channel cross-sections (see figure 1 of the supplementary
materials). The error is generally not discernible across the entire
channel width with the exception of a one discretization step thick
($\sim0.01$ for solutions reported here) region at the channel center.
\item The 1st iteration of the mean stress gradient, $|\nabla p^{(1)}(x)|$,
is then found from applying the global continuity condition $\left\langle v_{x}\right\rangle =1$,
where the gap-averaged velocity can be evaluated with the help of
(\ref{rheo}) as 
\begin{equation}
\left\langle v_{x}\right\rangle =-\int_{0}^{1}\sigma_{n}'\,\mathcal{I}\left(\frac{|\nabla p|\, y}{-\sigma_{n}'}\right)y\,\text{d}y\label{<v>}
\end{equation}
In order to solve the resulting integral equation for $|\nabla p|$
we use a sub-iterative procedure
\begin{equation}
|\nabla p_{\text{(next)}}|=\frac{1}{\left\langle v_{x}\right\rangle _{\text{(prev.)}}}|{\normalcolor \nabla p_{\text{(prev.)}}|}\label{sub}
\end{equation}
where ``next'' and ``prev.'' refer to successive sub-iterations
on $|\nabla p|$, and $\left\langle v_{x}\right\rangle _{\text{(prev.)}}$
is given by (\ref{<v>}) evaluated at $|\nabla p|=|\nabla p_{\text{(prev.)}}|$.
\item Once $|\nabla p^{(1)}|$, $\tau^{(1)}=|\nabla p^{(1)}{|}y$
and $\mu^{(1)}=-\tau^{(1)}/\sigma_{n}'$ are at hand, we recover the
1st iterations of the particle concentration, $\phi^{(1)}(x,y)$ from
the rheological relations (\ref{mu_alt})-(\ref{I_alt}), and of the
axial velocity $v_{x}^{(1)}(x,y)$ (by integrating (\ref{rheo})),
respectively. The 1st iteration of the cross-velocity $v_{y}^{(1)}(x,y)$
follows from integrating in $y$ the mixture continuity (\ref{mix1}),
where the relative cross-flux is given by $q_{y}^{(1)}=-\kappa(\phi^{(1)})(\partial\sigma_{n}'^{(1)}/\partial y)$. 
\end{itemize}
The above three steps are repeated until the iterations converge.
We assess the convergence using the gap-average of the particle concentration
in the fully-developed part of the flow (at large enough distances
$x$ from the entrance), specifically requiring $|\left\langle \phi\right\rangle ^{(i)}-\left\langle \phi\right\rangle ^{(i-1)}|<10^{-4}\times\phi_{o}$
to stop the iterations.  We found that using a weighted average between
the last two iterations (e.g., $0.75v_{x}^{(i-1)}+0.25v_{x}^{(i-2)}$)
to compute the next ($i\mbox{th}$) iteration improves the iterations
stability and convergence rate. Similarly, in sub-iterative procedure
of step 2, we have settled on a similar ``weighted'' modification
of (\ref{sub}). We found a typical number of iterations required
for the convergence to vary from the minimum of 2 to the maximum of
12-14, with larger numbers corresponding to larger values of the entrance
concentration.

The described numerical method is readily transportable to pipe flow.
In the following, we therefore present results pertaining to both
channel and pipe flow development.

\subsection{Examples}

\begin{figure}[t]
\noindent \begin{centering}
\includegraphics[scale=0.8]{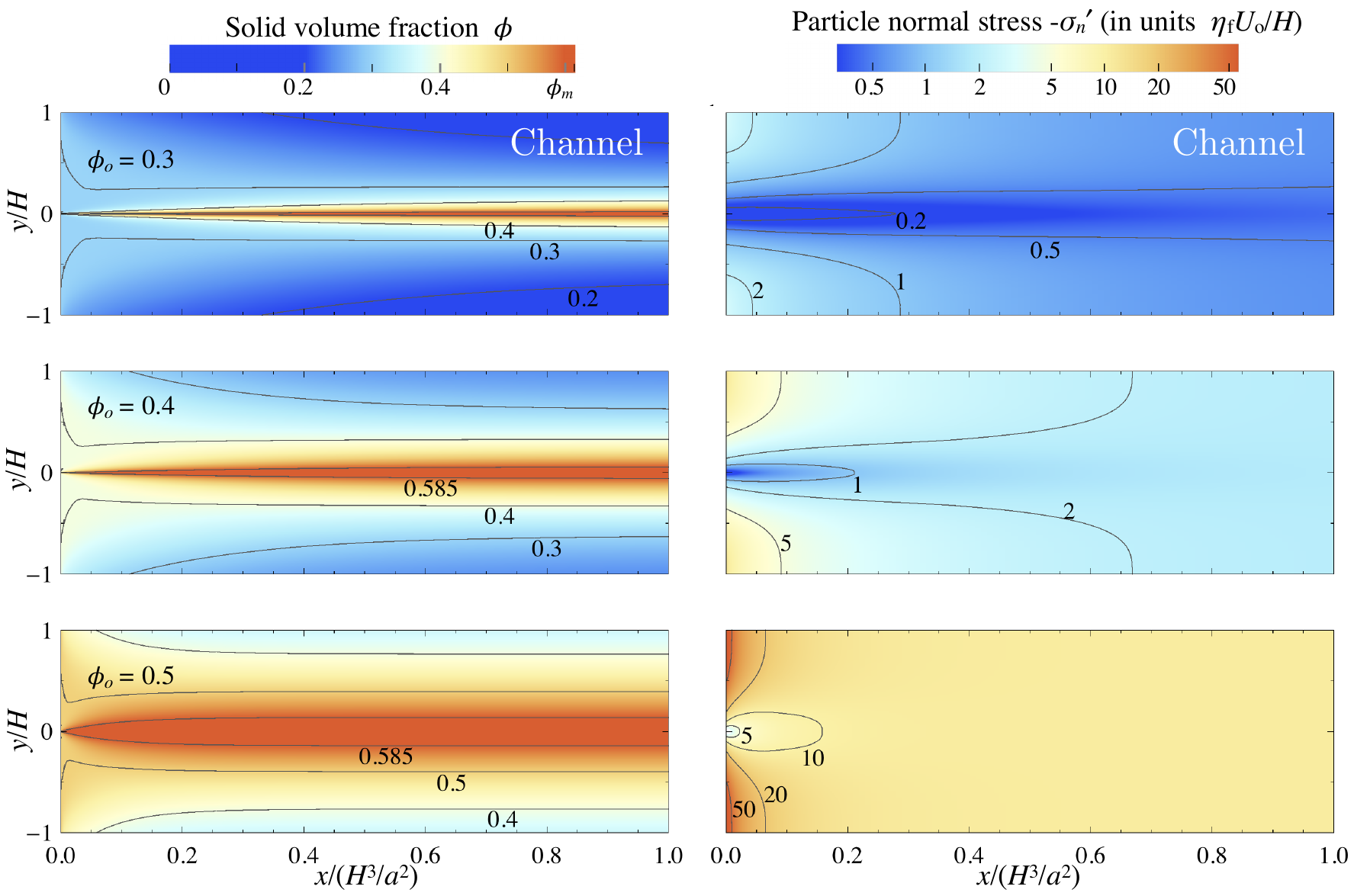}
\par\end{centering}

\caption{Development of the flow in a channel of half-width $H$ for three
different entrance values of the solid volume fraction, $\phi_{o}=0.3,0.4,$
and $0.5$. (left) Evolution of the solid volume fraction. (right)
Evolution of the  particle normal stress $-\sigma_{n}'$
in units of $\eta_{f}U_{o}/H$. \label{fig:dev-slot}}
\end{figure}

\begin{figure}[t]
\noindent \begin{centering}
\includegraphics[scale=0.8]{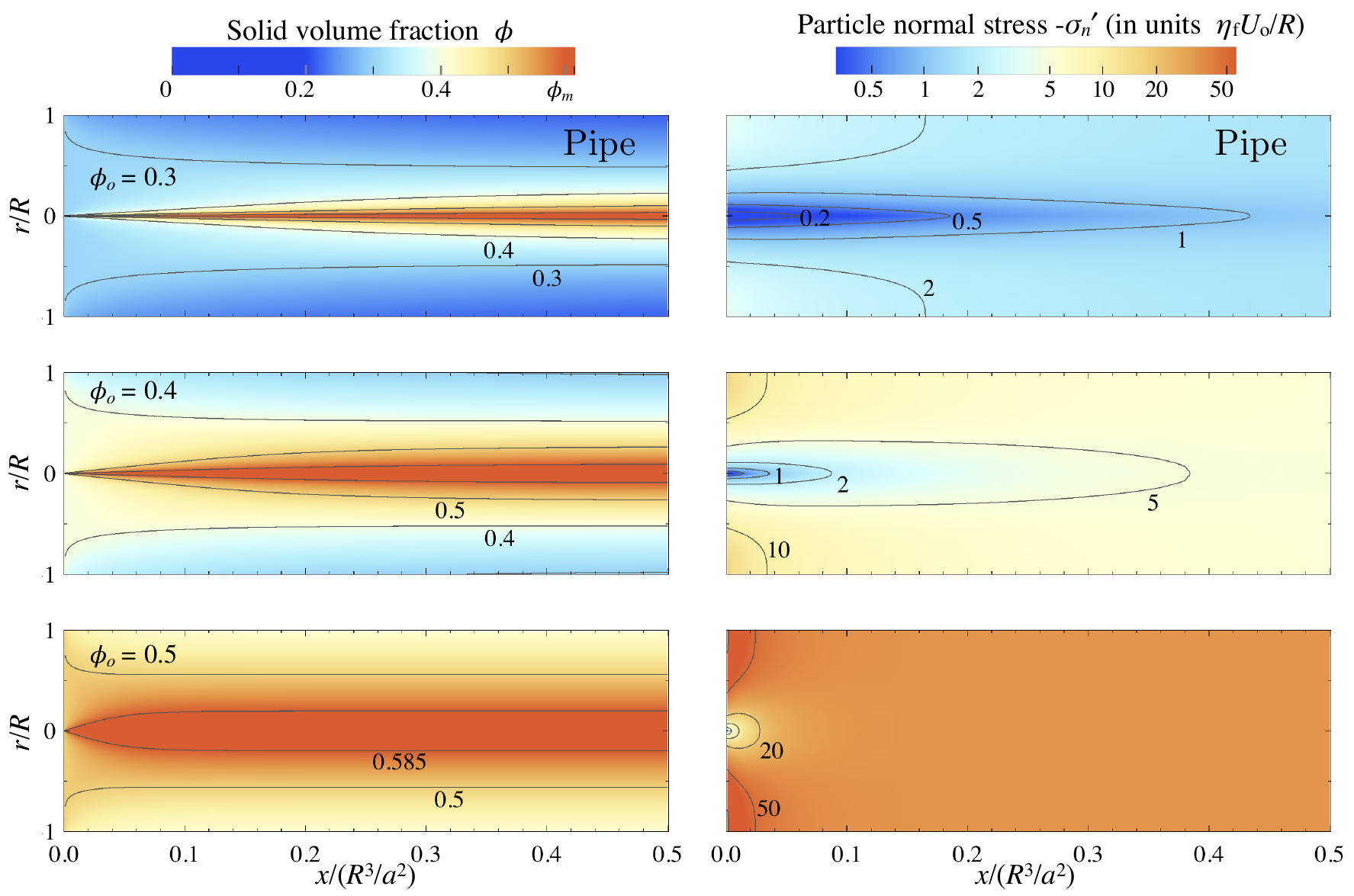}
\par\end{centering}

\caption{Development of the flow in a pipe of radius $R$ for three different
entrance values of the solid volume fraction, as in figure \ref{fig:dev-slot}
for channel flow. \label{fig:dev-pipe}}
\end{figure}

As stipulated earlier, we use the constitutive rheology (\ref{mu_alt})-(\ref{I_alt})
with parameters $\phi_{m}=0.585$, $\mu_{1}=0.3$, and $\beta=0.158$,
and Richardson-Zaki expression for the flow hindrance function with
laboratory-determined exponent $\alpha=5.1$ (i.e. $f(\phi)=(1-\phi)^{5.1})$.
Corresponding numerical solutions for axial development of the solid
volume fraction $\phi$ and the particle normal stress $-\sigma_{n}'$
in a channel and a pipe are visualized on figures \ref{fig:dev-slot}
and \ref{fig:dev-pipe}, respectively, for three different values
of the entrance particle concentration ($\phi_{o}=0.3,$ 0.4, and
0.5). Corresponding profiles of $v_{x}$, $\phi$, and $-\sigma_{n}'$
in a number of flow cross-sections ($x=10^{-3},$ 0.01, 0.1, and 1)
are shown on figures 2, 3, and 4 of the supplementary materials. 


\begin{figure}
\noindent \begin{centering}
\includegraphics[scale=0.55]{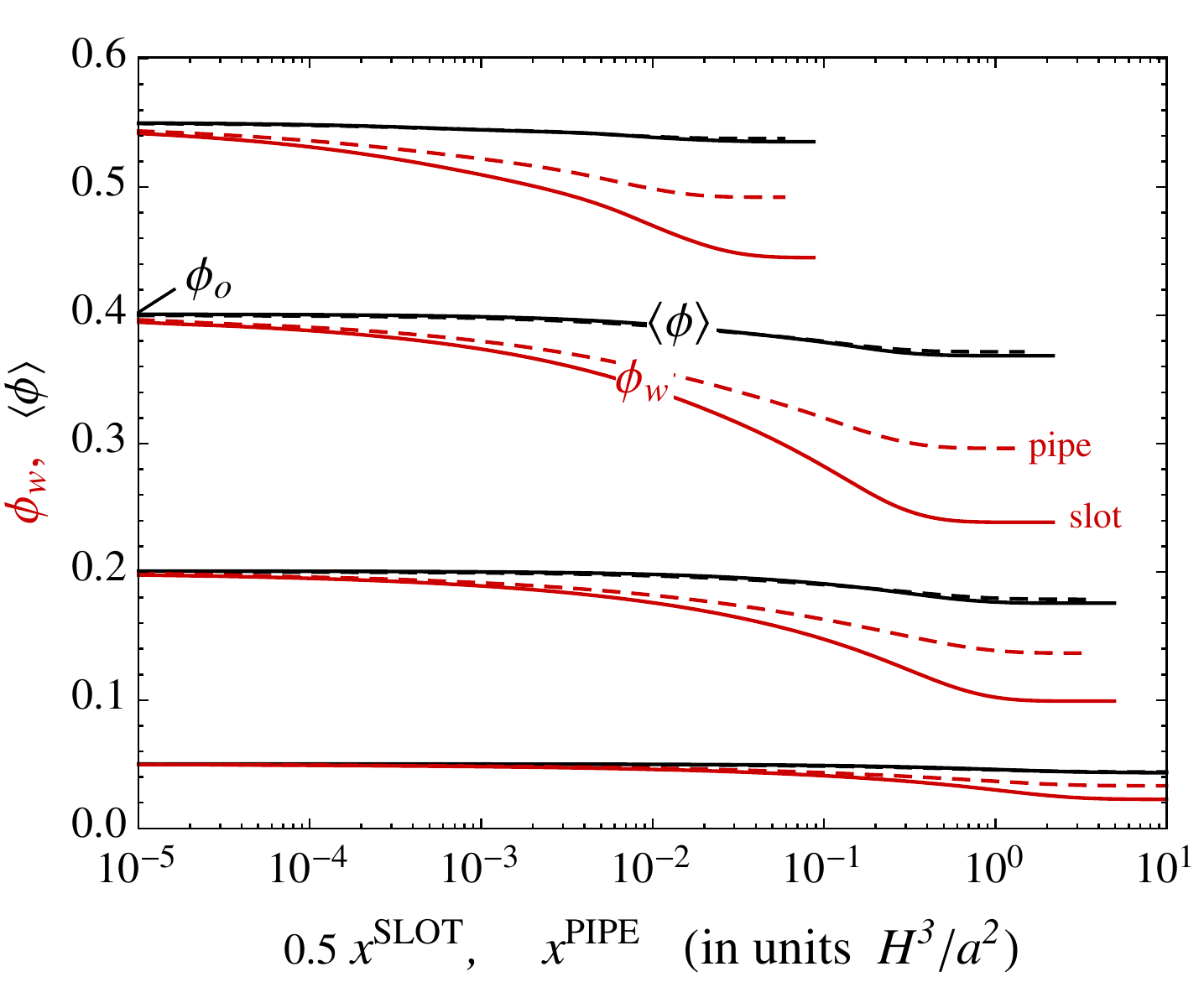}\includegraphics[scale=0.55]{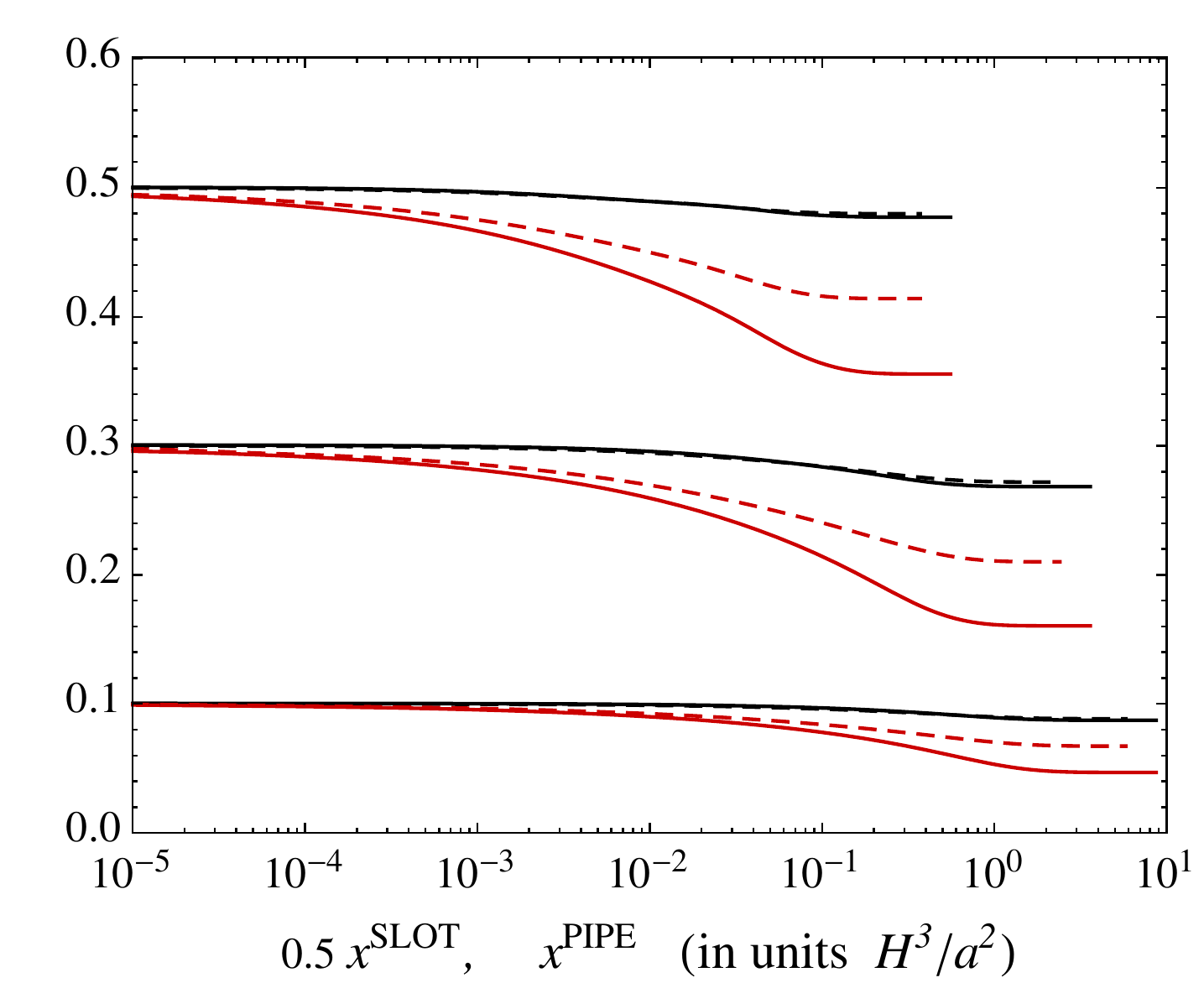}
\par\end{centering}

\caption{Development of the wall and gap-average values of the solid volume
fraction with axial distance from the entrance in the channel (solid
line) or pipe (dashed line) flow. Distance $x$ is in units of ($H^{3}/a^{2}$
or $R^{3}/a^{2}$), and $x^{\text{SLOT}}$ is scaled by prefactor
1/2. It is evident that the particle concentration in the pipe flow
with radius $R$ develops approximately twice as fast as it does in
the slot with half width $H=R$ for identical particle size $a$.
Results are for values of entrance concentration $\phi_{o}=$0.05,
0.2, 0.4, 0.55 (left) and $\phi_{o}=$0.1, 0.3, 0.5 (right), corresponding
to the intercepts with the $\phi$- axis. \label{fig:dev-phi}}
\end{figure}
\begin{figure}
\begin{centering}
\includegraphics[]{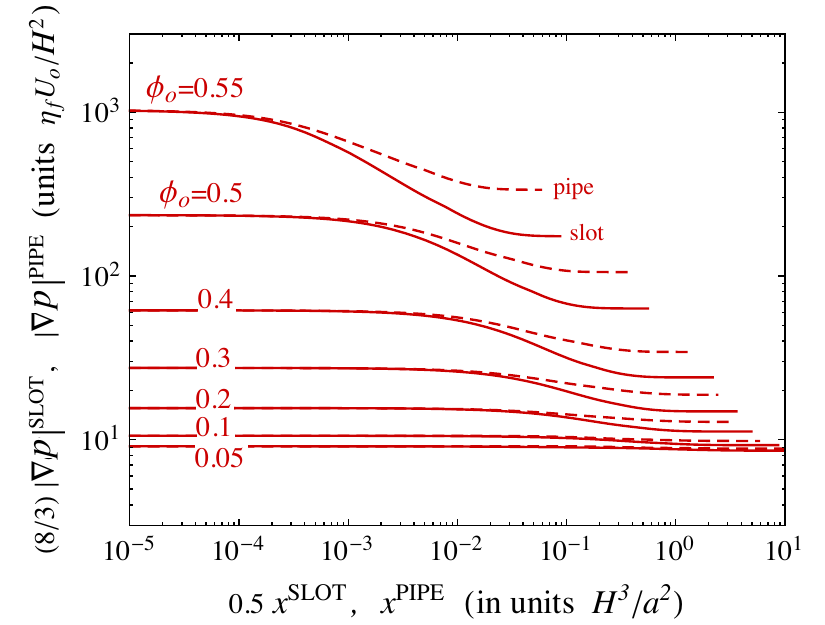}
\par\end{centering}

\caption{Development of the total pressure gradient in units of $\eta_{f}U_{o}/H^{2}$
or $\eta_{f}U_{o}/R^{2}$ with axial distance from the entrance in
the channel (solid line) or pipe (dashed line) flow. The total pressure
gradient in the channel $|\nabla p|^{\text{SLOT}}$
is scaled by prefactor $8/3$, such that $(8/3)|\nabla p|^{\text{SLOT}}=|  \nabla p |^{\text{PIPE}}$
at the channel/pipe entrance when $H=R$. Distance $x$ is in units
of $H^{3}/a^{2}$ or $R^{3}/a^{2}$, and $x^{\text{SLOT}}$ is scaled
by prefactor 1/2. \label{fig:dev-stress}}
\end{figure}

Figure \ref{fig:dev-phi} shows development of the solid volume fraction,
the gap-averaged and wall values thereof, for various entrance conditions,
and figure \ref{fig:dev-stress} shows similar development of the
mixture pressure gradient driving the suspension flow. Flow in a pipe
is seen to develop significantly faster than flow in a channel of
equivalent half-width (i.e. with $H=R$). This observation is further
quantified below.

\subsection{Development length\label{Sec:dev_length}}

We expect the \emph{normalized} axial flow development length to scale
with the gap-averaged normalized axial velocity ($v_{o}=1$) divided
by the normalized diffusivity coefficient $D$ (equation (\ref{diff1})),
e.g.,
\[
L_{\text{dev}}=\frac{1}{4\left\langle D\right\rangle },\qquad D=\frac{{ (-\sigma_{n}')}\kappa(\phi)}{S(\phi)}
\]
To specialize the latter, we choose to evaluate $\left\langle D\right\rangle $
at the channel entrance, where solid volume fraction is uniform and
the corresponding expressions for the particle normal stress and the
axial velocity are given by (\ref{ini}) and (\ref{ini1}), yielding
the normalized development lengthscale expression 
\begin{equation}
L_{\text{dev}}=\frac{1}{6}\frac{I(\phi_{o})S(\phi_{o})}{\kappa(\phi_{o})}\label{Ldev}
\end{equation}
in terms of the normalized permeability $\kappa(\phi)=2f(\phi)/9\phi$,
inelastic storage coefficient $S(\phi)=-(\mu/\phi)(\mbox{d}\phi/\mbox{d}\mu)$
(equation (\ref{S})), and the viscous number $I(\phi)$. 

Corresponding \emph{dimensional} diffusivity and development lengthscale
can be readily recovered from the normalized expressions above using
units $H^{2}/t_{*}=U_{o}H^{3}/a^{2}$ (diffusivity), $\tau_{*}=\eta_{f}U_{o}/H$
(mean particle stress), and $L=H^{3}/a^{2}$ (axial length), (equations
(\ref{s0}-\ref{s4})):
\begin{equation}
L_{\text{dev}}=U_{o}\frac{H^{2}}{4\left\langle D\right\rangle _{o}}=\frac{H^{3}}{a^{2}}\frac{I(\phi_{o})S(\phi_{o})}{6\,\kappa(\phi_{o})},\qquad D=\frac{a^{2}\kappa(\phi)}{\eta_{f}}\frac{(-\sigma_{n}')}{S(\phi)}\label{Ldev'}
\end{equation}

\begin{figure}
\noindent \begin{centering}
\includegraphics[scale=0.6]{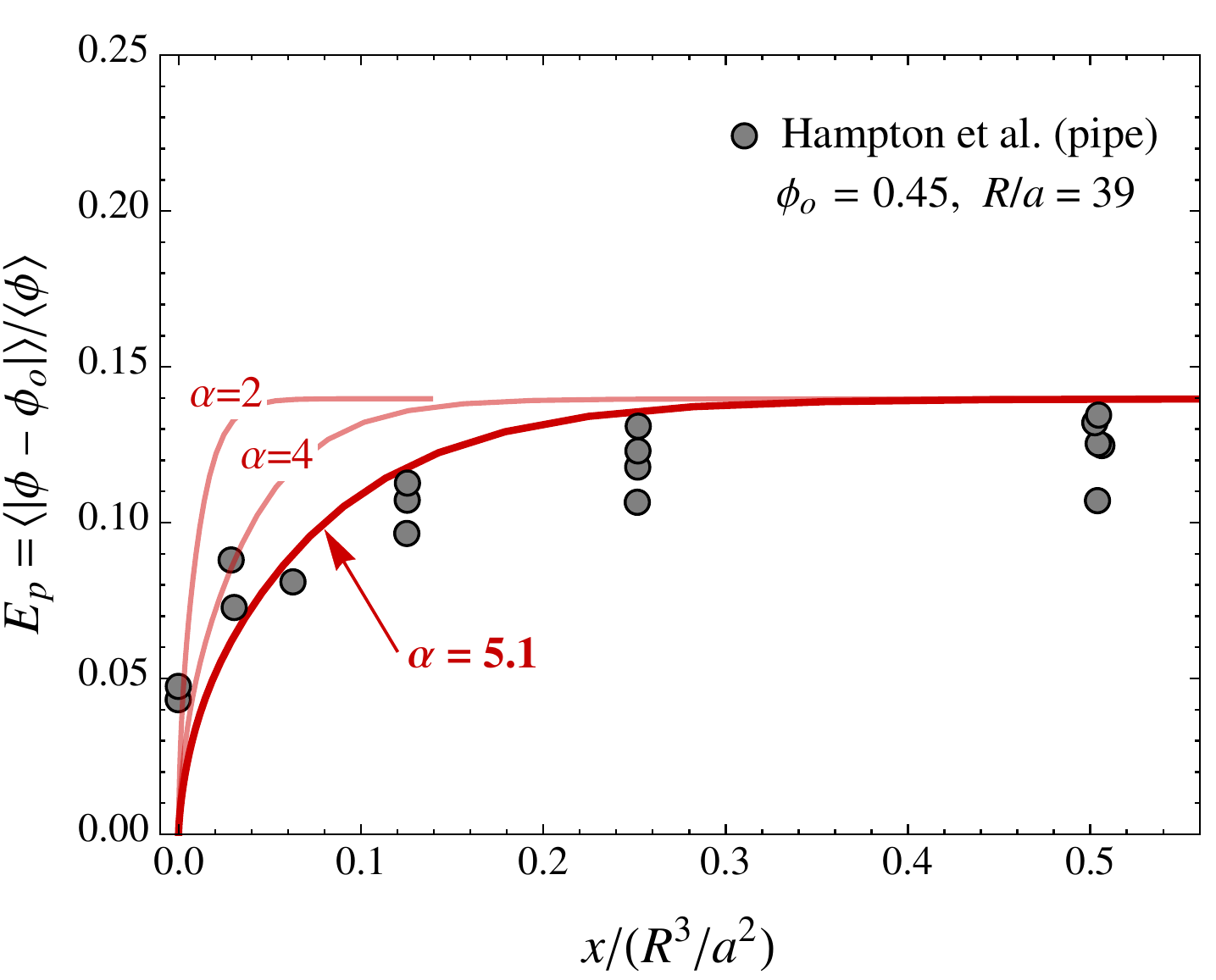}
\par\end{centering}

\caption{Numerical predictions for the evolution of number $E_{p}$ quantifying
the non-uniformity of solid volume fraction profile (maximum $E_{p}$
corresponds to the fully-developed flow) for $\phi_{o}=0.45$, contrasted
with the experimental data of \citet{HaMa97} for the system with
$a/R=1/39$. The solid line corresponds to the predictions for the
Richardson-Zaki permeability function with exponent $\alpha=5.1$
derived from numerous independent sedimentation experiments (see figure
\ref{fig:PermModels}, and \citet{GaAl77,DaAv85}), while opaque lines
show predictions for other values of $\alpha$ used in the previous
numerical modeling of suspension flow development.\label{fig:Ep_dev}}
\end{figure}

To track flow development, we make use of a measure \citep{HaMa97}
\begin{equation}
E_{p}(x)=\left\langle \left|\phi(x,y)-\phi_{o}\right|\right\rangle /\phi_{o}\label{Ep1}
\end{equation}
of the non-uniformity of the solid volume fraction profile, which
varies from zero at the flow entrance to the maximum, fully-developed
value away from the entrance. Figure \ref{fig:Ep_dev} contrasts \citeauthor{HaMa97}'s
measurements of $E_{p}$ for a suspension system with $\phi_{o}=0.45$
at various stages of the flow development to our numerical predictions
for Richardson-Zaki ($\alpha=5.1$) permeability. The match between
the theory and the experiment is remarkable, in that no constitutive
parameters have been adjusted from their values, as determined from
independent sets of rheological experiments. 

\begin{figure}
\noindent (a)\hspace{-1em}\includegraphics[scale=0.55]{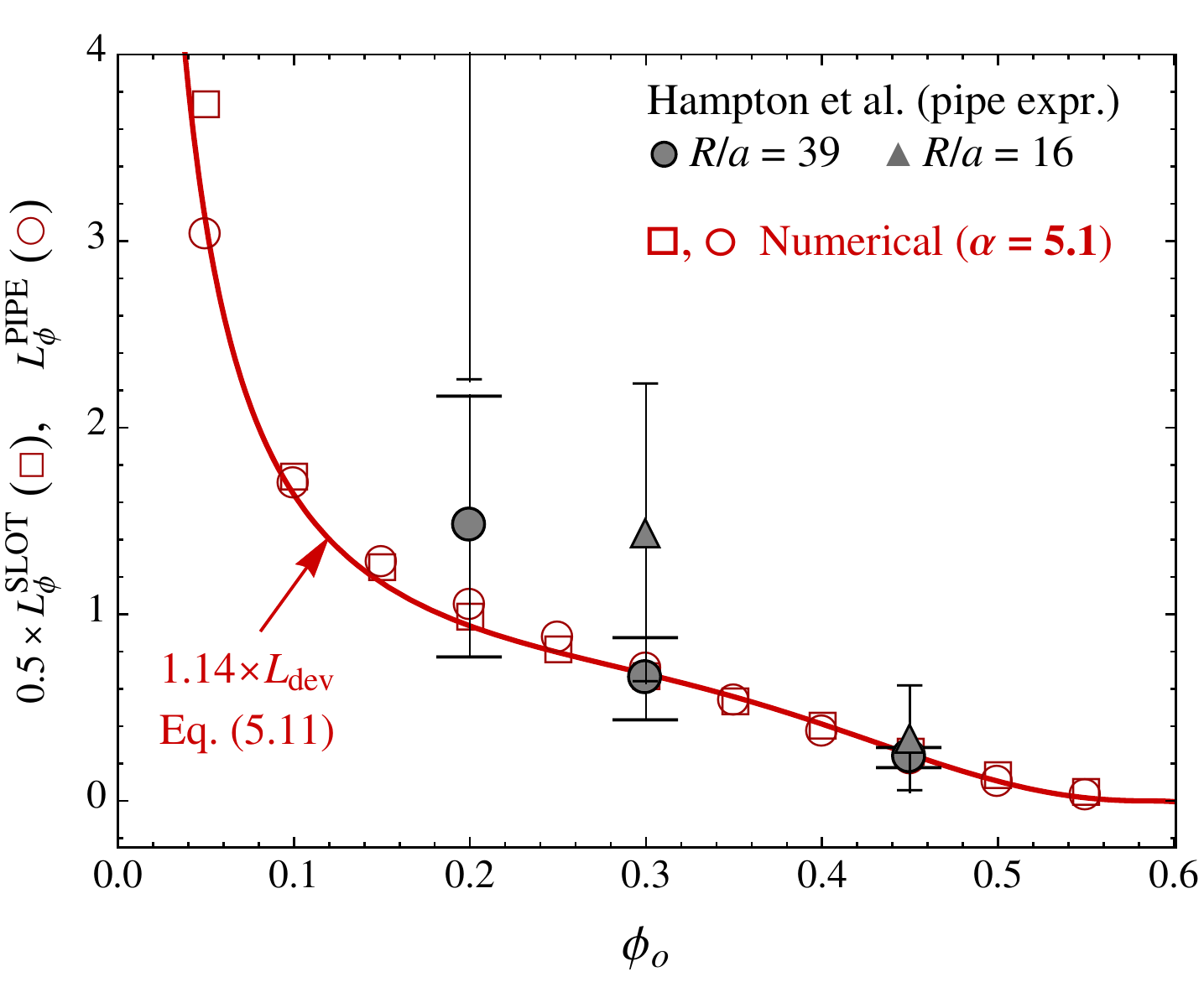}\enskip{}(b)\hspace{-1em}\includegraphics[scale=0.58]{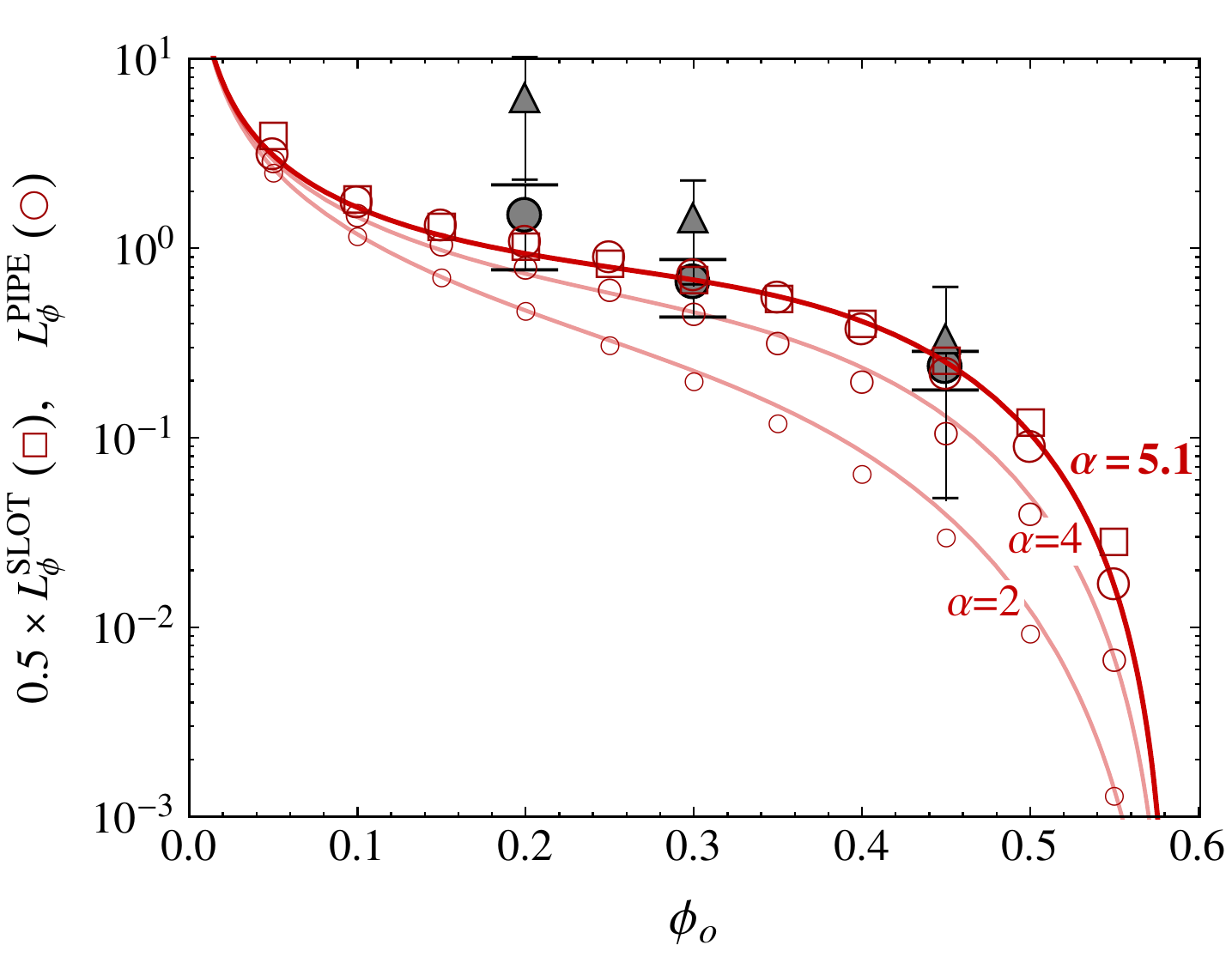}

\caption{(a) Development length for the particle concentration profile, $L_{\phi}$
in units of ($H^{3}/a^{2}$ or $R^{3}/a^{2}$), as a function of entrance
particle concentration $\phi_{o}$. (Richardson-Zaki exponent $\alpha=5.1$.)
The development length for the pipe flow is approximately half of
that for the slot flow. The open symbols show the numerical solutions,
and the solid line shows an approximation based on the apparent diffusivity,
$L_{\phi}^{\text{pipe}}\approx0.5\times L_{\phi}^{\text{slot}}\approx1.136\, L_{dev}(\phi_{o})$
(equation (\ref{Ldev})). The filled circles and triangles show the
data of \citet{HaMa97} for two systems, with smaller $R/a\approx39$
and larger $R/a\approx16$ particles, respectively. (b) Same as (a)
but in log-linear scale, to examine vanishing development length in
the limit of large solid volume fraction. Development length of the
numerical solutions for pipe flow characterized by other (than $\alpha=5.1$)
values of Richardson-Zaki exponent and corresponding approximation
using equation (\ref{Ldev}) are also shown by small open circles
and opaque lines, respectively. \label{fig:dev-len}}
\end{figure}

Now we seek a similar comparison for the development length. Following
\citet{HaMa97} (see also \citet{MiMo06}), we define the 95\% development
length, $L_{\phi}$, for the particle concentration profile as the
minimum distance from the flow entrance where $E_{p}$ is within 5\%
of its fully-developed value. We show on figure \ref{fig:dev-len}a
that, for Richardson-Zaki ($\alpha=5.1$) permeability, the numerical
solutions for the entire range of solid volume fraction are well approximated
by 
\[
L_{\phi}^{\text{pipe}}\approx0.5\times L_{\phi}^{\text{slot}}\approx1.136\, L_{\text{dev}}
\]
 with $L_{dev}$ given by (\ref{Ldev'}) in which $R$ replaces $H$
for the case of a pipe. \citeauthor{HaMa97}'s development data for
pipe flow for the system with a smaller particle size (filled circles)
is in excellent agreement with numerical model predictions. The experimentally
observed development length for suspensions of large particles (filled
triangles) deviates upward of the theoretical prediction whenever
the predicted central plug width is smaller than the particle size
(see figure \ref{fig:Pipe-Ha-Phi} for comparison of predicted values
of plug width to a particle size), or, in other words, when the continuity
approximation fails on the scale of the plug.

Figure \ref{fig:dev-len}b reproduces the results of figure \ref{fig:dev-len}a
but in semi-log scale, which allows to better track vanishing development
length in the dense regime, when the entrance concentration $\phi_{0}$
approaches its maximum flowing value.  Development length predictions
for other values of Richardson-Zaki exponent ($\alpha=2$ and $\alpha=4$),
as used in some previous studies of suspension flow \citep{MoBo99,MiMo06},
are also shown on figure \ref{fig:dev-len}b for comparison. For example,
the predicted development length based on an artificially-high permeability
with $\alpha=2$ is only a small fraction $\approx(1-\phi)^{3.1}$
of the development length based on the experimentally-validated permeability
with $\alpha=5.1$. The underestimation by the former is particularly
severe in the dense regime, e.g. by factor of $\approx9$ for $\phi_{o}=0.5$.

\begin{figure}
\noindent \begin{centering}
\includegraphics[]{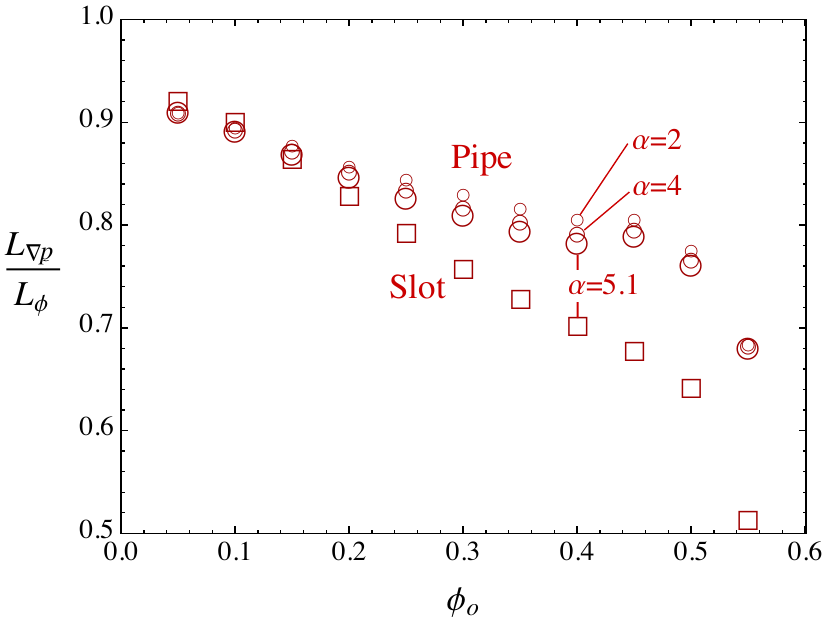}
\par\end{centering}

\caption{The ratio $L_{\nabla p}/L_{\phi}$ of the total pressure gradient
development length and the SVF development length, respectively, as
a function of entrance SVF $\phi_{o}$. \label{fig:dev-len-ratio}}
\end{figure}

Examination of similarly defined 95\% development length $L_{\nabla p}$
for the driving total pressure gradient (shown as a fraction of $L_{\phi}$
on figure \ref{fig:dev-len-ratio}) suggests that the total pressure
gradient develops somewhat faster than the concentration profile,
consistent with the previous prediction with suspension balance models
\citep[e.g.][]{MiMo06}.


\section{Discussion}

The theoretical predictions for the fully-developed flow based on
the frictional suspension rheology are in excellent agreement with
the experimentally measured fully-developed velocity and volume fraction
profiles for both pipe and channel geometries \citep{LyLe98a,HaMa97}.
In particular, the radius / half-width of the central plug is very
well predicted in all cases. These predictions of the experimental
responses are to be compared with the ones obtained on the same set
of experimental data using suspension balance models.  Among those,
\citet{MiMo06} and \citet{FaMa02} predict no discernible plug, which
is likely due to an unrealistically large (in excess of the random
close packing limit) value of $\phi_{m}$ assumed in these studies.
A finite plug, albeit still of a smaller size than that observed in
the experiments of \citet{HaMa97}, is predicted from suspension balance
modeling by \citet{Rama13}, who used a more plausible (below the
random close packing limit) value of $\phi_{m}$.  

Another interesting point that arises from the comparisons of the
predicted and experimentally observed fully-developed profiles relates
to the limit of validity of the continuum (macroscopic) assumption.
As already mentioned in section \ref{sec:Comp}, the experimental
solid volume fraction in the plug are in agreement with the continuum
theory as long as the plug width is larger than at least one particle.
Similarly for the case of the pipe flow (imaged by NMR), the predictions
of the solid volume fraction close to the wall are in good agreement
with experimental values at a distance from the wall larger than one-to-two
particles. In other words, the continuum theory can resolve the flow
relatively accurately at a scale of a single particle. It is clear
that in dilute cases (for low $\phi_{o}$), the size of the plug is
eventually getting smaller than the particle size. In this case, one
would have, therefore, to introduce the latter as an internal lengthscale
in the constitutive description, effectively making it non-local \citep[e.g.,][]{MiMo06}.

It is worthwhile to recall that all of the experiments investigated
here are characterized by a value of the gap width of at least ten
particle wide, and by a plug size of at most 20\% of the gap width
(as in the case of $\phi_{o}=0.5$, see figure \ref{fig:Channel_Phi05}).
From the quality of the agreement obtained between the predictions
based on a local rheology and the experiments, we can conclude that
non-local effects may not be important in the considered cases. These
effects may become important for larger values of the particle-to-gap
or/and plug-to-gap  ratios, i.e. for bulk concentration values $\phi_{o}$
closer to $\phi_{m}$ than those studied experimentally by \citet{LyLe98a}
and \citet{HaMa97}. It would be particularly interesting to further
test our proposed extension of the frictional rheology to the non-flowing
state, in the form of the linear relation between the stress ratio
$\mu$ and the solid volume fraction $\phi$, to cases where $\phi_{o}$
is closer to $\phi_{m}$ (e.g., for $\phi_{o}=0.55$ and higher),
as well as for larger values of the particle-to-gap ratio.

To our knowledge, we predict for the first time the axial development
of the flow (e.g. ``entrance'' length) and the fully-developed flow
measured in pipe and channel experiments \citep{HaMa97,LyLe98a} using
a model which is \emph{devoid of any fitting parameters}. Specifically,
this model is completely defined by the normalized permeability $\kappa(\phi)$
(or hindrance $f(\phi)$), friction $\mu(\phi)$, and viscous number
$I(\phi)$ functions derived from independent experimental data sets:
\citet{GaAl77} and \citet{DaAv85} for the hindrance function, and
\citet{BoGu11} for frictional rheology. In some previous studies,
\citet{MiMo06} used a suspension-balance model with an artificially
large value of the jamming solid volume fraction ($\phi_{m}=0.68$
instead of $\sim0.6$), artificially low value of the near jamming
friction coefficient ($\mu_{1}\approx0.13$ instead of $\sim0.3$)
and an artificially high permeability (corresponding to the hindrance
function of Richardson-Zaki form with an exponent $\alpha=2$ instead
of the experimental value $\sim5$) to solve for axial flow development.
The permeability/hindrance function they have used is $1/(1-\phi)^{3}$
times higher than the experimentally measured one. For example, for
$\phi=0.45$, this permeability exaggeration factor is $6.$ On the
other hand, the inelastic storage factor $d\phi/d\mu$ in their model
is also exaggerated. The two exaggerations partially ``neutralize''
each other in the expression for the diffusivity and therefore in
their predicted development length.

\section{Conclusions}

Using a local continuum formulation based on the frictional constitutive
law similar to the one proposed by \citet{BoGu11}, we have revisited
confined, pressure-driven Stokesian suspension flow in a channel and
a pipe. We have obtained an analytical solution for the fully developed
flow which exhibits the transition from Poiseuille to plug flow with
increasing solid concentration, thanks to the particle pressure dependent
yield stress of the frictional rheology. The theoretical fully-developed
solid volume fraction and velocity profiles agree very well with experimental
data available in the literature for these flows without any adjustment
of the constitutive parameters obtained from independent rheological
experiments in an annular shear cell \citep{BoGu11}. Slight mismatches
of the solid volume fraction profile are observed when the size of
the predicted plug is lower than about one particle diameter, i.e.
when the continuum approximation is breaking down in the jammed part
of the flow. It is particularly striking that a continuum description
can resolve the flow down to the scale of one particle.

A modification of the original constitutive law of \citet{BoGu11}
has been proposed in order to avoid an unphysical behavior of the
plastic compressibility coefficient $\beta=-d\phi/d\mu$ close to
the jamming limit. The proposed modification resolves a slight inconsistency
in \citeauthor{BoGu11}'s formulation by ensuring the dominance of
the contacts over the hydrodynamics contributions to the macroscopic
friction in the dense regime. This modification does not affect the
fully developed solution (figure \ref{fig:Transition-Poiseuille-Plug})
to any significant degree for injected volume fraction lower than
$0.55$. 

We also proposed to extend the linear plastic ``compressibility''
between the solid volume fraction $\phi$ and the stress ratio $\mu$
into the jammed part of the flow (i.e. when $\phi>\phi_{m}$ and $\mu<\mu_{1}$),
and obtained $\phi\approx\phi_{rcp}$ when $\mu\approx0$. This type
of linear compaction in non-flowing regions appears to be present
within the central plug in the existing pipe and channel flow experiments.
It may be enabled by microscopic velocity and particle pressure fluctuations
(similar to tapping or cyclic deformation applied to compacting static
granular packs \citep[e.g.][]{Knight95,Pouliquen03}) originating,
in this case, from the surrounding flowing material. This stress-ratio-dependent
compaction in the \emph{non-flowing} part may be linked, by extension,
to the dilation property of \emph{flowing} (sheared) suspensions and
dry granular materials, where $\phi$ decreases with the increase
in the applied stress ratio $\mu>\mu_{1}$. We conjectured, that this
dilation property is preserved in the non-flowing part, where the
stress ratio is below the flow threshold ($\mu<\mu_{1}$), if an external
energy source, e.g. in the form of velocity/pressure fluctuations
originating from the surrounding flowing material, is present in order
to facilitate microscopic, ``in-cage'' particle rearrangements,
within the jammed pack. It is worthwhile to note that such an extension
of the frictional rheology to the jammed state as well as the solution
framework for the fully-developed flow can be directly transferred
to the case of a dry, frictional granular media.

The compaction in the plug impacts the suspension velocity for values
of the entrance solid volume fraction above $0.55$, allowing flow
of denser suspensions with maximum gap-average exceeding the jamming
threshold $\phi_{m}$, i.e. $\max\left\langle \phi\right\rangle =(\phi_{m}+\phi_{\mathrm{rcp}})/2$
for the channel and $(2\phi_{m}+\phi_{\mathrm{rcp}})/3$ for the pipe
flow. 

The axial development of the flow has been solved numerically. The
entrance length of the flow in channel and pipe is a function of the
fluid permeability / sedimentation hindrance function governing the
relative phase slip and of the compressibility coefficient. Our numerical
results compare well with existing experimental data for pipe flow
when using the accepted values of the parameters of known phenomenological
models (e.g., Richardson-Zaki phenomenology). We notably predict that
the entrance length is longer for dilute suspensions, and about twice
longer in slot compared to pipe flow. Departure from the continuum
assumption in experimental flows with larger particles size (which
we ascertain to be the case when the predicted plug size is smaller
than one particle diameter) is manifested by an increase of the experimentally-observed
development length over the prediction. Another mechanism which may
potentially contribute to longer-than-predicted development length
corresponds to the relaxation / compaction timescale for the jammed
plug (which can be related to the number and magnitude of velocity
fluctuations required to compact the plug, analogous to the number
and magnitude of taps required to relax a static granular pack).

More experimental investigations are needed in order to further test
the continuum description of these confined flows. In particular,
the case of a higher entrance solid volume fraction $(\phi_{o}>0.55)$
needs to be investigated experimentally in order to further check
whether the proposed extension of the plastic compressibility in the
jammed state $(\phi>\phi_{m})$ is indeed relevant. Finally, the regime
of a larger ratio of the particle size over the channel width, $a/H$,
which is relevant for some applications, also needs to be further
addressed both experimentally and theoretically.

\paragraph*{Acknowledgments}

We would like to thank Schlumberger for support to D.G. and for the
permission to publish this work. 


\appendix

\section{Fully-developed flow in a pipe \label{sec:Pipe-Geometry}}

\subsection{Scaling and normalized solution}

There is no particular difficulty in extending the solution for channel
flow to the pipe geometry. In particular, the zero-order problem is
very similar. We consider a pipe of radius $R$ and characteristic
axial length $L$. We are interested in the case where $\delta=R/L$
is small. As before, the coordinates in the $x$ and $r$ direction
are scaled with respect to $L$ and $R$ respectively. We assume an
axi-symmetric flow independent of the azimuthal position $\theta$.
The scaling of the velocity and stress component are similar to that
of the channel:
\begin{equation}
t_{*}=\frac{L}{U_{0}}\qquad x_{*}=L\qquad r_{*}=R\qquad(v_{x})_{*}=U_{0}\qquad(v_{r})_{*}=\delta U_{0}\qquad\dot{\gamma}_{*}=\frac{U_{0}}{R}\label{s1-1}
\end{equation}
shear stress ($\tau_{*}$), particle stress ($p_{*}'$), fluid pressure
($p_{*}^{f}$), and mixture pressure ($p_{*}$) scale
\begin{equation}
\tau_{*}=p_{*}'=\frac{\eta_{f}U_{0}}{R}\qquad p_{*}=p_{*}^{f}=\frac{\tau_{*}}{\delta}\label{s2-1}
\end{equation}
and relative phase flux scale
\begin{equation}
q_{*}\equiv\frac{a^{2}}{\eta_{f}}\frac{p_{*}^{f}}{L}=\left(\frac{a}{R}\right)^{2}U_{0}\label{s3-1}
\end{equation}
 It is possible to show that the balance and continuity equations
have similar form as in the channel case, accounting for the proper
differential operator in polar coordinates and assuming negligible
normal stress difference $\sigma_{rr}'-\sigma_{\theta\theta}'\approx0$.
The latter assumption approximately holds for dilute suspensions,
but breaks down for concentrated ones \citep[e.g.][]{Couturier11,ZaHi00}.
The fully-developed solution framework developed here can be extended
to account for the normal stress difference, and will be pursued elsewhere.

The similar arguments as those for the channel geometry can be made
for the axial development length in the pipe flow by looking at the
first order terms in $\delta$, which lead to the choice of development
lengthscale $L=R^{3}/a^{2}$. 

The shear rate is simply $\dot{\gamma}=\left|\partial v_{x}/\partial r\right|$
and the shear stress $\tau=\tau_{xr}$. Following the same method
as for the channel flow, we obtain after use of the symmetry and boundary
conditions:
\begin{eqnarray*}
\tau & = & \frac{1}{2}\left|\frac{\partial p}{\partial x}\right|r\\
p & = & p(x)\qquad p^{f}=p^{f}(x)
\end{eqnarray*}
In the flowing part, we can write ${\tau/(-\sigma'_{n})}=\mu_{\mathrm{w}}r$,
where 
\[
\mu_{\mathrm{w}}=\mu(\phi_{\text{w}})=\frac{1}{2}{\displaystyle {\color{black}\frac{\left|\partial p/\partial x\right|}{{\normalcolor -\sigma'_{n}}}}}
\]
is the wall friction and $\phi_{\text{w}}$ is the wall value of the
solid volume fraction. The friction and solid volume fraction profiles
then follow in the form identical to that of the channel: 
\begin{equation}
\mu(\phi(r))=\mu_{\mathrm{w}}r\label{I(r)}
\end{equation}
The shear rate can then be expressed as 
\[
\frac{\partial v_{x}}{\mbox{\ensuremath{\partial}}r}=-\frac{1}{2}\frac{I(\phi(r))}{\mu_{\mathrm{w}}}\times\frac{\partial p}{\partial x}
\]
which upon integration using the no-slip condition at the wall yields
\begin{equation}
v_{x}(r)=-\frac{h(\phi(r))}{2}\times\frac{\partial p}{\partial x}\label{eq:VelocityProfile-1}
\end{equation}
where the function $h(\phi)$ is exactly the same function as for
the channel flow (see Eq. (\ref{h})).

\subsection{Cross-sectional averages}

The cross-section averages are slightly different than for channel
flow due to the difference in the flow geometry: $\left\langle \cdot\right\rangle =2\int_{0}^{1}(\cdot)r\mbox{ d}r$.
The average velocity becomes:
\[
\left\langle v_{x}\right\rangle =-\frac{\left\langle h\right\rangle }{2}\times\frac{\partial p}{\partial x}
\]
where, similarly to the channel case, integrating separately over
the plug and the flowing part, and using substitution $\text{d}r=\text{d}\mu/\mu_{w}$
in the latter, we can obtain 
\begin{equation}
\left\langle h\right\rangle =\frac{1}{\mu_{\mathrm{w}}^{4}}\int_{\phi_{m}}^{\phi_{\text{w}}}I(\phi)\,\mu^{2}(\phi)\frac{\mbox{d}\mu}{\mbox{d}\phi}\mbox{ d}\phi\label{eq:BigH_Pipe}
\end{equation}
The gap-average of the solid volume fraction is obtained as:
\begin{equation}
\left\langle \phi\right\rangle =\frac{2}{\mu_{\mathrm{w}}^{2}}\int_{\phi_{rcp}}^{\phi_{\mathrm{w}}}\phi\mu(\phi)\frac{\mbox{d}\mu}{\mbox{d}\phi}\mbox{ d}\phi=\frac{\mu_{1}^{2}}{\mu_{\mathrm{w}}^{2}}\left\langle \phi\right\rangle _{\text{plug}}+\frac{2}{\mu_{\mathrm{w}}^{2}}\int_{\phi_{m}}^{\phi_{\mathrm{w}}}\phi\mu(\phi)\frac{\mbox{d}\mu}{\mbox{d}\phi}\mbox{ d}\phi,\qquad\left\langle \phi\right\rangle _{\text{plug}}=\phi_{m}+\frac{\beta\mu_{1}}{3}\label{<phi>-Pipe}
\end{equation}

Finally, we evaluate the entrance concentration as a function of the
wall friction in the fully-developed pipe flow 
\begin{equation}
\phi_{o}=\frac{\left\langle \phi v_{x}\right\rangle }{\left\langle v_{x}\right\rangle }=\frac{2}{\mu_{\text{w}}^{2}\left\langle h\right\rangle }\int_{\phi_{rcp}}^{\phi_{\mathrm{w}}}\phi h(\phi)\mu(\phi)\frac{\mbox{d}\mu}{\mbox{d}\phi}\mbox{ d}\phi\label{ratio-Pipe}
\end{equation}
where $h(\phi)$ and $\left\langle h\right\rangle $ are given by
(\ref{h}) and (\ref{eq:BigH_Pipe}), respectively.
\end{document}